\documentclass[review]{elsarticle}

\usepackage{lineno,hyperref}
\modulolinenumbers[5]
\usepackage[all]{hypcap}

\usepackage{amssymb}
\usepackage[fleqn]{amsmath}		
\usepackage{graphicx}
\usepackage{comment}

\newcommand*\diff{\mathop{}\!\mathrm{d}}

\usepackage{color,soul}

\usepackage[greek,english]{babel}

\usepackage[normalem]{ulem}

\biboptions{numbers,sort&compress}

\usepackage[utf8]{inputenc}
\usepackage{cleveref}
\crefname{section}{§}{§§}
\Crefname{section}{§}{§§}

\usepackage[draft]{todonotes}

\usepackage{adjustbox,lipsum}

\journal{International Journal of Hydrogen Energy}

\begin{document}

\begin{frontmatter}

\title{Multi-Component High Aspect Ratio Turbulent Jets Issuing from Non-Planar Nozzles}

\author[uvic]{Majid Soleimani nia\corref{cor1}}
\ead{majids@uvic.ca}
\author[uvic]{Peter Oshkai}
\author[uvic]{Ned Djilali}

\cortext[cor1]{Corresponding Author}
\address[uvic]{Department of Mechanical Engineering and Institute for Integrated Energy Systems, University of Victoria, PO Box 1700 STN CSC, Victoria, Canada, V8W 2Y2}

\begin{abstract}
Fundamental insight into the physics of buoyant gas dispersion from realistic flow geometries is necessary to accurately predict flow structures associated with hydrogen outflow from accidental leaks and the associated flammability envelope. Using helium as an experimental proxy, turbulent buoyant jets issuing from high-aspect-ratio slots on the side wall of a circular tube were studied experimentally applying simultaneous particle image velocimetry (PIV) and planar laser-induced fluorescence (PLIF) techniques. Two slots with an aspect ratio of 10 were considered in this study.  The effects of buoyancy, asymmetry, jet densities and Reynolds numbers on the resulting flow structure were studied in both vertical and horizontal orientations. Significant discrepancies were found between the evolution of current realistic jets issuing from curved surfaces and those conventional high-aspect-ratio jets originating from flat surfaces. These realistic pipeline leak-representative jets were found to deflect along the jet streamwise axis. It was found that increases in aspect ratio caused a reduction in the angle of deflection, jet centreline decay rates and the width growth on both velocity and scalar fields compared to their non-planar round jet counterparts, most notably in the far field.  
\end{abstract}


\end{frontmatter}



\section{Introduction}
Global reliance on fossil fuels has resulted in unprecedented build up of atmospheric carbon dioxide and global warming. Achieving clean, safe and sustainable energy is key to reducing carbon emissions and mitigating greenhouse effects. Hydrogen, a carbon-free energy vector, is being considered as a clean alternative to traditional hydrocarbon based fuels for transportation and energy storage applications. However, due to a wide range of ignition limits (between 4\% and 75\% by volume) \cite{Lewis1961}, modern safety standards for hydrogen infrastructure must be assured before widespread public use can become possible. It is therefore of paramount interest to gain insight into the physics of hydrogen outflow from accidental leaks and the associated flammability envelope to accurately develop codes and standards. In the current investigation, we attempt to quantify the dispersion of high-aspect-ratio turbulent jets experimentally, as they emerge from a realistic pipeline geometry, using state-of-the-art experimental imaging techniques. It is the extension of previous investigations on non-planar round jets \cite{Soleimaninia2018IJoHE, Soleimaninia2018Ae} from the same pipeline configuration to emphasize on importance of nozzle geometry and aspect ratio on the evolution of realistic jets.

Non-circular jets are found in a wide range of applications in nature and engineering systems. These jets have higher entrainment rates than their axisymmetry round jet counterparts, and as a result, more effective mixing occurs \cite{Gutmark1999ARoFM239}. The enhanced mixing is believed to be associated with the higher degree of three-dimensionality in the coherent structures of the flow, which is attributed either to the non-uniform curvature of the nozzle perimeter, or to the instabilities originating at the sharp perimeter of the nozzle. The three-dimensionality, which is highly sensitive to the initial conditions, becomes the main characteristic of non-circular jet flows, with the asymmetrical streamwise and azimuthal vorticity playing the main role in entraining ambient fluid. As the jet spreads, dynamic deformation of the asymmetric vortices yields a complex topology, with interaction of streamwise and azimuthal vortices and energy transfer between them. This ``axis-switching'' phenomenon has been observed in the evolution of non-circular jets \cite{Gutmark1999ARoFM239,Mi2010FTaC583}, whose cross-section can frequently develop into shapes similar to those of the origin nozzle but with axes sequentially rotated at angles characteristic of the nozzle geometry.   

Among non-circular jet flows, plane jets received extensive investigation in the last couple of decades due to their two-dimensional characteristics, which made measurement and numerical simulations along with statistical analyses much easier  \cite{Everitt1978JoFM563,Gutmark1976JoFM465,Deo2007ETaFS825,Deo2008PoF75108}. It was found that initial conditions organized coherent structures in the near field but their effects were also noticed in the self-similar far field region of the plane jet. It was observed that an increase in the Reynold number ($Re$), resulted in shortening the potential core length and increased the near field velocity spreading rate, while the far field rates of mean velocity decay and spread showed the reverse dependency, and decreased with increasing the $Re$ \cite{Deo2008PoF75108}.

Non-circular three-dimensional jets (i.e. rectangular, elliptic, triangle, and other nozzle shapes) have been studied extensively, both through experiments \cite{Gutmark1987TPoF3448,Quinn1992ETaFS203,Zaman1996A,Mi2010FTaC583} and numerical simulations \cite{Miller1995CF1,Tam1993JoFM425,Grinstein1995PoF1483}. In general, due to the three-dimensional nature of the jet's initial configuration, the near-field decay rates of the mean velocity and turbulence intensity are much greater compared to the axisymmetric jet. In the near field, jets that experience the axis-switching phenomenon exhibited a higher decay rate of the centreline velocity. Regardless of the nozzle shape, a change in the nozzle type (sharp-edged orifice plate (OP) and smooth contraction (SC)), results in shorter potential core length in OP jets compared to SC jets but the nozzle type does not affect the far-field centreline velocity decay rate \cite{Mi2010FTaC583,Quinn2007EJoM-B583,Gutmark1999ARoFM239}. Like other jet flows, the development of non-circular jets is significantly influenced by the jet initial conditions, even in the self-similar far field region. Previous studies of rectangular and elliptic jets found that the distance from the orifice, where axis-switching occurs, increases with the nozzle aspect ratio \cite{Hussain1989JoFM257,Quinn1992ETaFS203}. Mixing in the near field is also enhanced with increasing the nozzle aspect ratio \cite{Quinn1992ETaFS203}.  

It is noteworthy that all aforementioned studies on non-circular jets have been limited to the jet flow emerging through flat surfaces, where the direction of the jet mean flow was aligned with the flow origin. However, in practical engineering applications (i.e. pipe lines or storage facilities), any accidental gas leakage would not be limited to flows through flat surfaces, and leaks through cracks in the side walls of circular pipes or storage tanks should also receive attention. The common belief was that within sufficient distance from the nozzle, non-circular jet evolution follows that of a round axisymmetry jet, only a few studies on high-aspect-ratio jet have proven that this assumption is not necessarily correct. These limited studies \cite{Meares1998,Wakes2002JoHM1}, investigated the effects of the orifice shape and gas pressure on gas dispersion behavior from a failed flange joint, by the means of flow visualization and pressure measurements along the jet centreline. These effects persisted to distances up to 250 slot widths away from the orifice, and the centreline velocity decay rate could not be approximated by a round axisymmetry jet assumption. To explore the realistic gas leakage from a curved surface in more detail, recent studies investigated vertical and horizontal buoyant jet evolutions through round holes from curved surfaces, numerically and experimentally \cite{Soleimaninia2018IJoHE,Soleimaninia2017,Maxwell2017,Soleimaninia2018,Soleimaninia2018Ae}. Through these recent studies, significant discrepancies were found between the evolution of axisymmetric round sharp-edged orifice plate (OP) jets through flat surfaces and those originating from curved surfaces.  Most notably, jet deflection from the jet axis, and asymmetric dispersion patterns were always observed.  Also, in horizontal jets, buoyancy effects were dominated much sooner than expected in axisymmetric round jets. To our knowledge, however, the dispersion of high-aspect-ratio jets from curved surfaces have not yet been investigated, by the means of simultaneous velocity and scalar measurements.

This work was motivated by questions related to the safety and dispersion of combustible gases (hydrogen in particular). Using helium as an experimental proxy, the effects of asymmetry and buoyancy on the evolution of high-aspect-ratio jets issuing from realistic pipeline geometries were investigated experimentally. In addition, jet release experiments were also conducted with air to provide further insight into buoyancy effect on outflow structures. Flow patterns and dispersion of gas through a curved surface originated from a source whose original velocity components were nearly perpendicular to the direction of the ensuing jets were studies. From now on, this jet configuration is referred as a 3D jet. Two slots with the same aspect ratio ($AR=10$) as possible crack geometries were considered in this study, although their configuration was aligned in parallel to and perpendicular with the direction of flow inside the tube. These realistic high-aspect-ratio 3D jets were also compared to their 3D round jet counterparts as well as high-aspect-ratio jets issuing through flat surfaces.  Particle imaging velocimetry (PIV) and planar laser-induced fluorescence (PLIF) were simultaneously used to measure high-resolution instantaneous velocity and concentration fields, respectively. The purpose of this investigation was to identify and characterize departures from non-circular jets emerging through flat surfaces, and to highlight the effects of initial conditions, buoyancy and the asymmetric nature of the 3D jets, which ensued from a practical geometry arrangement.

%
%
\section{Experimental system and techniques}

\begin{figure}
	\centering
	\raggedright \underline{\textbf{Slot 2:}}	\  \ \ \ \ \ \ \ \ \ \ \ \ \ \ \ \ \ \ \ \ \ \ \ \ \ \ \ \ \ \ \ \ \ \ \ \ \ \ \ \underline{\textbf{Slot 3:}}\\
	a)\includegraphics[scale=0.3]{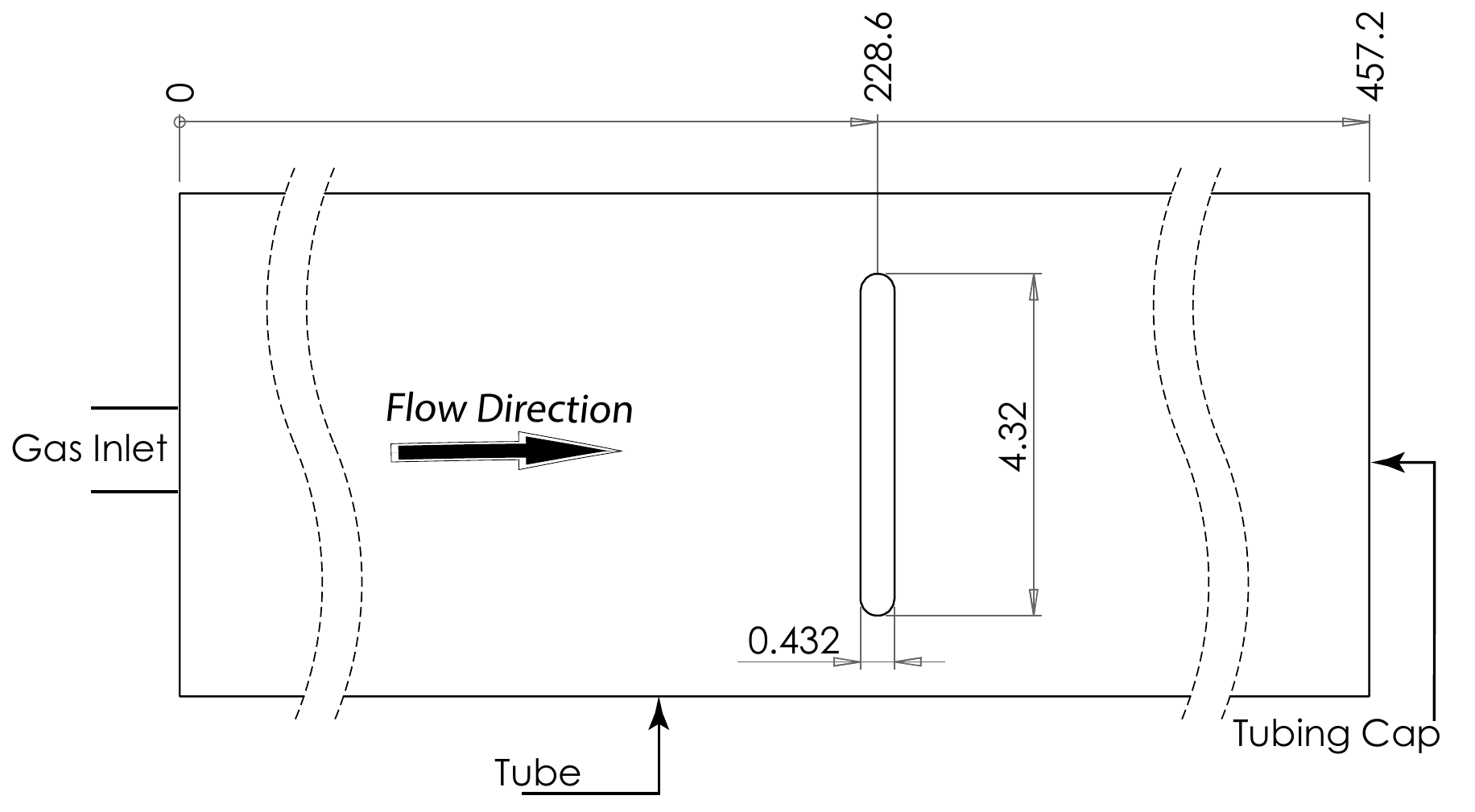}
	b)\includegraphics[scale=0.3]{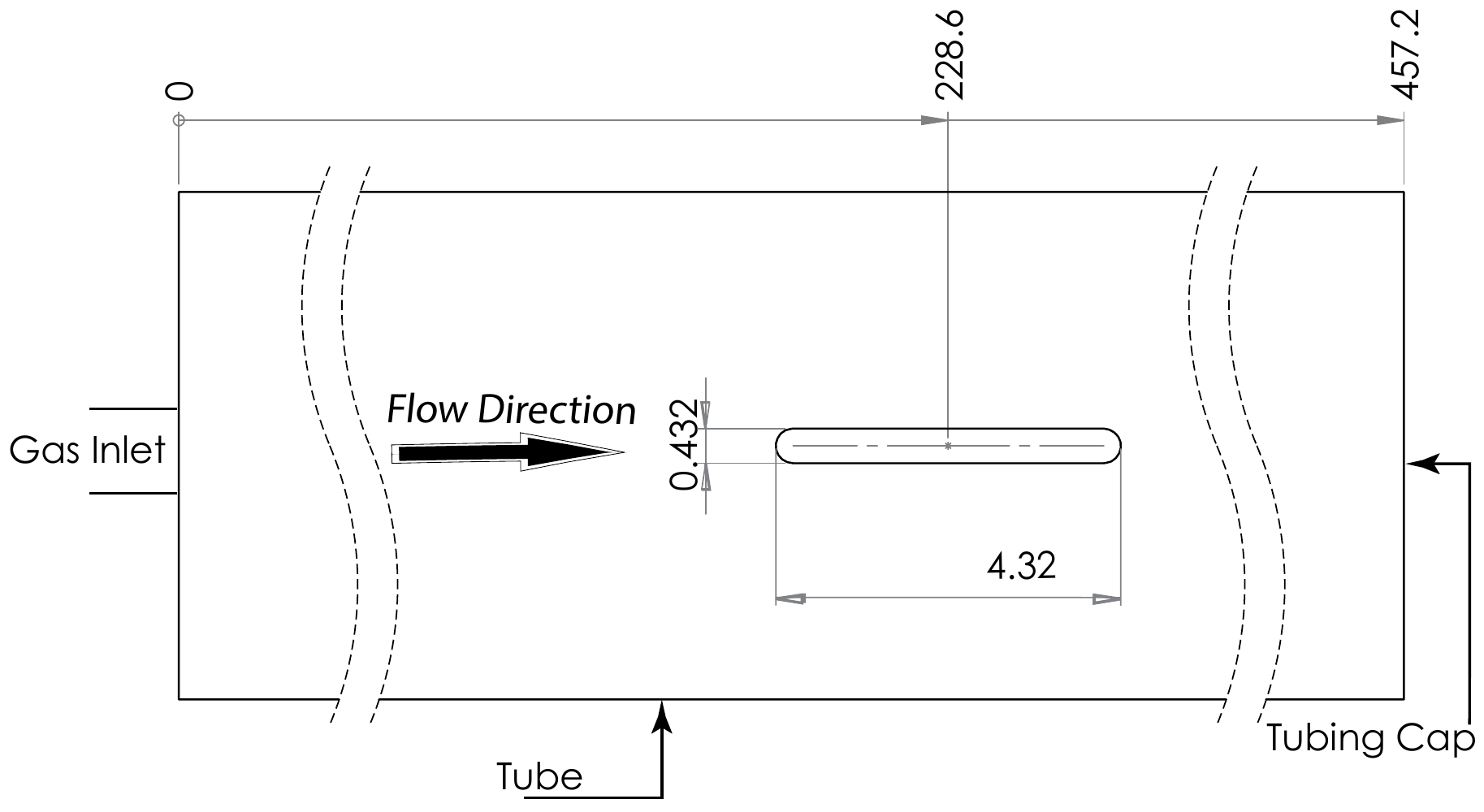}\\
	c)\includegraphics[scale=0.125]{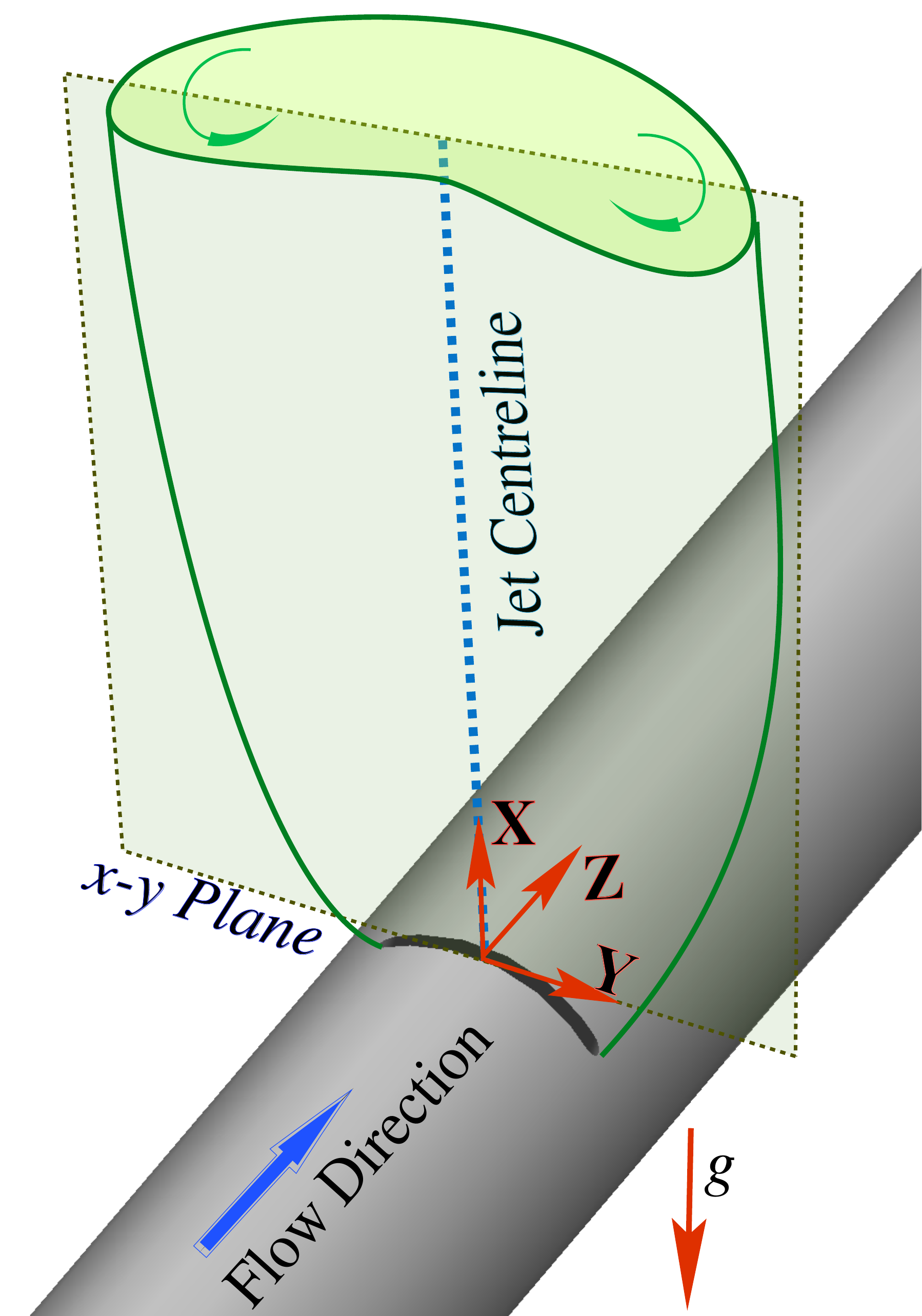}
	d)\includegraphics[scale=0.125]{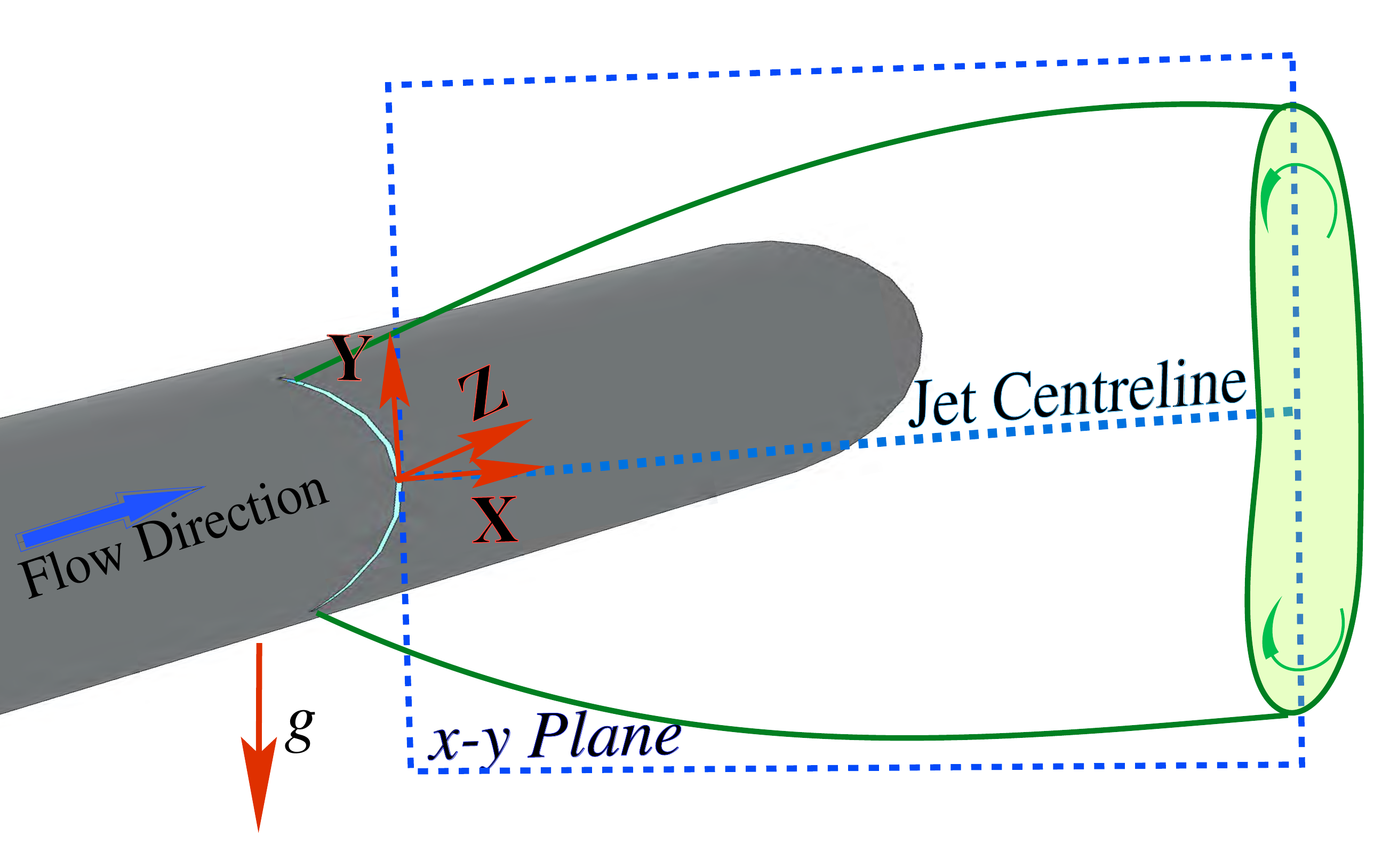}
	e)\includegraphics[scale=0.125]{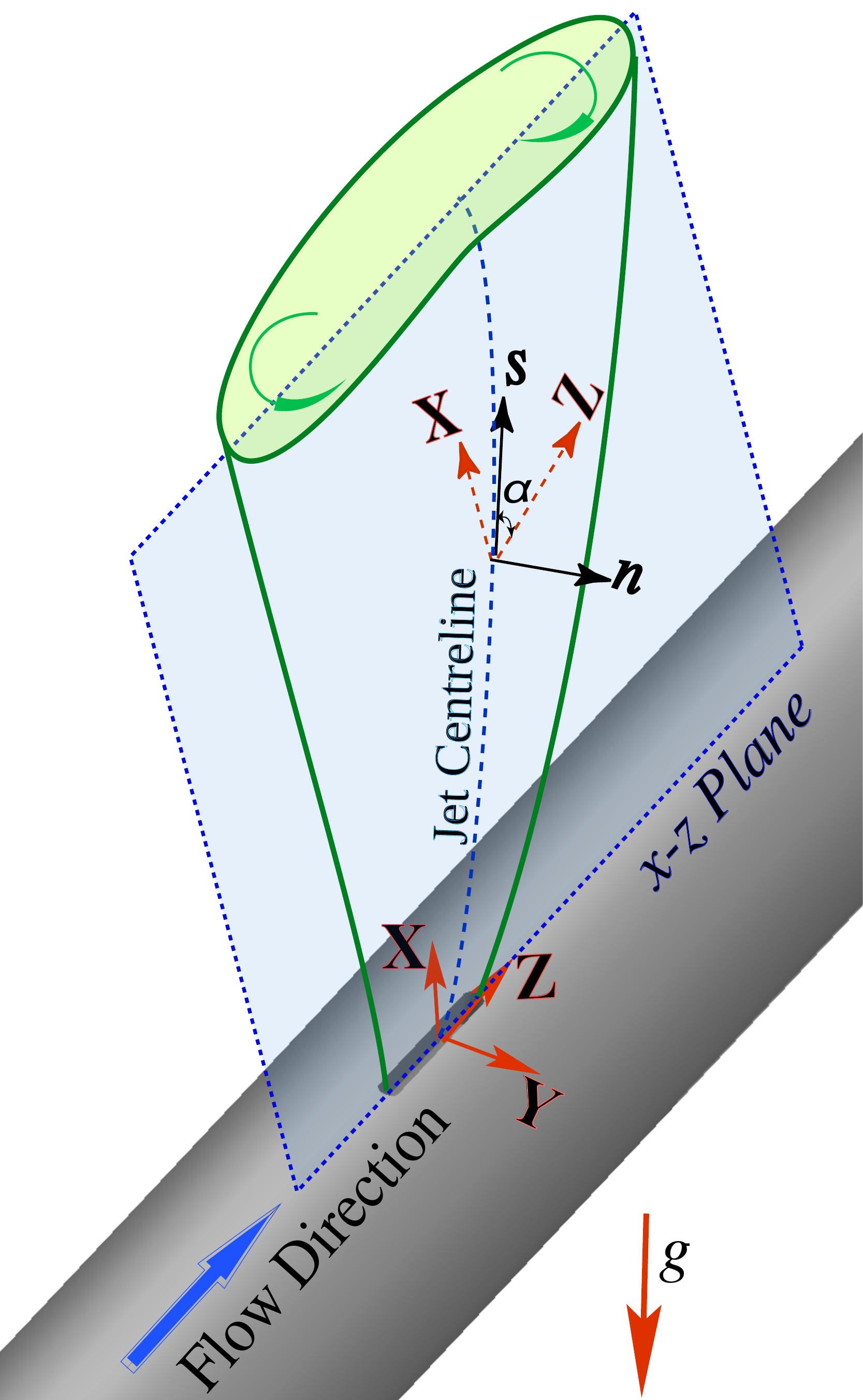}
	f)\includegraphics[scale=0.125]{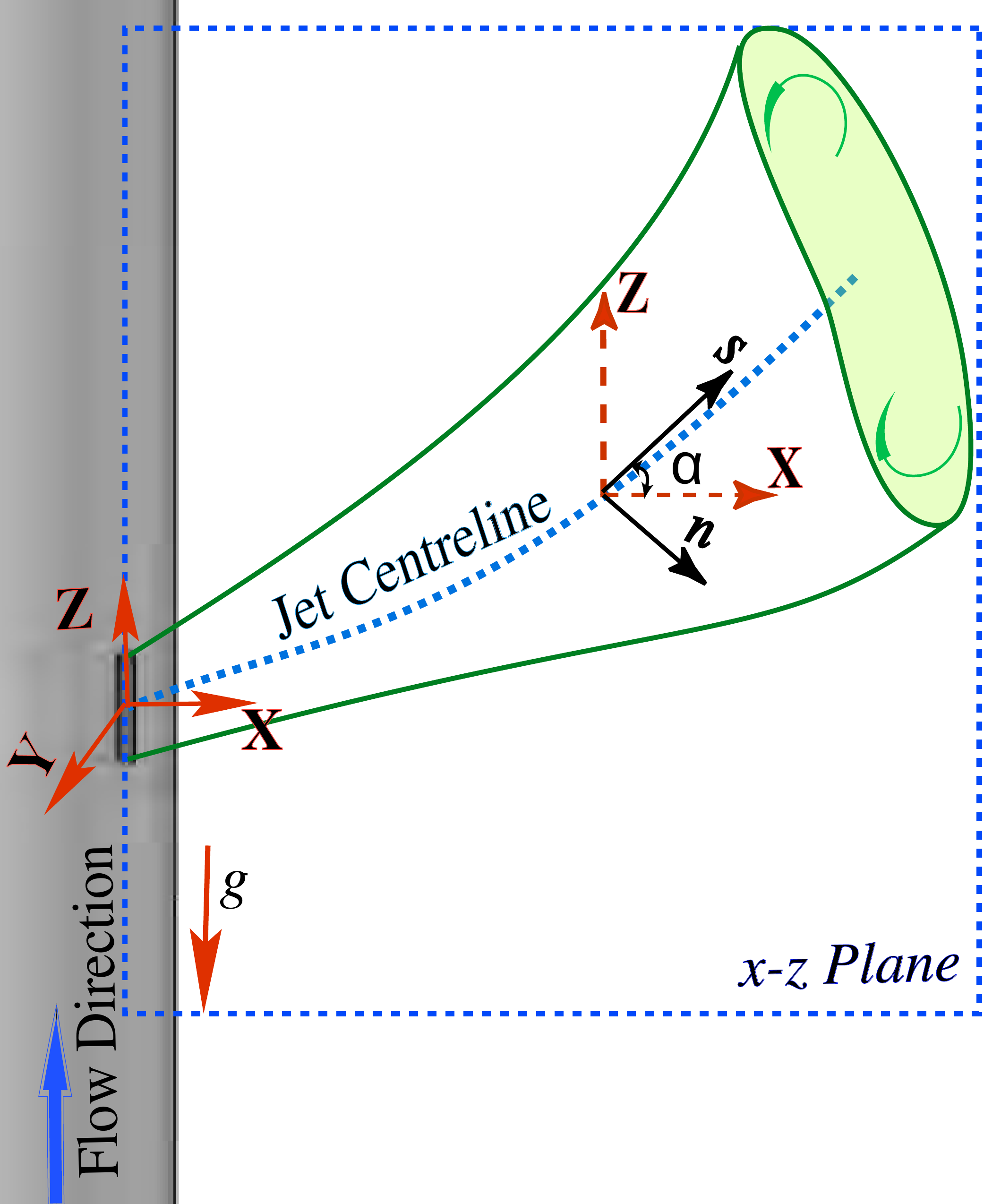}	
	\caption{a-b) Schematic of slot 2 and 3 geometries. c-f) schematic of vertical (c \& e) and horizontal (d \& f) 3D slot jets flow measurement areas. All dimensions are in mm.}
	\label{fig.Experimental_Layout3}
\end{figure}

Figure \ref{fig.Experimental_Layout3} illustrates the orifice geometries along with the evolution of the jet flow from the orifices considered in this study, for both air and helium. Two high-aspect-ratio slots, with the same aspect ratio ($AR=10$), were machined into the side of a round tube (closed at one end) with an outer diameter of 6.36 mm and 0.82 mm wall thickness. Aspect ratio is defined as the ratio of the two symmetrical (long/short) axes  of the slot geometry. These high-aspect-ratio 3D slots were aligned parallel and perpendicular to the direction of the flow inside the tube, and also compared to their 3D round ($AR=1$) jet counterparts \cite{Soleimaninia2018IJoHE,Soleimaninia2018Ae}.  Hereafter, slots 1, 2 and 3 refer to the round orifice, the slot perpendicular to, and the slot parallel with the direction of flow inside the tube, respectively. The new length-scale, equivalent diameter ($D_{eq}$), was defined to adequately characterize the jet flow \cite{Hussain1989JoFM257}. Here, $D_{eq}$, is the diameter of an equivalent circle with the same area of the nozzle. Due to the curvature of the tube surface, and also to keep $AR$ identical for both slots ($AR=10$), slightly different equivalent diameters ($D_{eq}$) were obtained for the slots. The range of $D_{eq}$ for the current study are 2, 1.6 and 1.53 mm for the slots 1, 2 and 3, respectively. To capture development of the jet, measurements were obtained on the two-dimensional planes aligned with the major axis of the slots. Here, the two-dimensional measurements planes are denoted as $x$-$y$ and $x$-$z$ (as indicated in Figs. \ref{fig.Experimental_Layout3} c-f), for the slots 2 and 3, respectively.  Also shown in the figure is the jet centreline, which acts as a reference from which measurements are later obtained.  Owing to potential deviation of the jet  (slot 3) from the orifice axis ($x$-axis), the jet centreline tangent and normal lines are shown as $s$ and $n$ coordinates in the figure, respectively. It should be noted that the jet centreline, for slot 2, is considered at geometrical centre of the orifice ($y=0$) in the $x$-$y$ measurement plane. 

The experiments were conducted in a controlled stagnant environment, at room temperature and pressure (${T_o}\sim22^{\circ \textrm{C}}$, ${p_o}\sim100$ kPa). Flow controllers (Bronkhorst, EL-FLOW series) were used to control mass flow rates of dry air and pure scientific grade helium to the system, with a high accuracy (standard $\pm0.5\%$ of reading plus $\pm0.1\%$ full scale) and precision (within $0.2\% $ of the reading). After the test gas was mixed and seeded with the PIV and PLIF tracer particles, gas flow entered the test section of the tube. Isothermal and isobaric conditions were ensured in all measurements. Further specific details of the flow facility used in the current experiments can be found in \cite{Soleimaninia2018IJoHE,Soleimaninia2018Ae}. The orifice, through which the gas dispersed, was located sufficiently downstream along the tube length to ensure fully developed flow inside the tube at the orifice location.  Within the tube, flow controllers were used to ensure fully developed subsonic and turbulent flow inside the tube.

\begin{table}[t]
	\centering
	\caption{Flow properties of 3D jet experiments}
	\begin{adjustbox}{max width=\textwidth}
		{\small
			\begin{tabular}{|l|l|l|l|l|l|l|l|l|l|l|l|}
				\hline 
				
				Slot& Jet & Orien- &$AR$&$D_{eq}$&  $Q$ & $\langle{\boldsymbol{u}_{j}}\rangle_{c}$ &$\rho_{{j}}$ &$\nu_{{j}}$  &${M} $ &  $Fr$ & $Re_{\delta}$\\[1ex]
				&& tation&& [m]&  [L/min] & [m/s]  & [Kg/$\textrm{m}^{3}$] &  [$\textrm{m}^{2}$/s]  & [N/m]    & & \\						
				\hline
				1 & Air& H &1& $2\times10^{-3}$&15&147.5 & 1.17 & $1.54\times10^{-5} $&50.9  & - &19,000 \\			
				\hline
				1 & Air& V &1& $2\times10^{-3}$& 15&147.5 & 1.17 & $1.54 \times 10^{-5} $&50.9  & - &19,000 \\
				\hline
				2 & Air& H &10& $1.6\times10^{-3}$&15&169.7 & 1.17 & $1.54 \times 10^{-5} $&51.7  & - &20,300 \\
				\hline
				2 & Air& V &10& $1.6\times10^{-3}$&15&170.2 & 1.17 & $1.54 \times 10^{-5} $&51.8  & - &20,300 \\
				\hline
				3 & Air& H &10& $1.53\times10^{-3}$&15&209.2 & 1.17 & $1.54 \times 10^{-5} $&53.3  & - &21,000 \\
				\hline
				3 & Air& V &10& $1.53\times10^{-3}$&15&208.6 & 1.17 & $1.54 \times 10^{-5} $&53.2  & - &21,000 \\
				\hline
				1 & He & H &1& $2\times10^{-3}$&35& 399.5& 0.165 &$1.21 \times 10^{-4} $&51.3 &$1.34\times10^{6} $&51,500 \\		
				\hline
				1 & He & V &1& $2\times10^{-3}$&35& 399.7& 0.165 &$1.21 \times 10^{-4} $&51.4 &$1.34\times10^{6} $&51,500 \\			
				\hline
				2 & He& H &10& $1.6\times10^{-3}$&35&468.8 & 0.165 &$1.21 \times 10^{-4} $&52.2  &$2.4\times10^{6} $&50,800 \\
				\hline
				2 & He& V &10& $1.6\times10^{-3}$&35&469.1 & 0.165 &$1.21 \times 10^{-4} $&52.3  &$2.4\times10^{6} $&50,800 \\
				\hline
				3 & He& H &10& $1.53\times10^{-3}$&35&511.4 & 0.165 &$1.21 \times 10^{-4} $&52.9  &$2.8\times10^{6} $&51,600 \\
				\hline
				3 & He& V &10& $1.53\times10^{-3}$&35&510.7 & 0.165 &$1.21 \times 10^{-4} $&52.7  &$2.8\times10^{6} $&51,600 \\
				\hline
		\end{tabular}}
		\label{tab.Flow properties3}
	\end{adjustbox}
\end{table}

In order to compare the behaviour of both test gases, for each experimental setup, the averaged momentum flux (${M}$) at the jet exit was estimated and matched for all slot 1 cases. This matching was achieved, iteratively, by varying the volumetric flow rate (${Q}$) in the system, after which time, the same $Q$ was considered for both slots 2 and 3 experiments.  Here,  ${M}$ was calculated by first obtaining the time-averaged jet exit velocity from two-dimensional PIV measurements.  The two-dimensional momentum flux, in units of [N/m], was then calculated from
\begin{equation}
{M}=\int_{-{D}/2}^{{D}/2} {{\rho}_{{j}}  \langle{\boldsymbol{u}({r})}\rangle^{2}  \diff{{r}}}
\label{eqn.momentum_flux_experiment}
\end{equation}
where the subscript `$j$' refers to the conditions at the nozzle, the angle brackets `$\langle{}$  ${}\rangle$' refers to the time-averaged quantity, and also $\rho$ and $r$ refer to density and radius, respectively. Table \ref{tab.Flow properties3} shows the flow properties used in this study, for both the horizontal and vertical high-aspect-ratio 3D jet configurations (Slot 2 \& 3), as well as vertical and horizontal 3D round jets (Slot 1) which have been used for comparison \cite{Soleimaninia2018IJoHE,Soleimaninia2018Ae}. Here, the subscript `$c$' refers to the conditions at the jet centreline, $Fr$ is the Froude number, and H \& V refer to horizontal and vertical orientations, respectively. In all cases, the jets were characterized by the outer-scale Reynolds number, $Re_{\delta}= \langle{\boldsymbol{u}_{j}}\rangle\delta/{\nu}_{\infty}$. Where, ${\nu}_{\infty}$ is the ambient fluid kinematic viscosity and $\delta$ is the width of the mean axial velocity profile, evaluated from limits of 5\% of the centreline velocity at $x\simeq0$. 

Particle imaging velocimetry (PIV) was used to capture the two-dimensional velocity flow field information. A dual-head Nd: YAG pulsed laser (New Wave's SOLO III  15 HZ) was used to illuminate a two-dimensional cross-section of the flow, which was seeded with Di-Ethyl-Hexyl-Sebacate (DEHS), with a typical diameter of less than 1 $\mu$m, which acted as a tracer particle.  The light sheet had an approximate height of 5 cm and a thickness of 1 mm.  The camera's field of view (PIV CCD) was a 40$\times$30 $\textrm{mm}^{2}$ window with an approximate pixel size of 6.5 $\mu$m in physical space. This resolution was estimated to be comparable to the finest scale of the flow, with respect to the Nyquist criterion \cite{Su2003JoFM1}. Each pair of images were then processed using LaVision DaVis 8.4 software to calculate the global instantaneous flow velocity field. Following the PIV uncertainty propagation method\cite{Sciacchitano2016MSaT84006}, conservative uncertainty was estimated as 3\% and 6\% in the time-averaged velocity and Reynolds shear stress profiles, respectively.

To measure the gas concentration, we applied planar laser-induced fluorescence (PLIF). To simultaneously apply PLIF with PIV, the flow was also seeded, at consistent rate of $\sim10\%$ by volume, with acetone vapour.  A Pulsed Nd: YAG laser (Spectra-Physics INDI-40-10-HG) was used in order to excite the acetone molecules in a light sheet with an approximate height of 5 cm and a thickness of 350 $\mu$m, which was then recorded with a PLIF CCD camera. The camera's field of view for all cases corresponded to a 38$\times$28 $\textrm{mm}^{2}$ window with an approximate pixel size of 6.5 $\mu$m. The images were taken at a frequency of 5 Hz and then processed using LaVision DaVis 8.4 software.  After correcting for errors associated with background noise, fluctuations in cross-sectional laser beam intensity, and laser energy per pulse deviations, one can assume the remaining non-uniformity of the scalar field is due to  signal to noise ratio ($S/N$). The error in the $S/N$ can be estimated using the standard deviation of this ratio in a uniform low signal region of the flow field.   Based on this data, and the uncertainty propagation method, the uncertainty in the time-averaged and variances of the concentration field was estimated as conservative values of 4\% and 7\%, respectively. For each experimental case, a total of 500 images were acquired to determine the time-averaged molar concentration ({$\langle{X}\rangle$}) and variances (${X^{\prime^2}}$).  Further details of the experimental procedure can be found in \cite{Soleimaninia2018IJoHE,Soleimaninia2018Ae}.

\begin{figure}[h!]
	\centering
	\raggedright \underline{\textbf{Slot 2:}}\\
	a)\includegraphics[scale=0.080]{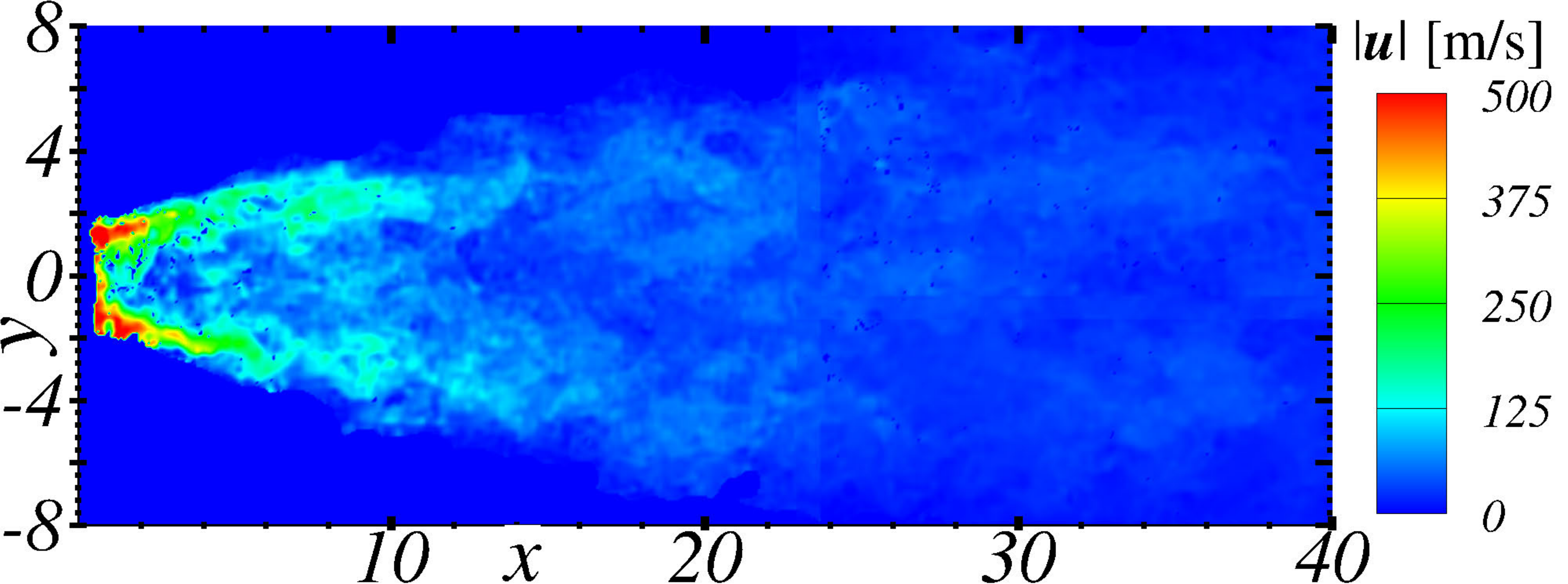}
	b)\includegraphics[scale=0.080]{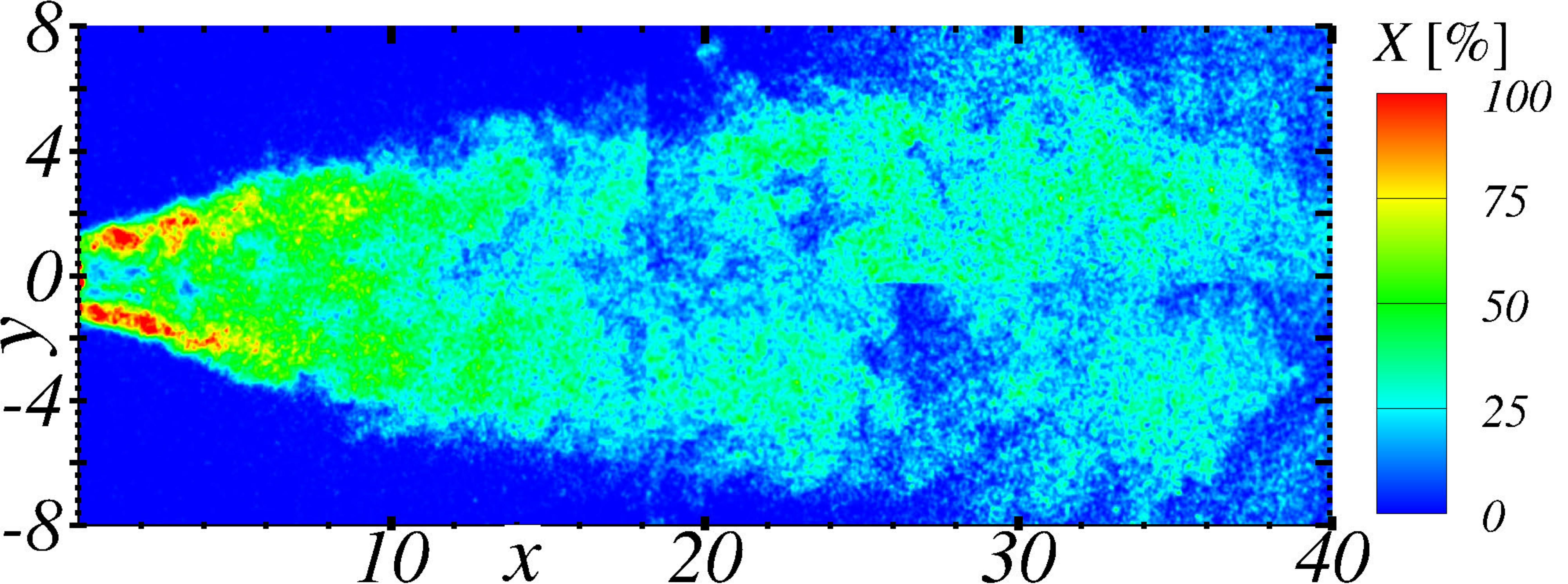}\\
	\raggedright \underline{\textbf{Slot 3:}}\\
	a)\includegraphics[scale=0.080]{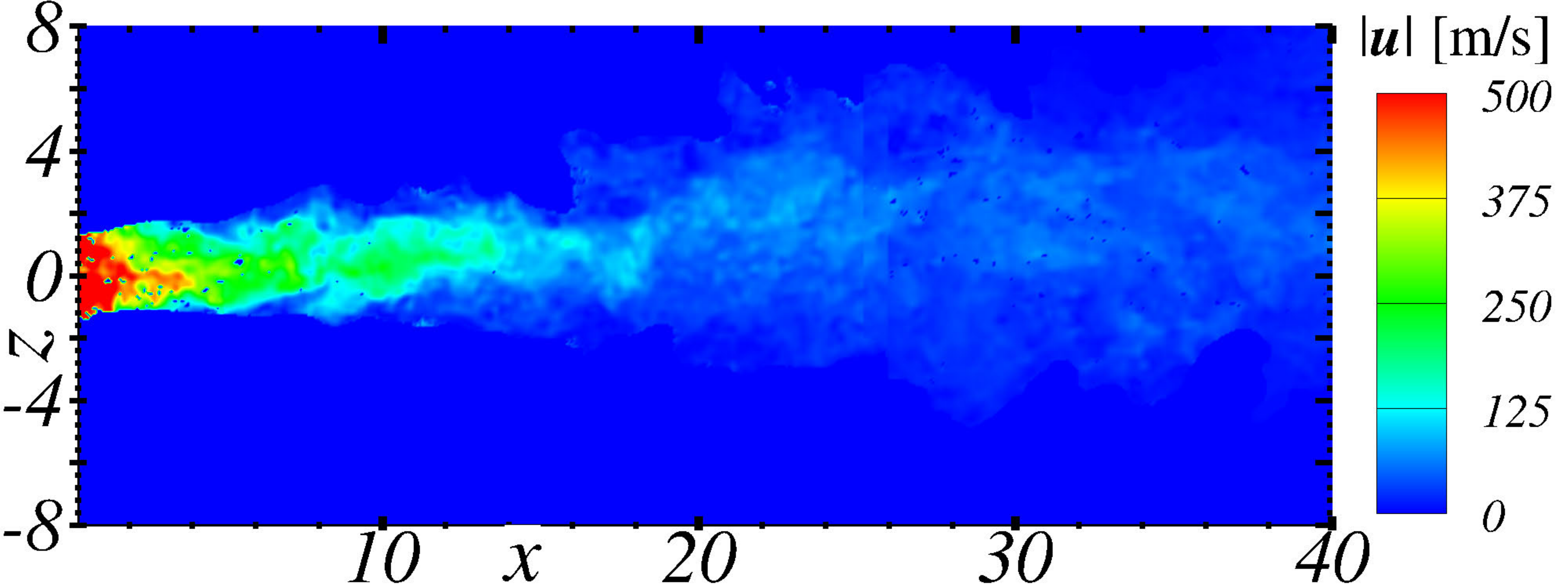}
	b)\includegraphics[scale=0.080]{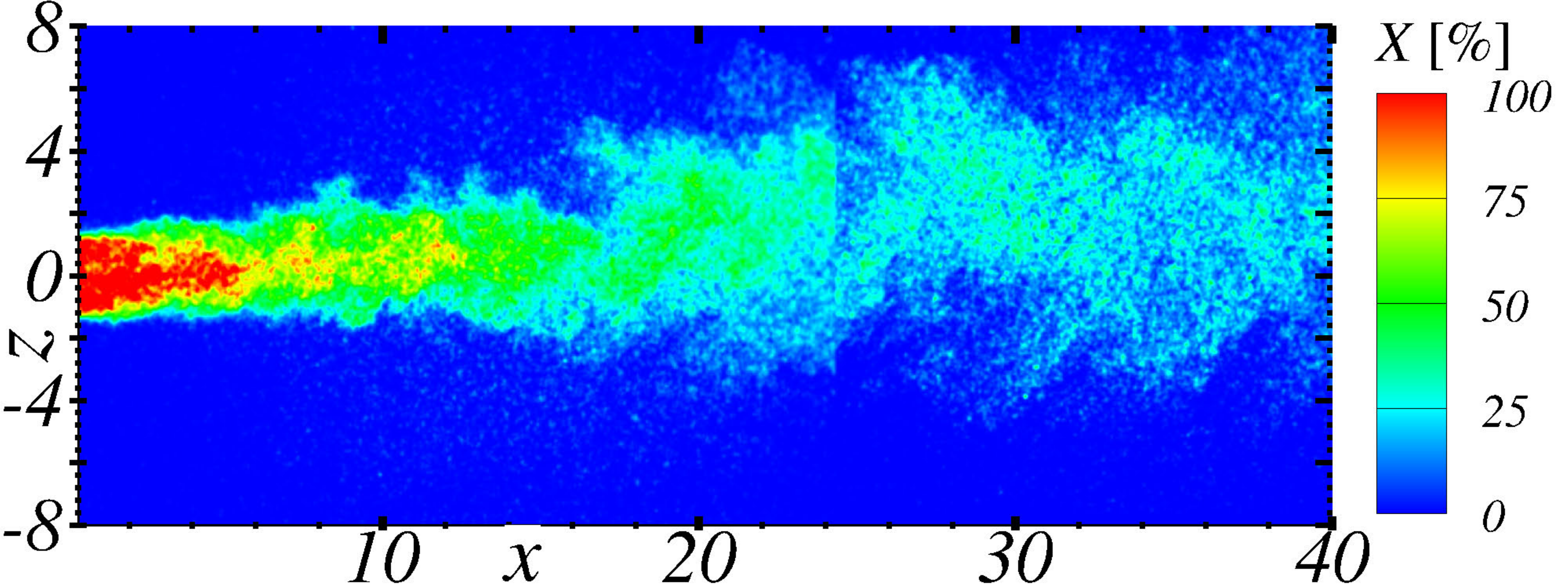}\\
	\caption{Instantaneous a)  velocity and b) molar concentration fields obtained from helium slots 2 \& 3 in $x$-$y$ and $x$-$z$ planes, respectively. Note: The flow fields were constructed from up to three different experiments, where individual imaging windows have been stitched together.}
	\label{fig.Instantaneous_plots3}		
	\end{figure}

Finally, to retain the spatial resolution of the flow field, the full measurement region is covered by up to four individual imaging windows (depends on the slot geometry and jet orientation) with at least a 20\% overlap between each window. Figure \ref{fig.Instantaneous_plots3} shows an example of the instantaneous velocity and concentration fields, for helium slots 2 and 3 in the $x$-$y$ and $x$-$z$ planes, respectively. It should be noted that the flow fields were constructed from up to three different experiments, where individual imaging windows have been stitched together.

Distances reported here have been normalized such that
\begin{gather}
x = \frac{\textrm{\bf{X}}}{D_{eq}}, \;\;\;\; \;\;\;\; y = \frac{\textrm{\bf{Y}}}{D_{eq}}, \;\;\;\; \;\;\;\; z = \frac{\textrm{\bf{Z}}}{D_{eq}}, \;\;\;\; \;\;\;\; s = \frac{\textrm{\bf{s}}}{D_{eq}}, \;\;\;\; \;\;\;\; n = \frac{\textrm{\bf{n}}}{D_{eq}}
\label{eqn.Non-Dimensional}
\end{gather}%
where ${D_{eq}}$, the equivalent diameter of the orifice, is taken as a reference length scale.

\section{Results}
\subsection{Time-averaged flow fields}
\begin{figure}[h!]
	\centering
	\raggedright \underline{\textbf{air:}}\\
	a)\includegraphics[scale=0.037]{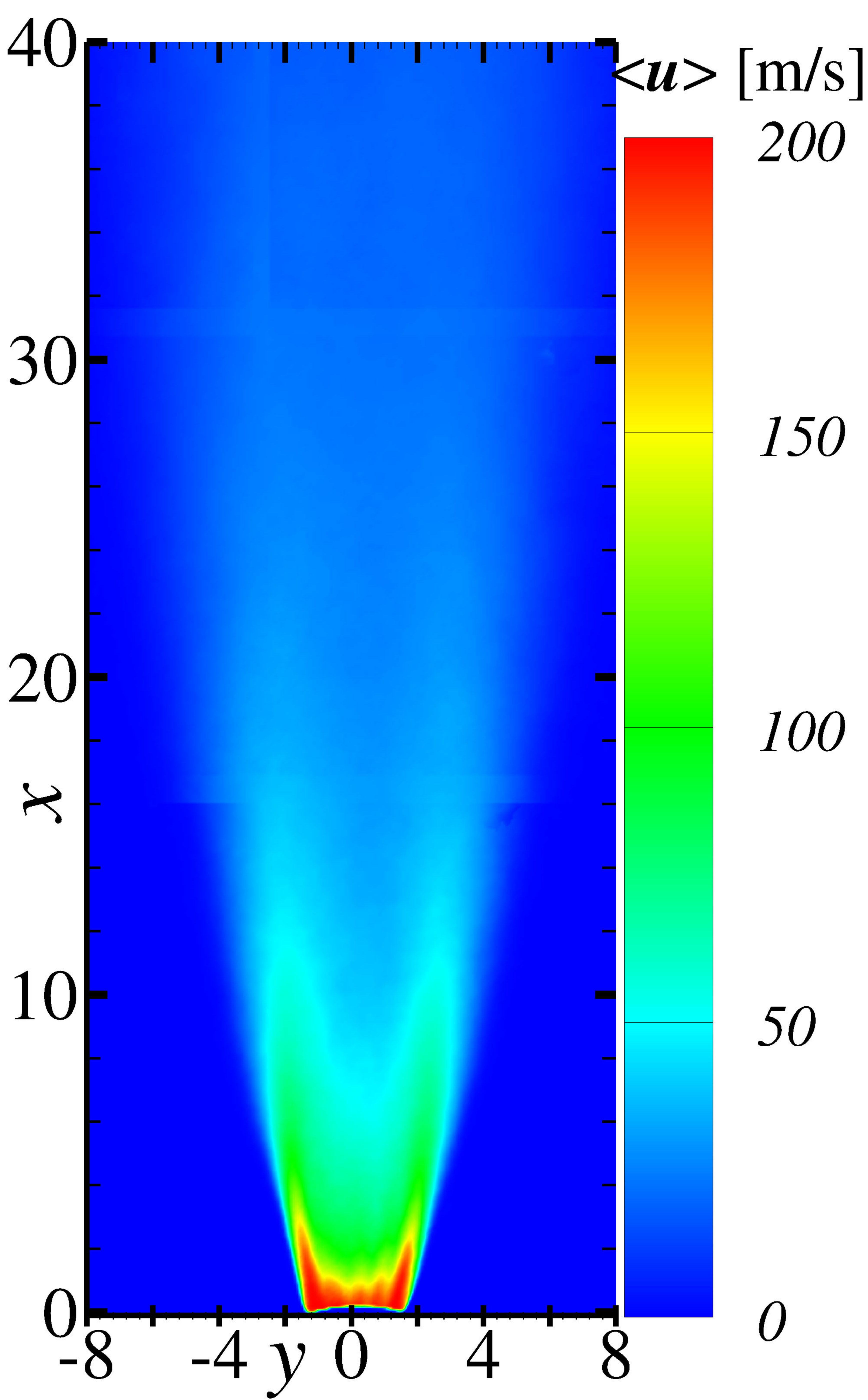}
	b)\includegraphics[scale=0.037]{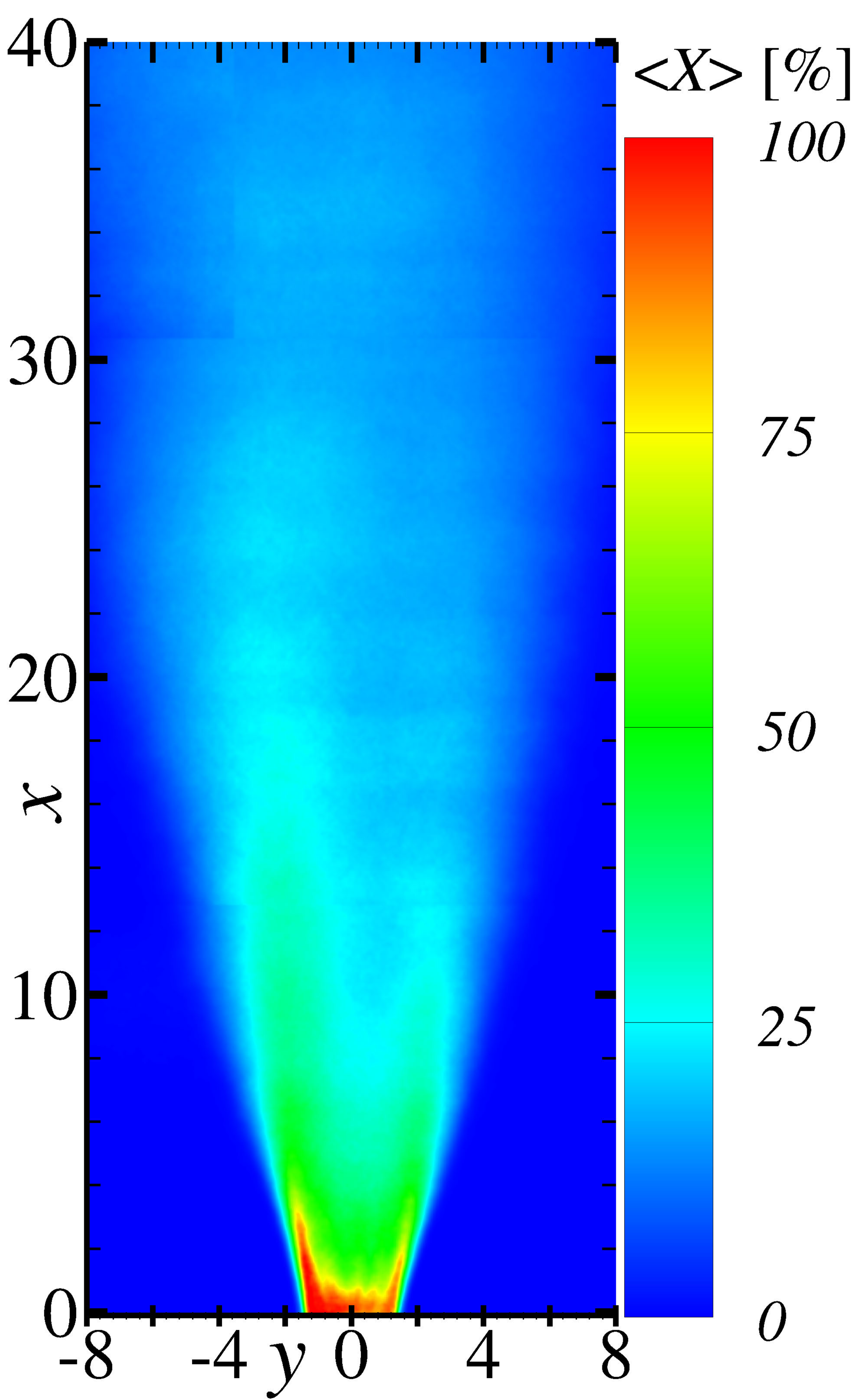}
	c)\includegraphics[scale=0.037]{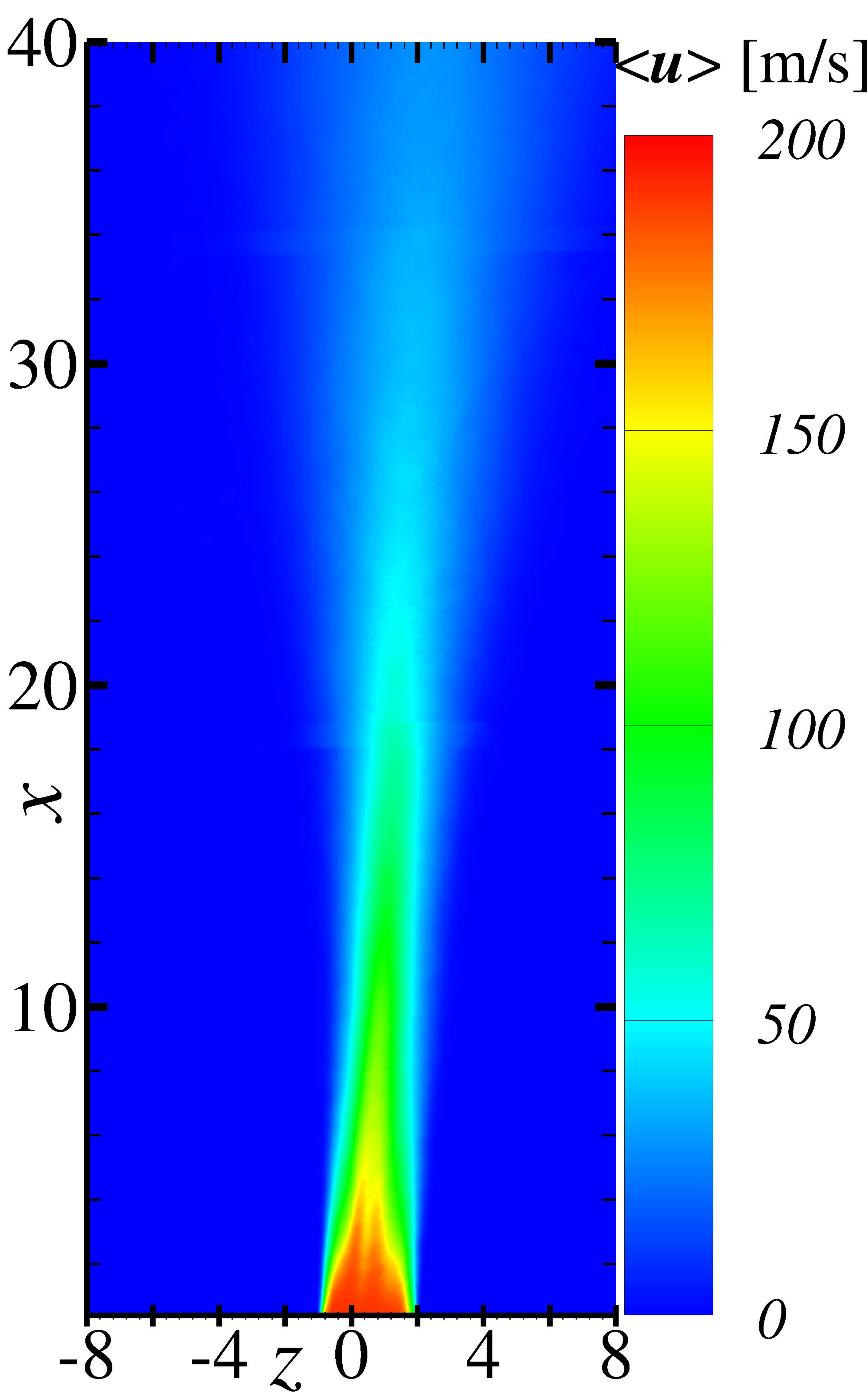}
	d)\includegraphics[scale=0.037]{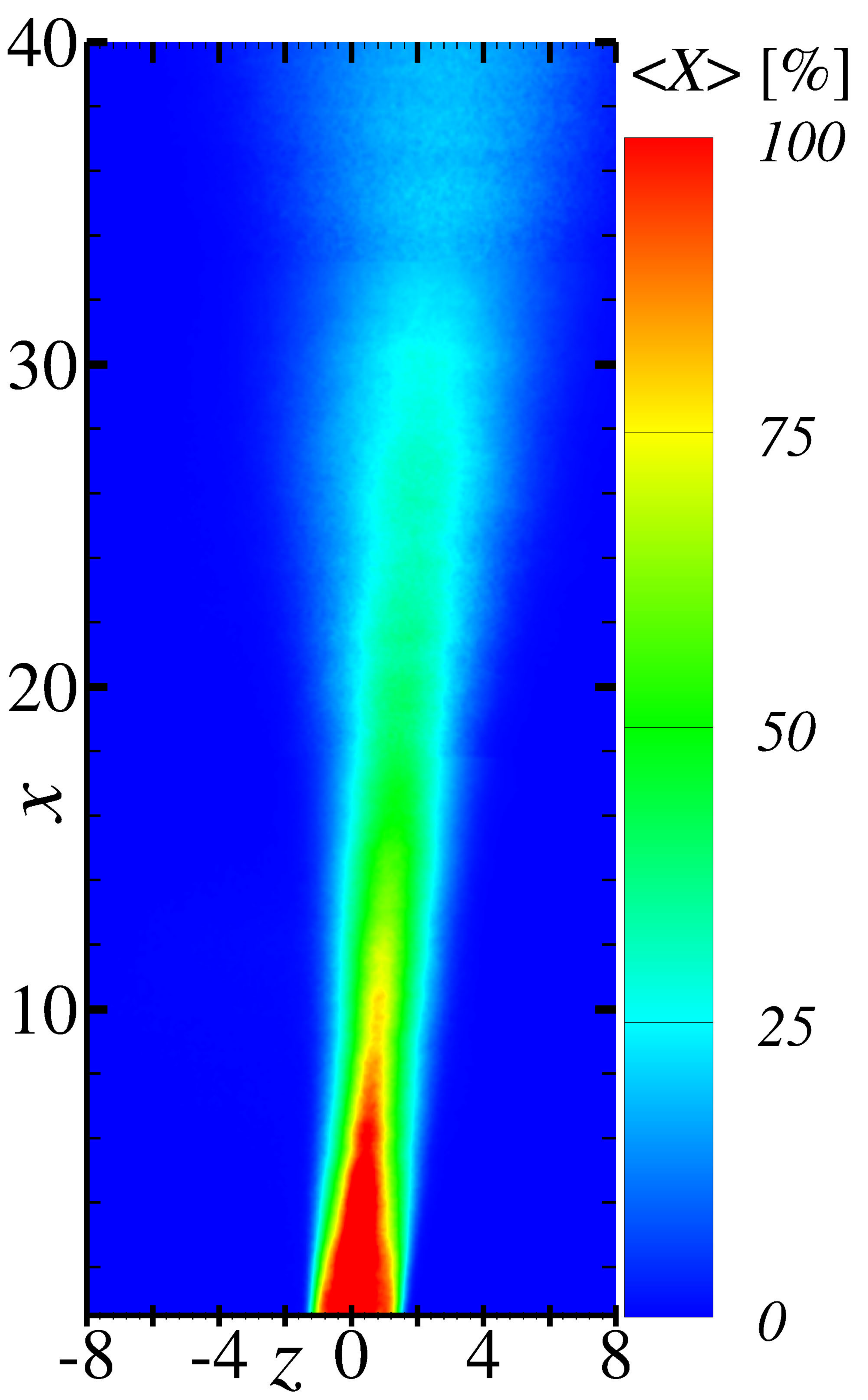}\\
	\raggedright \underline{\textbf{helium:}}\\
	a)\includegraphics[scale=0.037]{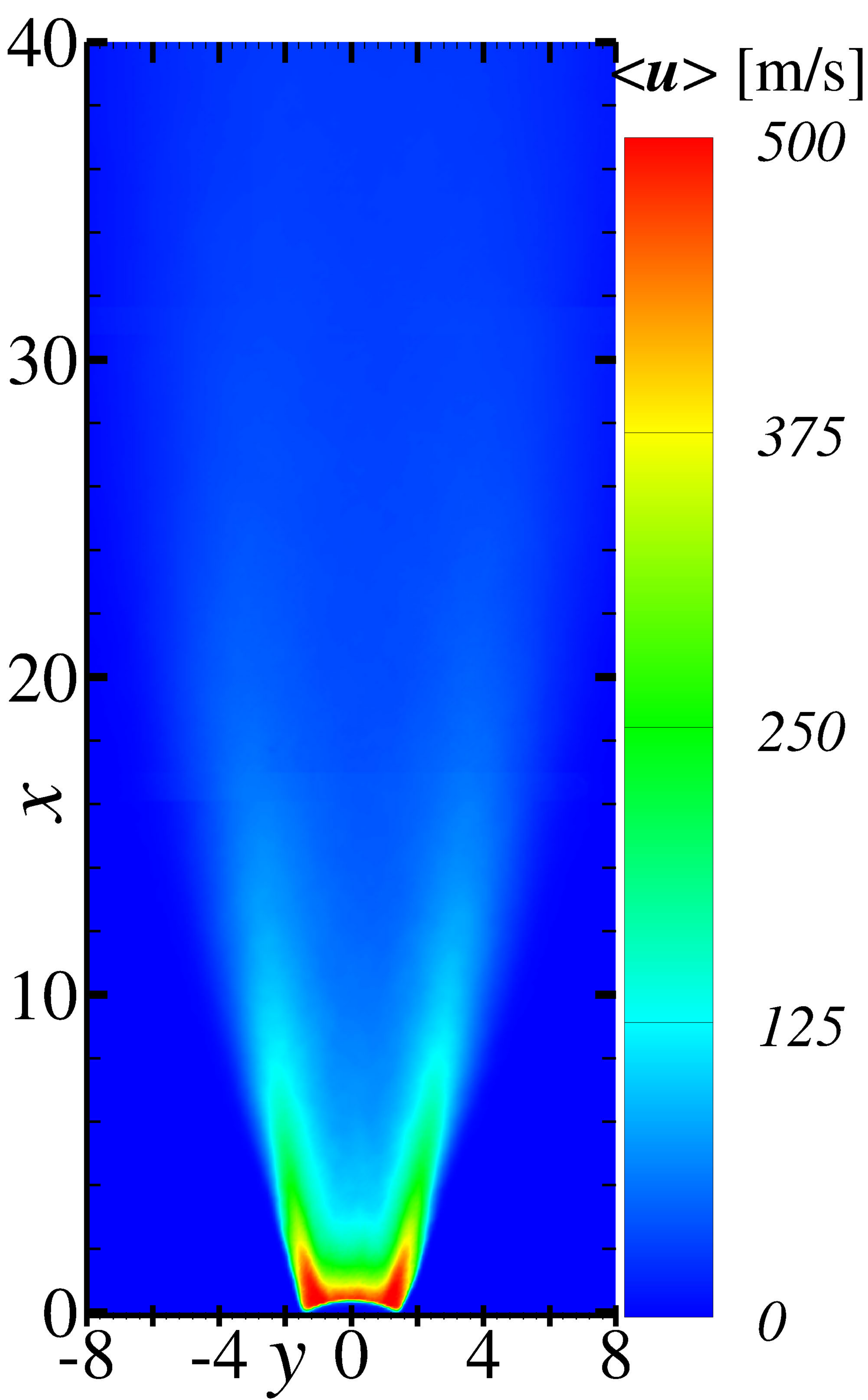}
	b)\includegraphics[scale=0.037]{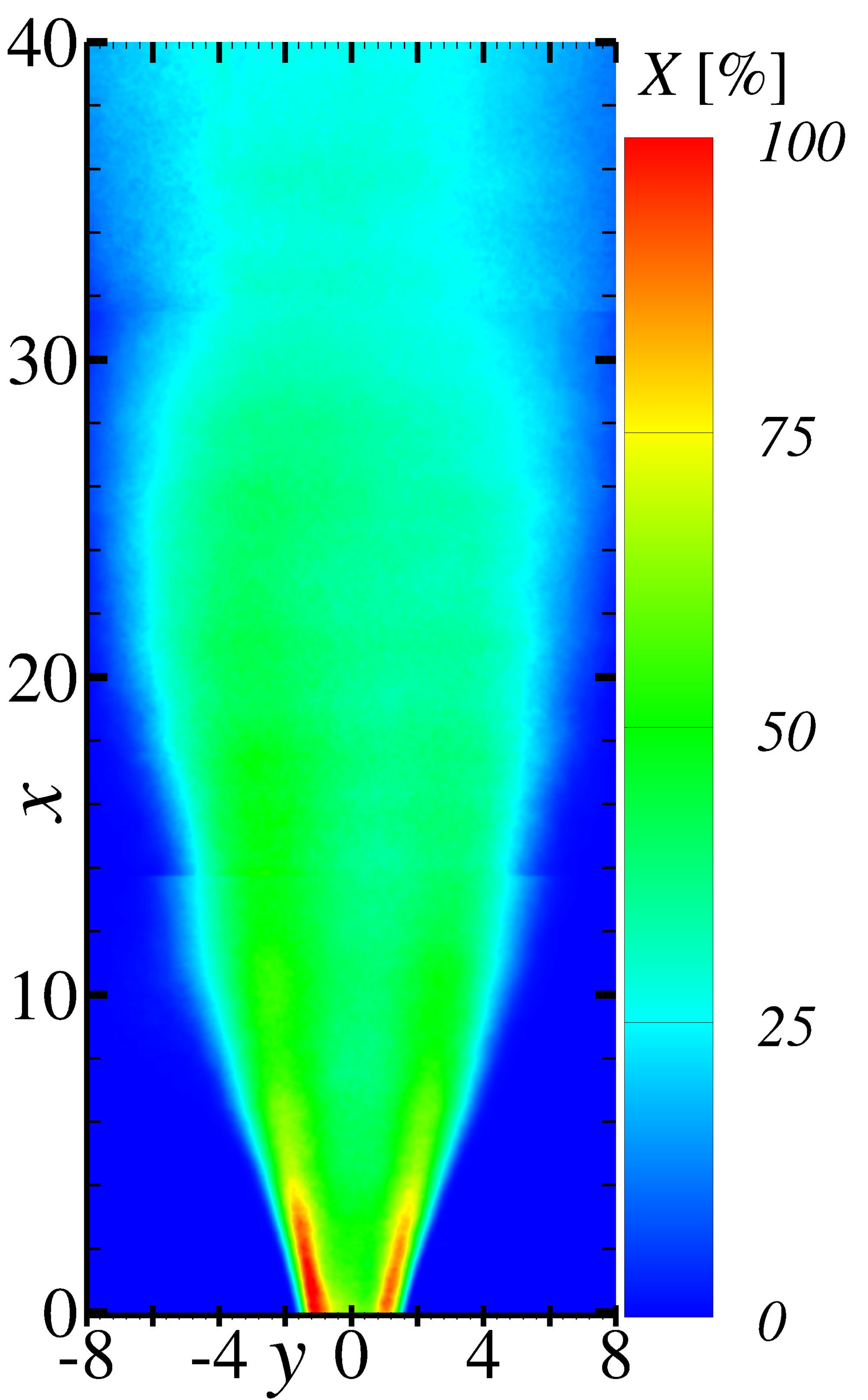}
	c)\includegraphics[scale=0.037]{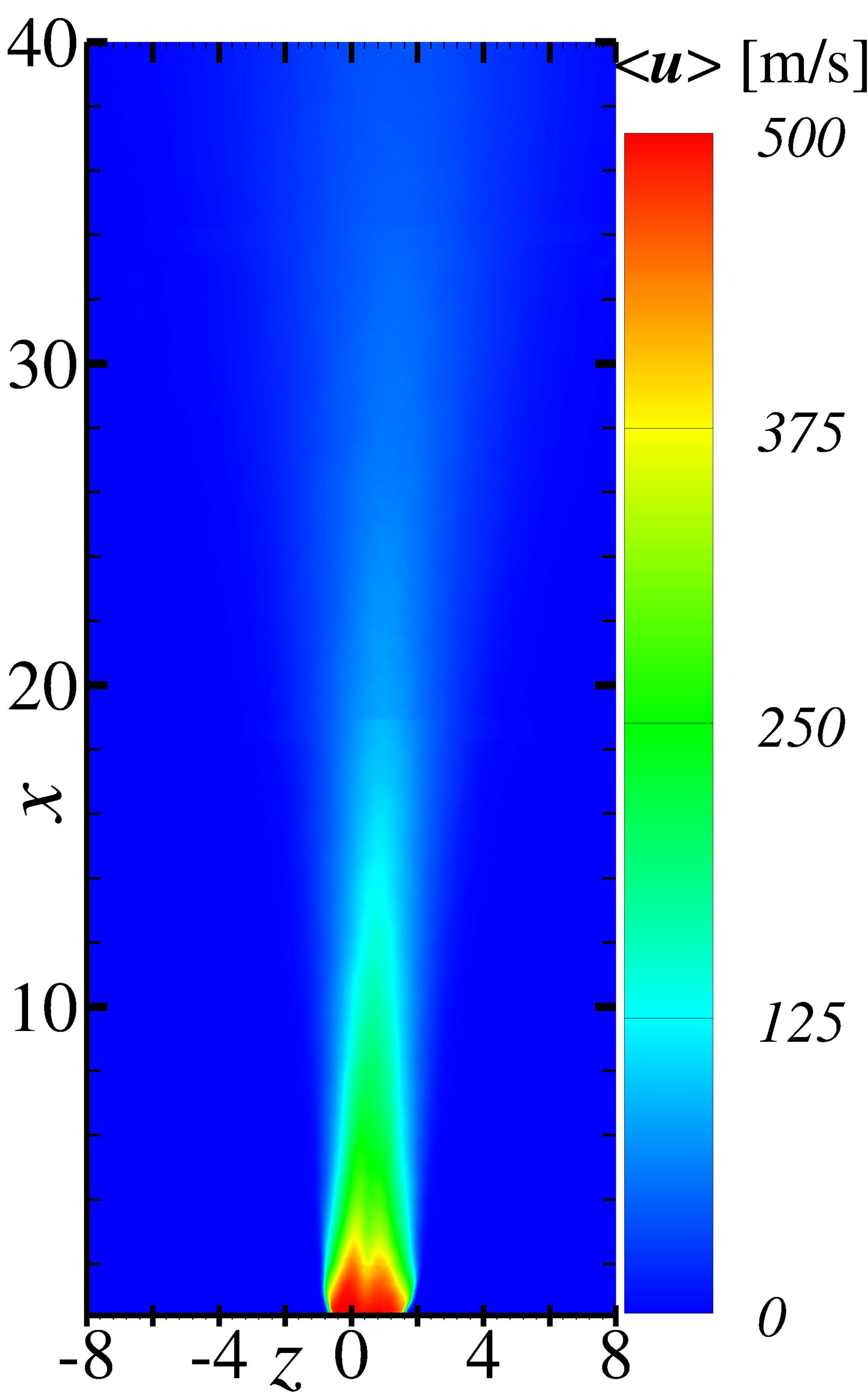}
	d)\includegraphics[scale=0.037]{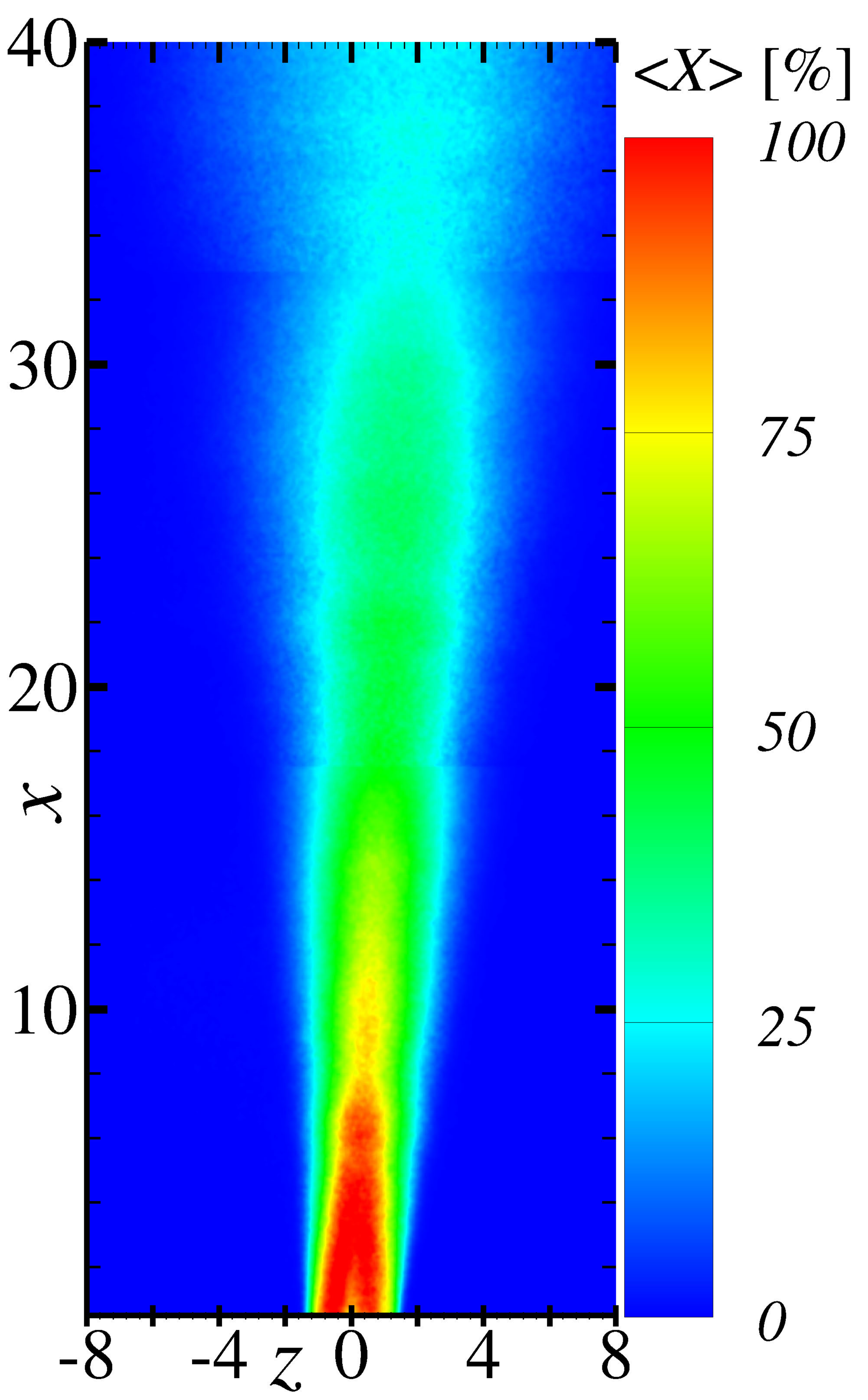}\\
	\caption{Time-averaged velocity and molar concentration contours from vertical high-aspect-ratio slot jet on the side of the tube (3D slot jet) for air and helium, obtained from a) slot 2 velocity contours in $x$-$y$ plane, b) slot 2 molar concentration contours in $x$-$y$ plane, c) slot 3 velocity contours in $x$-$z$ plane and d) slot 3 molar concentration contours in $x$-$z$ plane.}
	\label{fig.Velocity_Concentartion_Contours_Slot2-3_V}
\end{figure}

\begin{figure}[h!]
	\centering
	\raggedright \underline{\textbf{air:}}\\
	a)\includegraphics[scale=0.08]{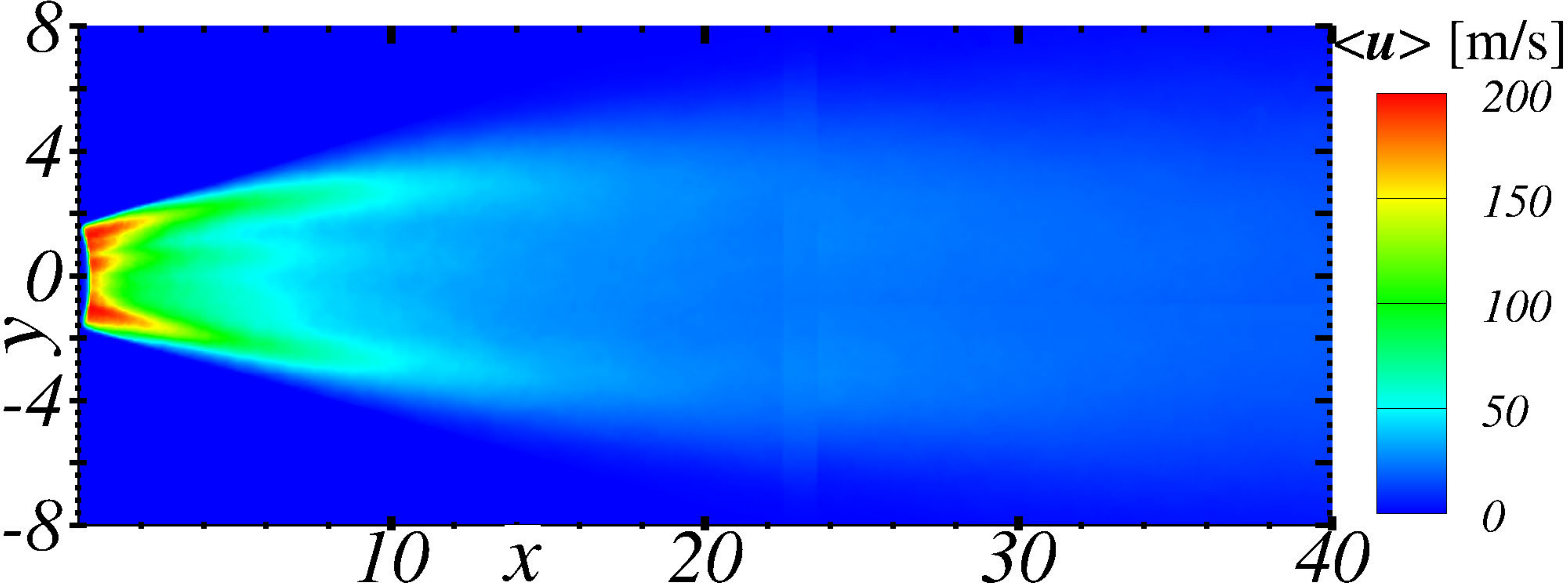}
	b)\includegraphics[scale=0.08]{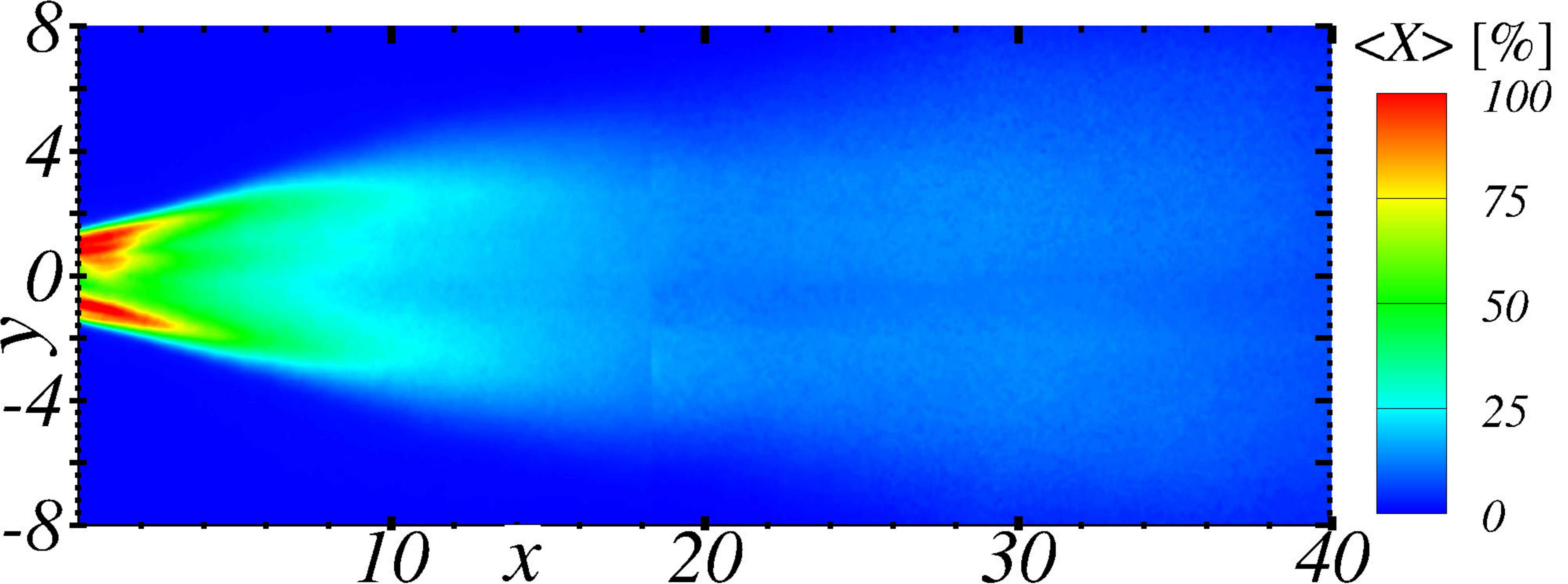}\\
	c)\includegraphics[scale=0.08]{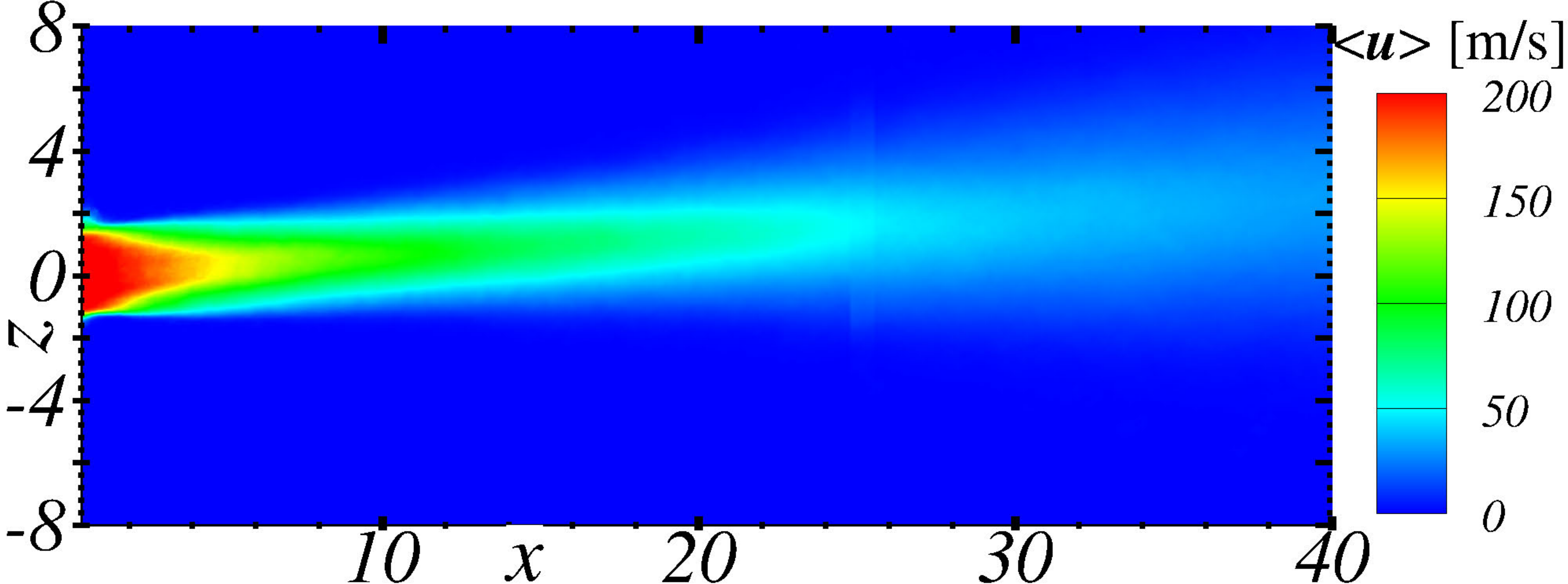}
	d)\includegraphics[scale=0.08]{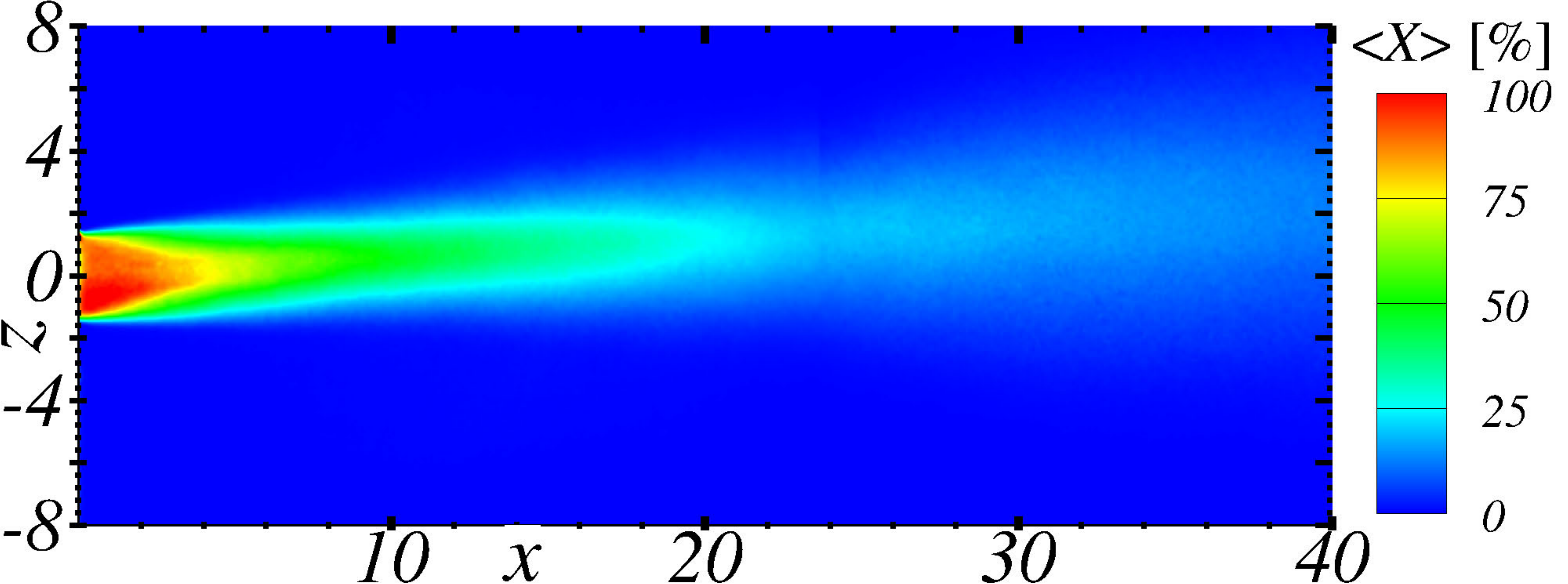}\\
	\raggedright \underline{\textbf{helium:}}\\
	a)\includegraphics[scale=0.105]{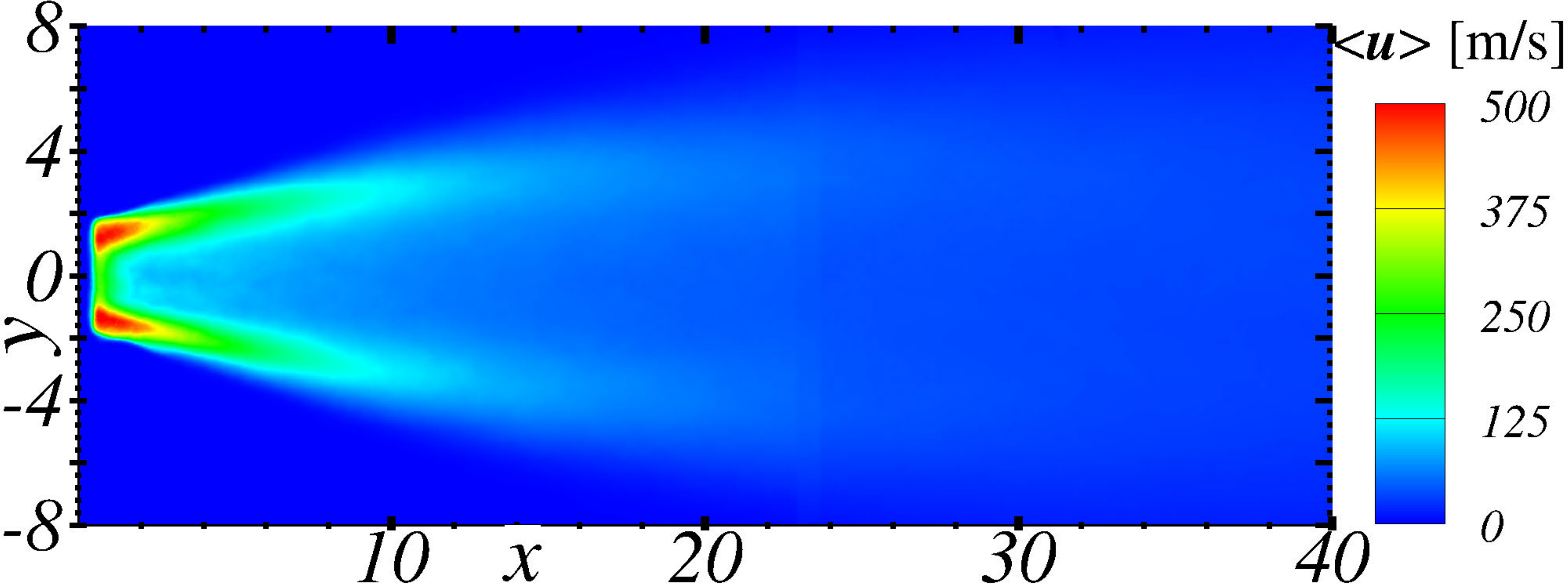}
	b)\includegraphics[scale=0.08]{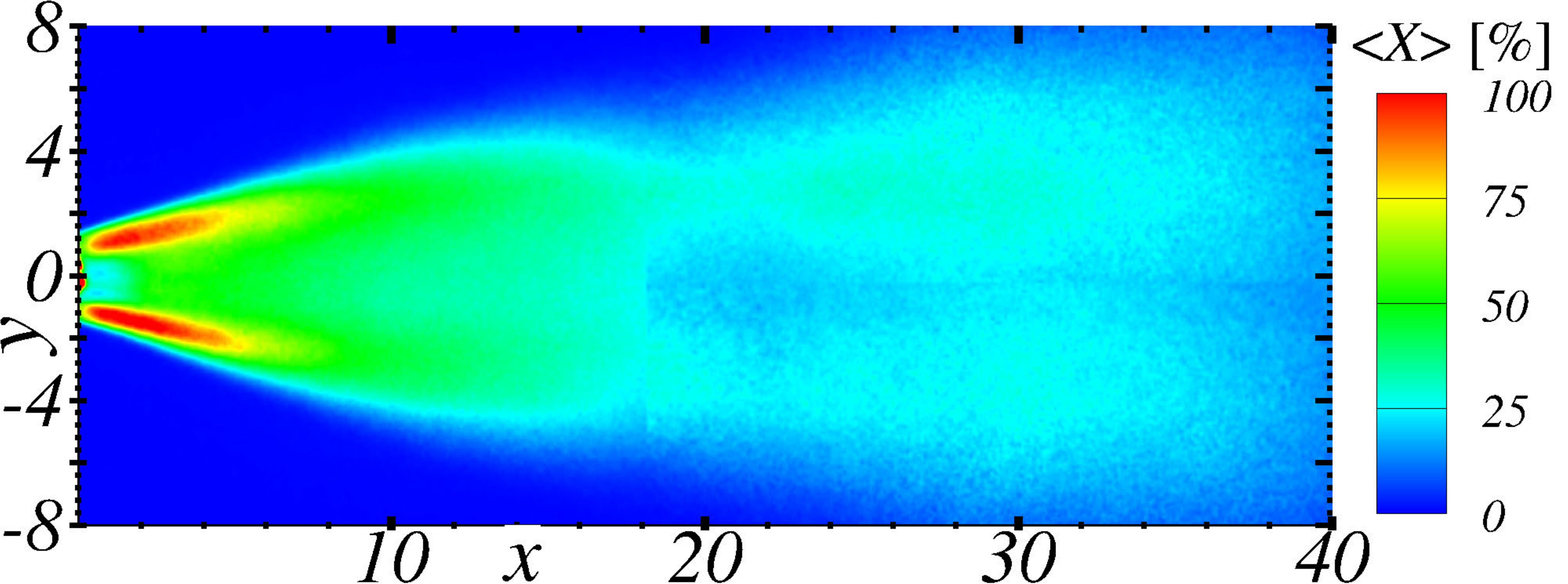}\\
	c)\includegraphics[scale=0.08]{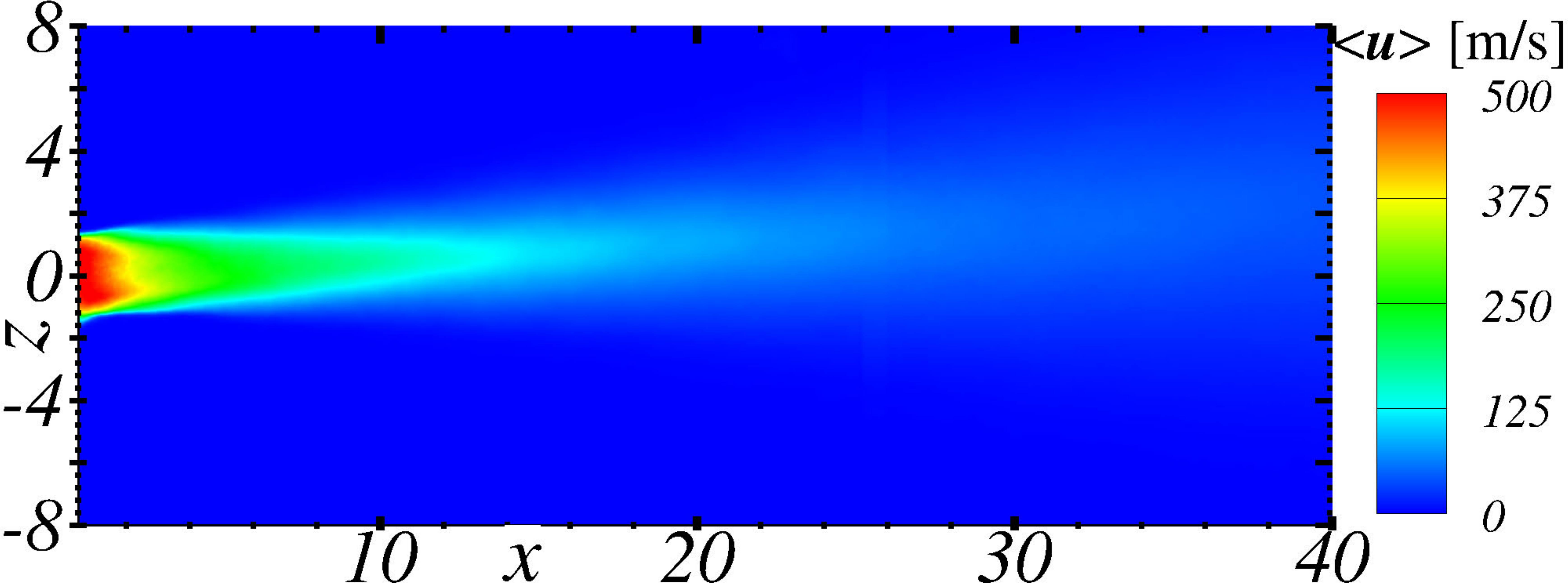}
	d)\includegraphics[scale=0.08]{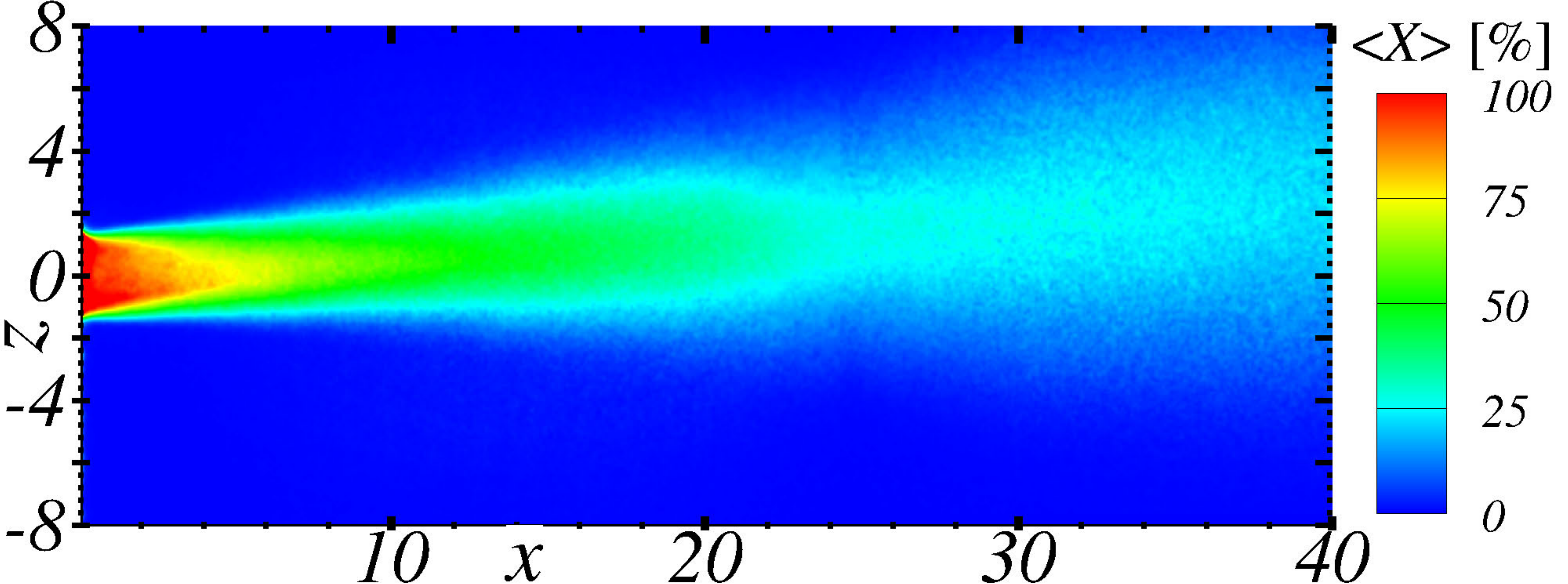}\\
	\caption{Time-averaged velocity and molar concentration contours from horizontal high-aspect-ratio slot jet on the side of the tube (3D slot jet) for air and helium, obtained from a) slot 2 velocity contours in $x$-$y$ plane, b) slot 2 molar concentration contours in $x$-$y$ plane, c) slot 3 velocity contours in $x$-$z$ plane and d) slot 3 molar concentration contours in $x$-$z$ plane.}
	\label{fig.Velocity_Concentartion_Contours_Slot2-3_H}
\end{figure}

The time-averaged velocity and molar concentration contours obtained for all vertical and horizontal 3D slot jet experiments are shown in Figs.\ \ref{fig.Velocity_Concentartion_Contours_Slot2-3_V} and  \ref{fig.Velocity_Concentartion_Contours_Slot2-3_H}, respectively. In vertical slot 2 for both gases, there were two high-velocity peaks (saddle-back behaviour), at about $y\pm1.5D_{eq}$, on each side of the $x$-axis toward the edge of the jet. Slightly shorter potential-core length was observed for helium ($\simeq2D_{eq}$), compared to air ($\simeq3D_{eq}$). This saddle-back behaviour was previously found to originate from a velocity deficit region which forms inside the orifice due to flow separation as the gas inside the tube encounters the edge of the 3D round orifice \cite{Soleimaninia2018IJoHE}. However, another possible reason is the sharp edge of the slot along with the curvature of the tube \cite{Mi2001JoFE878}. The initial saddle-back profile and shorter potential core length for helium compared to air, were also observed in both air and helium horizontal slots 2, presented in Fig.\ \ref{fig.Velocity_Concentartion_Contours_Slot2-3_H} a.

For all vertical and horizontal slot 3 experiments, air and helium jets were found to deviate from the jet streamwise axis ($x$-axis), in the direction of the initial flow inside the tube. There was also a shorter potential-core length for helium ($\simeq3D_{eq}$), compared to air ($\simeq4D_{eq}$).  Additionally, near the potential-core region, there was slightly more jet spreading on the positive $n$-coordinate of the jet centre compared to the opposite side. It should be noted that the $n$-coordinate refers to lines which are normal to the centreline, aligned with the opposite direction of flow inside the tube, and coplanar with the $x$-$z$ plane (see the coordinate system in Fig.\ref{fig.Experimental_Layout3} e-f). The spreading rate in the opposite direction of the flow within the tube, near the potential-core region, was previously found to be higher in the 3D round jets (Slot 1) \cite{Soleimaninia2018IJoHE,Soleimaninia2018Ae} compared to the slot 2 and slot 3 geometries of the current investigation.

In general, concentration profiles were qualitatively similar to the velocity profiles presented in Figs.\ \ref{fig.Velocity_Concentartion_Contours_Slot2-3_V} - \ref{fig.Velocity_Concentartion_Contours_Slot2-3_H}, with two exceptions. First, the potential core lengths in the vertical slot 3 jets were approximately $\simeq12D_{eq}$ and  $\simeq11D_{eq}$, for air and helium, respectively; in contrast, the potential core lengths in the horizontal slot 3 jets were approximately $\simeq5D_{eq}$ for both gases. Also for both horizontal and vertical experiments, much higher concentration levels and higher spreading rates were observed for helium in the far field compared to air.

\begin{figure}[h!]
	\centering
	\includegraphics[scale=0.33]{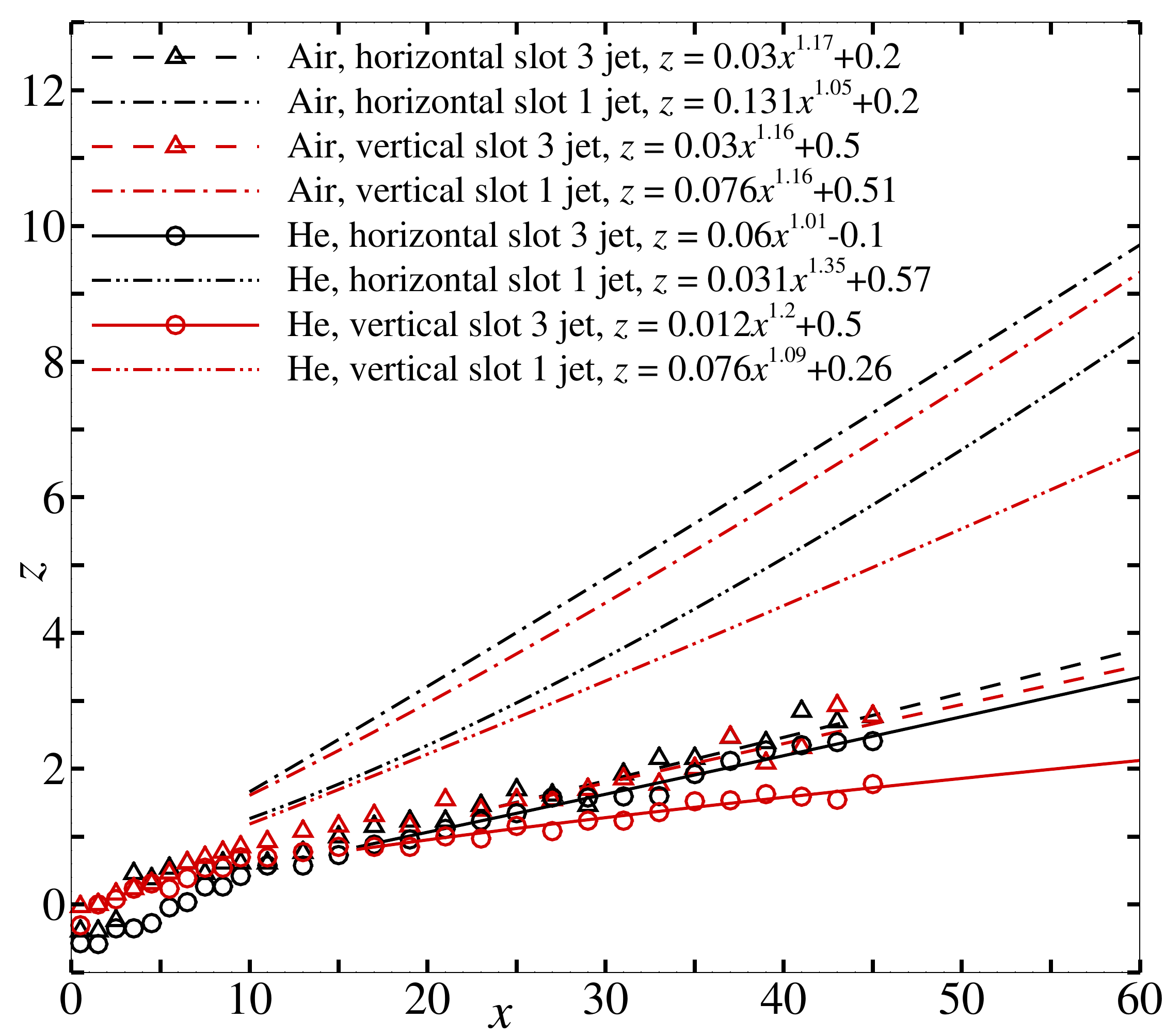}
	\caption{Jet centrelines taken along the location of maximum velocity magnitudes ($\lvert\langle{ \boldsymbol{u}}\rangle \rvert_{\textrm{max}}(x)$) in $x$-$z$ plane from slot 3 measurements. Also shown, for comparison are vertical and horizontal slot 1 jets experiments \cite{Soleimaninia2018IJoHE,Soleimaninia2018Ae}.}
	\label{fig.Jet_Centerline_Trajectory3}
\end{figure}
\subsection{The jet centreline trajectory}
In Fig.\ \ref{fig.Jet_Centerline_Trajectory3}, the jet centreline trajectories, obtained in the $x$-$z$ plane, are presented for all slot 3 jets. Here, the trajectories were determined by the maximum velocity magnitude, $\lvert\langle{ \boldsymbol{u}}\rangle \rvert_{\textrm{max}}(x)$ locations. Also shown for comparison are the jet centreline trajectories obtained in the far field (beyond $x\geq10D_{eq}$) from previous vertical and horizontal slot 1 jet experiments \cite{Soleimaninia2018IJoHE,Soleimaninia2018Ae}. In general, the centrelines obtained from slot 3 experiments followed a nearly linear upward trajectory, in the direction of flow within the tube, after the potential core lengths of  $x\sim3D_{eq}$ for helium and $x\sim4D_{eq}$ for air. Around $x\sim17D_{eq}$ for helium, and $x\sim25D_{eq}$ for air, a sudden change in the jet trajectory was observed.  These locations coincided with the extent of the potential-core regions as shown in Figs.\ \ref{fig.Velocity_Concentartion_Contours_Slot2-3_V} - \ref{fig.Velocity_Concentartion_Contours_Slot2-3_H}.  In order to determine the effect of buoyancy on the horizontal slot 3 jets after these locations, lines of best fit, using linear regression to power law expressions were presented for the far field and are also shown in Fig.\ \ref{fig.Jet_Centerline_Trajectory3}. Upon comparison of the centreline trajectories of air to helium, for both vertical and horizontal jets, it became clear that the buoyancy of the helium jet caused significant deflection from the horizontal axis, despite the high Froude number ($Fr= 2.8\times 10^6$). While such buoyancy effects were not observed in the vertical and horizontal air experiments, jet centreline trajectories were almost parallel to each other (even upon extrapolating the line of best fit to the far field, beyond the experimental data collected). In addition, air was found to deviate from the $x$-axis to a greater extent than helium, owing to its higher gas density and negligible buoyancy effect. The same remarks were previously observed for slot 1 jets \cite{Soleimaninia2018IJoHE,Soleimaninia2018Ae}. However, the jet trajectory for the horizontal helium slot 3 jet was found to be described by a nearly linear relation (i.e. power law exponent $\sim1$), whereas the horizontal helium slot 1 jet was previously found to have a power law exponent $\sim1.3$. Slot 1 jets experienced a significant upward deflection, in the direction of flow inside the tube, compared to slot 3 jets as shown in Fig.\ \ref{fig.Jet_Centerline_Trajectory3}. 
\subsection{Velocity decay and jet spreading rates}
\begin{figure}[h!]
	\centering
	a)\includegraphics[scale=0.245]{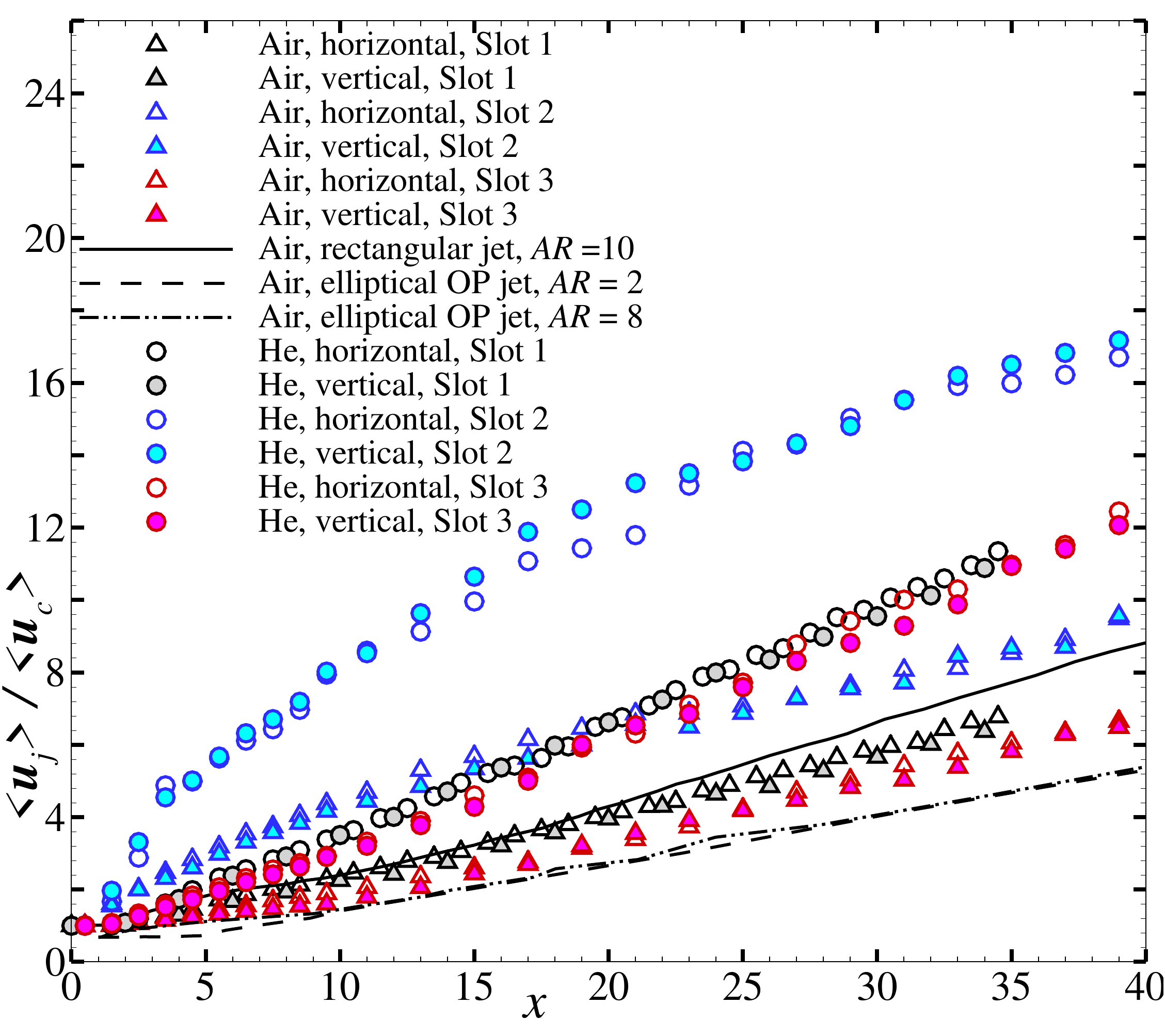} 
	b)\includegraphics[scale=0.245]{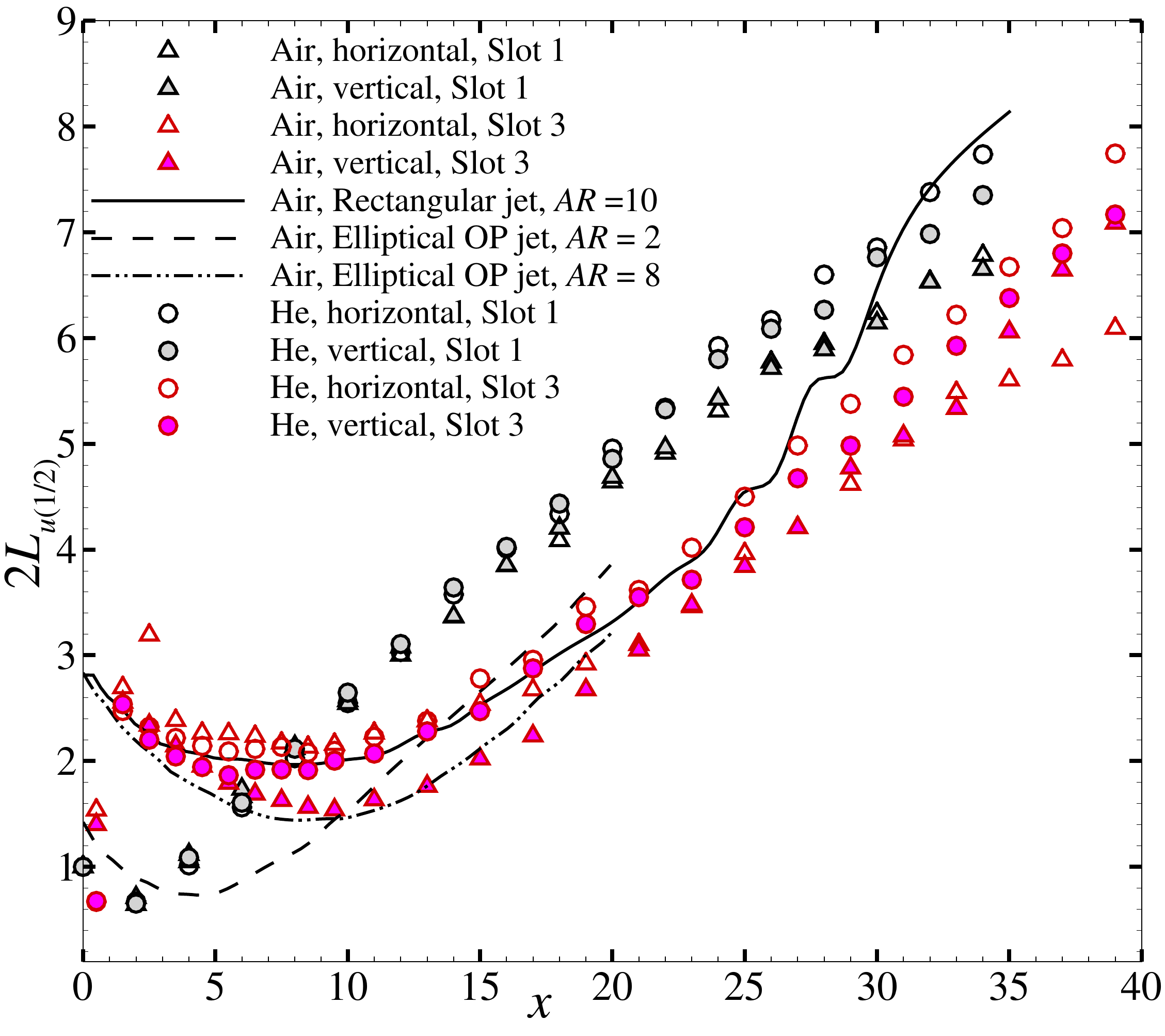} 
	\caption{a) Inverse time-averaged velocity decay and b) jet velocity widths (2$L_{u(1/2)}$) obtained along the $\lvert\langle{ \boldsymbol{u}}\rangle\rvert_{\textrm{max}}(x)$ centrelines, in $x$-$z$ plane from measurements. Note, $n$-coordinate refers to lines which are normal to the centreline, and coplanar with the $x$-$z$ plane (see the coordinate system in Fig.\ref{fig.Experimental_Layout3} e-f). Also shown, for comparison are vertical and horizontal 3D Slot 1 experiments \cite{Soleimaninia2018IJoHE,Soleimaninia2018Ae}, vertical air sharp-edged rectangular and OP elliptical jet measurements \cite{Quinn1992ETaFS203,Hussain1989JoFM257}.} 
	\label{fig.Velocity_JetDecay_SpreadingRate3}
\end{figure}

Fig.\ \ref{fig.Velocity_JetDecay_SpreadingRate3} a) shows the inverse time-averaged velocity decay ($\langle{ \boldsymbol{u}_j}\rangle/\langle{ \boldsymbol{u}_c}\rangle$) along the jet centrelines ($s$-coordinate illustrated in Fig.\ \ref{fig.Experimental_Layout3} e-f) for all experiments. Here, the subscript `$c$' refers to the conditions at the jet centreline, while the subscript `$j$' refers to the jet exit condition. Also shown, for comparison, are velocity decay rates obtained from vertical  and horizontal slot 1 experiments \cite{Soleimaninia2018IJoHE,Soleimaninia2018Ae}, vertical air sharp-edged rectangular ($AR=10$) \cite{Quinn1989PoFA1716} and OP elliptical jet ($AR=2$ \& $8$) \cite{Hussain1989JoFM257} measurements. Upon comparison of air slots 1 \& 3 to the sharp-edged rectangular and OP elliptical jets, both horizontal and vertical slot 3 jets ($AR=10$) show the same velocity decay rates to that of the OP elliptical jet with $AR=8$ in the near field ($x<10D_{eq}$). Within this range, a slightly higher decay rate was found for the centreline velocity of slot 3 jets compared to the OP elliptical jet with an $AR=2$. Beyond this range, velocity decay rates for air slot 3 jets were found to be higher than both OP elliptical jets. Furthermore, both air slot 3 jets exhibited a lower velocity decay rate compared to the sharp-edged rectangular jet ($AR=10$) within the entire domain. On the other hand, the horizontal and vertical slot 1 experiments ($AR=1$) were found to exhibit almost the same velocity decay rate as that of the sharp-edged rectangular jet ($AR=10$) up to $x\sim17D_{eq}$, and after this point, the sharp-edged rectangular jet was found to have a slightly higher decay rate for the rest of the measurement's domain. In contrast, the velocity decay rates for the slot 1 jets were found to be higher than those of both OP elliptical jets within the entire domain. Comparison of the velocity decay rate of slot 3 jets with slot 1 jets \cite{Soleimaninia2018IJoHE,Soleimaninia2018Ae} revealed a slightly higher decay rate in slot 1 jets, with helium jet decaying faster than air jet. In general, upon comparison between horizontal and vertical 3D slot jets, buoyancy was not found to significantly affect the velocity decay rates.

Among all slot experiments, the highest velocity decay rate was observed for slot 2 cases. This was due to the geometrical centreline (along the $y$=0 in $x$-$y$ plane) considered for this slot, where the measured velocity was essentially not the maximum value (see the initial saddle-back velocity profile observed in the time-averaged velocity contours in Figs.\ \ref{fig.Velocity_Concentartion_Contours_Slot2-3_V} - \ref{fig.Velocity_Concentartion_Contours_Slot2-3_H}). Also, it should be noted that the fixed  data acquisition plane for slot 2 cases ($x$-$y$ plane) may not accurately acquire gas dispersion in the far field region, due to deflection of the jet from the streamwise axis ($x$ -axis) in the $x$-$z$ plane. While, three-dimensional velocity measurement is not currently available for slot 2 experiments, it would be impossible to point out the exact location of jet deflection from the data acquisition plane; one may visually conclude that the measurement plane has correctly acquired the jet dispersion up to $x\sim15-20D_{eq}$ from the time-averaged velocity contours (Figs.\ \ref{fig.Velocity_Concentartion_Contours_Slot2-3_V} - \ref{fig.Velocity_Concentartion_Contours_Slot2-3_H}). However, the slot 2 experiments data will be presented as qualitative results rather than quantitative data, more specifically in the far field ($x\ge20D_{eq}$) where the result is not representative of the correct gas dispersion.  

For slot 1 \& 3 experiments in the $x$-$z$ plane, Figure \ref{fig.Velocity_JetDecay_SpreadingRate3} b presents the jet velocity widths (2$L_{u(1/2)}$) obtained by determining the locations where $\lvert\langle{ \boldsymbol{u}}\rangle\rvert=0.5\lvert\langle{ \boldsymbol{u}}\rangle\rvert_{\textrm{max}}(x)$ along lines orthogonal to the jet centrelines.  These orthogonal lines have been indicated previously as coordinate `$n$' in Fig.\ \ref{fig.Experimental_Layout3} e-f. For the 3D jets, in slot 3 cases, a large peak was observed in the jet widths within  $1D_{eq}<x<2.5D_{eq}$ distances from the orifice, whereas in the slot 1 experiments a slight contraction has been observed from $1D_{eq}<x<4D_{eq}$.  Beyond this point for slot 3, the jet spreading rates were observed to experience a slight contraction up to $x\sim15D_{eq}$. Where after this point, much greater spreading rates were observed in slot 1 compared to slot 3 jets for all cases.  Moreover, from slot 3 experiments, the jet spreading rate was found to be comparable for both air and helium gases, with exception of vertical air slot 3 case.  However, helium slot 3 jets exhibit higher spreading rates compared to air, in the far field (beyond $x\geq15D$). This trend was slightly more clearer upon comparison of the horizontal slot 3 cases between helium and air. It should be noted that this remark is also evident in slot 1 cases, where a higher spreading rate is found in helium compared to air beyond $x\geq13D$. Upon comparison of slot jets to rectangular \cite{Quinn1992ETaFS203} and OP elliptical \cite{Hussain1989JoFM257} jets, the spreading rate for slot 1 was found to be much higher than others in the range of  $8D_{eq}<x<30D_{eq}$. Beyond this range, the sharp-edged rectangular jet exhibited a slightly higher spreading rate.

\subsection{Scalar concentration decay and jet spreading rates}
\begin{figure}[h!]
	\centering
	a)\includegraphics[scale=0.245]{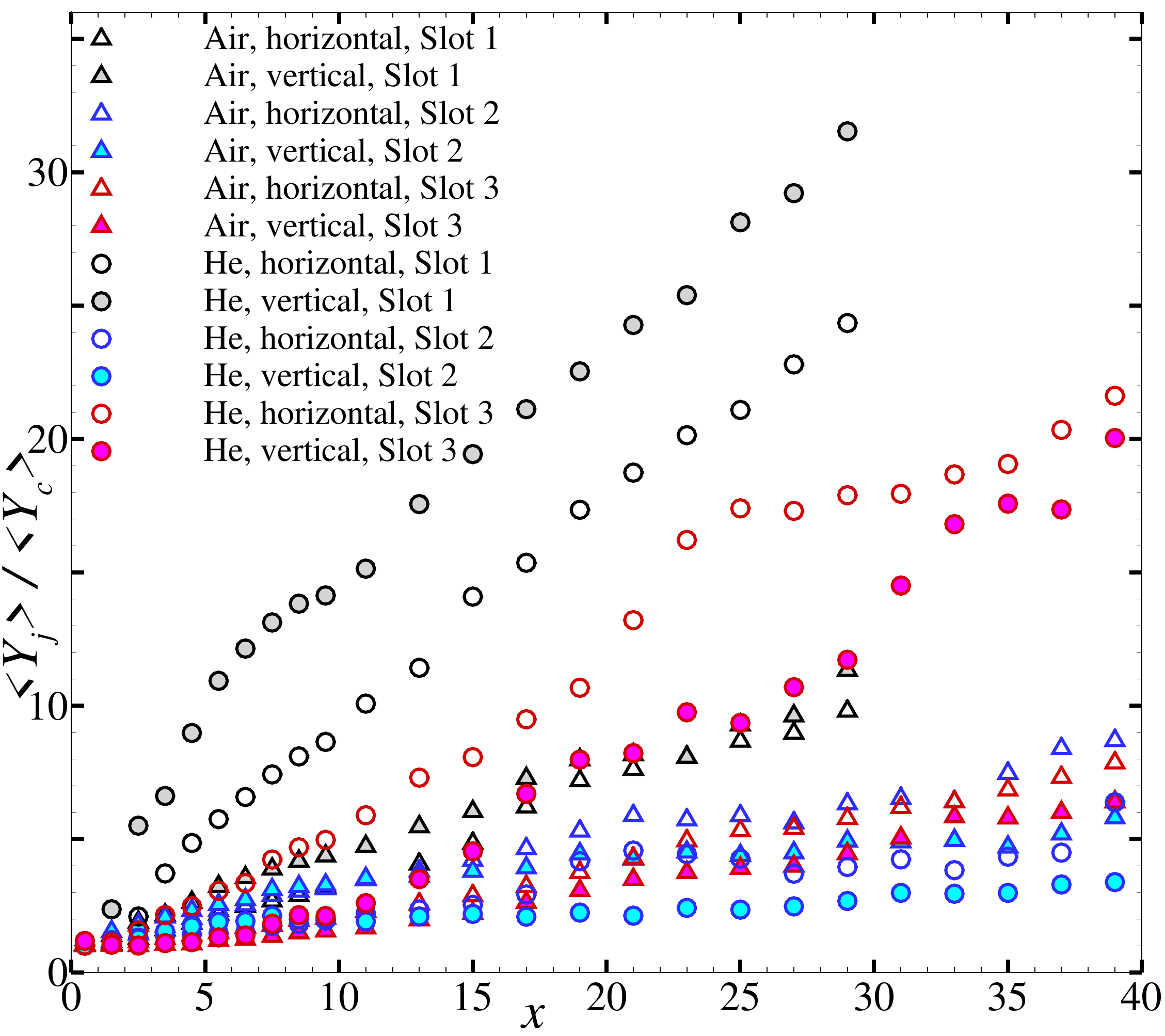} 
	b)\includegraphics[scale=0.245]{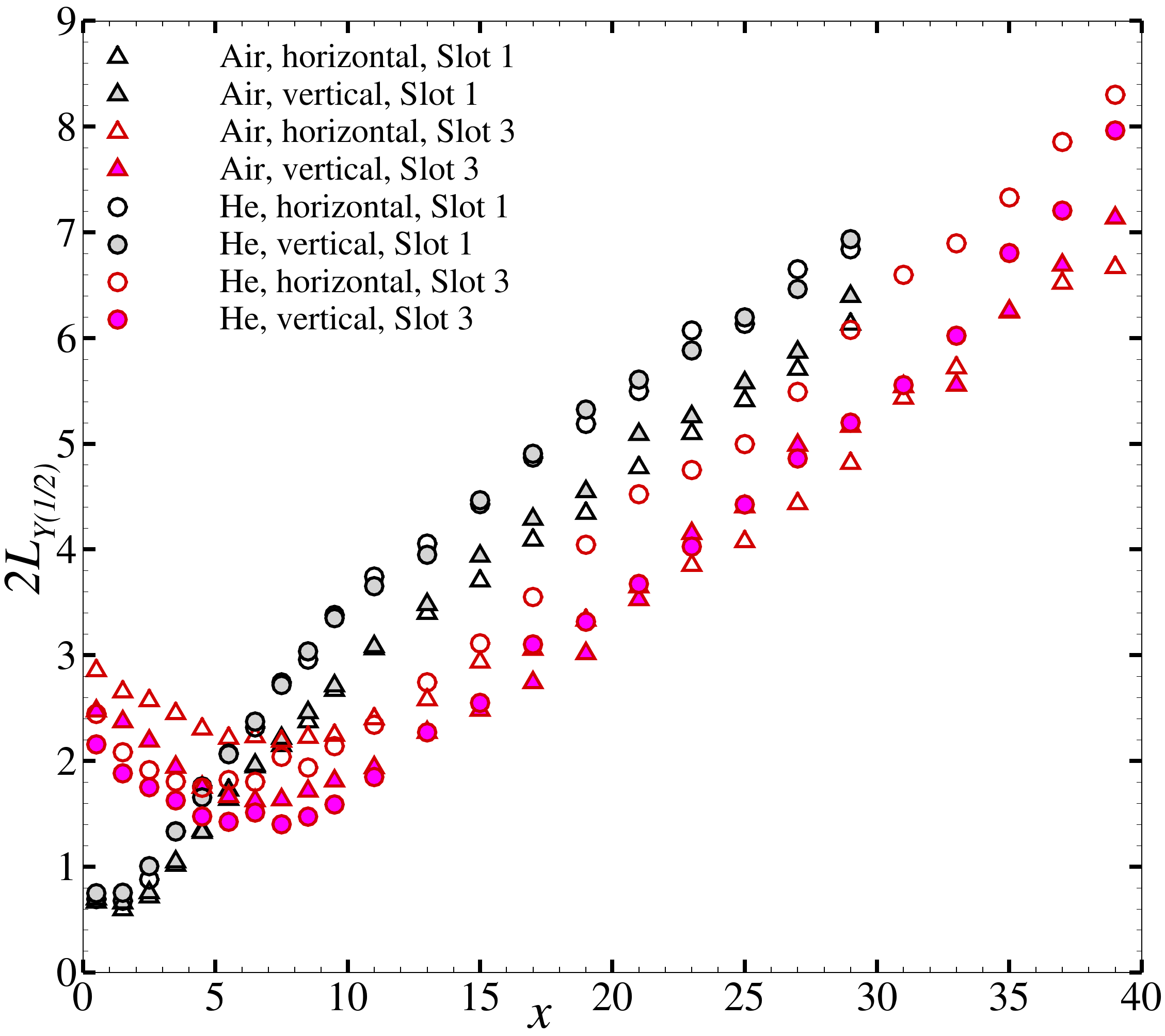}	
	\caption{a) Inverse time-averaged jet gas mass fraction decay and b) mass fraction widths (2$L_{Y(1/2)}$) obtained along the $\langle{Y}\rangle_{\textrm{max}}(x)$ centrelines, in $x$-$z$ plane from measurements. Also shown, for comparison {are} vertical and horizontal 3D slot 1 jet experiments \cite{Soleimaninia2018IJoHE,Soleimaninia2018Ae}.} 
	\label{fig.Concentration_JetDecay_SpreadingRate3}
\end{figure}

Figure\ \ref{fig.Concentration_JetDecay_SpreadingRate3}a shows centreline evolution of the inverse time-averaged jet gas mass fraction, $\langle{Y_j}\rangle/\langle{Y_c}\rangle$, for both air and helium measurements. Here, the jet gas mass fractions were determined from the measured mole fractions through
\begin{equation}
Y=\frac{X W_j}{\overline{W}}
\label{eqn.mass_fraction}
\end{equation}
where $X$ and $W_j$ refer to the mole fraction and molecular weight of the jet gas, respectively, and $\overline{W}$ refers to the mean molecular weight of the local jet gas-ambient air mixture given by
\begin{equation}
\overline{W}= X W_j + (1-X) W_{\textrm{air}}
\label{eqn.mean_molecular_weight}
\end{equation} 

Additionally shown for comparison, are the centreline mass fraction decay rates obtained from vertical and horizontal 3D  slot 1 jet experiments \cite{Soleimaninia2018IJoHE,Soleimaninia2018Ae}. In general for all 3D slot jet cases, except the slot 2, helium jets were always observed to decay faster than air jets. It is noted, however, that the lowest decay rate observed in mass fraction for the helium slot 2 jets are likely due to the geometrical centreline ($y=0$) where mass fraction obtained, and may not represent the location of jet centreline, as previously discussed in velocity decay profiles. Also, the centreline mass fraction decay rates observed in the experimental slot 1 jets were much faster compared to the slot 2 \& 3 jets. Moreover, upon comparison of the helium slot 1 jets ($AR=1$), the vertical orientation was found to have a faster mass fraction decay rate compared to the horizontal case.  Such differences in behaviour was not observed for the air slot 1 jets, suggesting that buoyancy has a remarkable affect on the mass fraction decay rates \cite{Soleimaninia2018Ae}. On the other hand, for helium slot 2 experiments ($AR=10$), a faster decay rate was observed for horizontal case compared to vertical orientation. Again, such remarks were not found to be valid for the air slot 1 jets, suggesting that the aspect ratio plays a more significant role on the mass fraction decay rates, compared to the buoyancy in the 3D high-aspect-ratio jets. Also, upon comparison to the velocity decay rates in Fig.\ \ref{fig.Velocity_JetDecay_SpreadingRate3}a, it is noted that the jet centreline mass fraction decays faster than the velocity for helium, owing to the low Schmidt number ($Sc<1$). This was also concluded in the recent experimental studies on the slot 1 jets \cite{Soleimaninia2018IJoHE,Soleimaninia2018Ae}.

As was done for the velocity field, the jet widths based on the jet gas mass fraction (2$L_{Y(1/2)}$) have been obtained for slot jet experiments; these are presented in Fig.\ \ref{fig.Concentration_JetDecay_SpreadingRate3}b. This was achieved by determining the locations where $\langle{Y}\rangle=0.5\langle{Y}\rangle_{\textrm{max}}(x)$ along orthogonal lines to the jet centreline, in the $x$-$z$ planes. Also shown, for comparison, are the jet widths based on the jet gas mass fraction obtained from vertical and horizontal slot 1 jet experiments \cite{Soleimaninia2018IJoHE,Soleimaninia2018Ae}. For all 3D jet cases, a slight contraction in the jet mass fraction widths were observed from $x<4D_{eq}$ and $x<15D_{eq}$ for slots 1 \& 3, respectively, as was previously observed for the jet widths based on velocity. Beyond this point, the jet scalar growth rates, along $n$, were observed to be much greater in slot 1 compared to slot 3 jets for all cases. The helium jet also exhibited a faster spreading rate compared to air, in both horizontal and vertical cases. 
\subsection{Jet centreline statistics}

\begin{figure}[h!]
	\centering
	\raggedright \underline{\textbf{Slot 2, vertical:}}\\
	a)\includegraphics[scale=0.245]{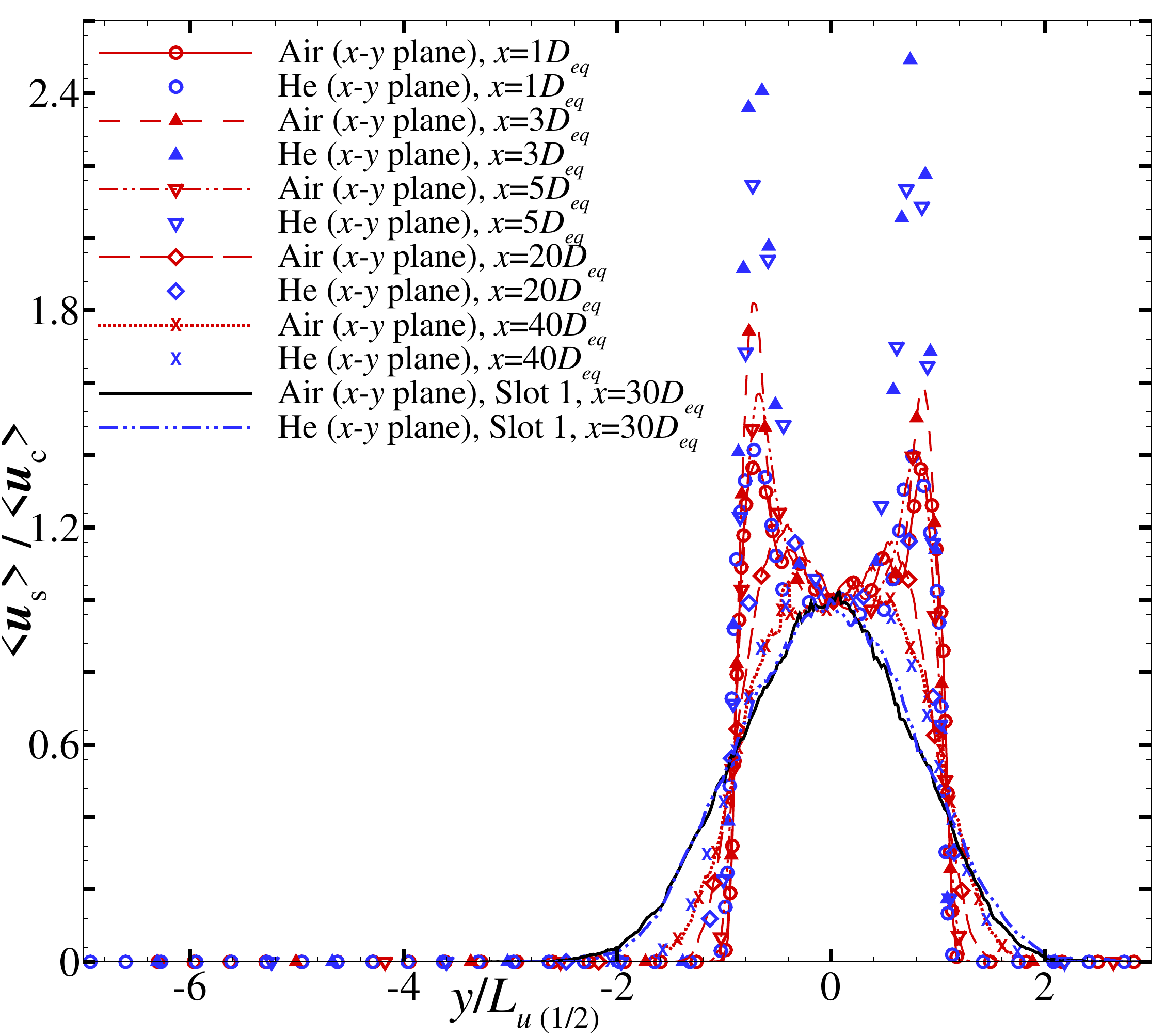} 
	b)\includegraphics[scale=0.245]{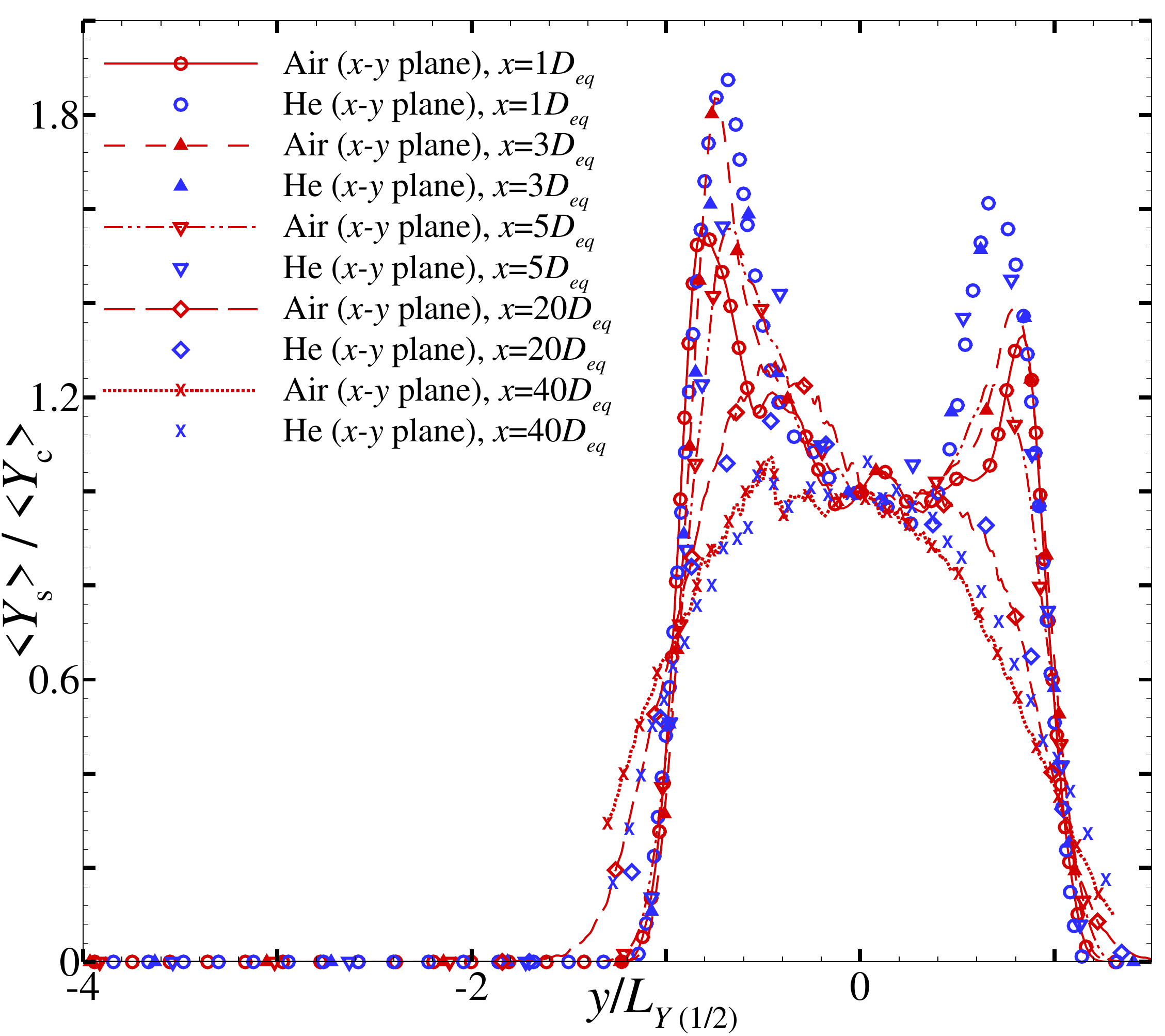}\\
	\raggedright \underline{\textbf{Slot 2, horizontal:}}\\
	a)\includegraphics[scale=0.245]{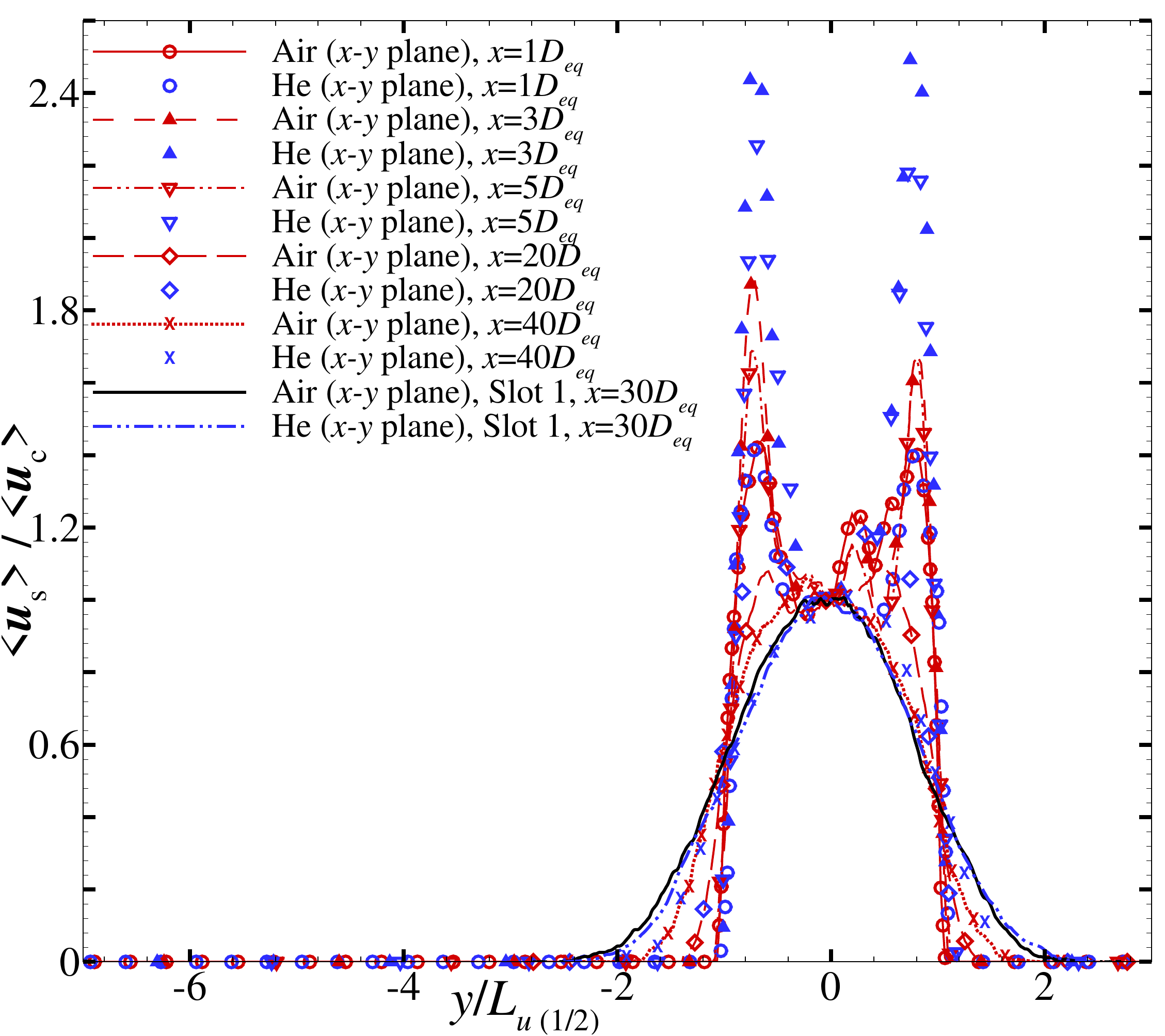} 
	b)\includegraphics[scale=0.245]{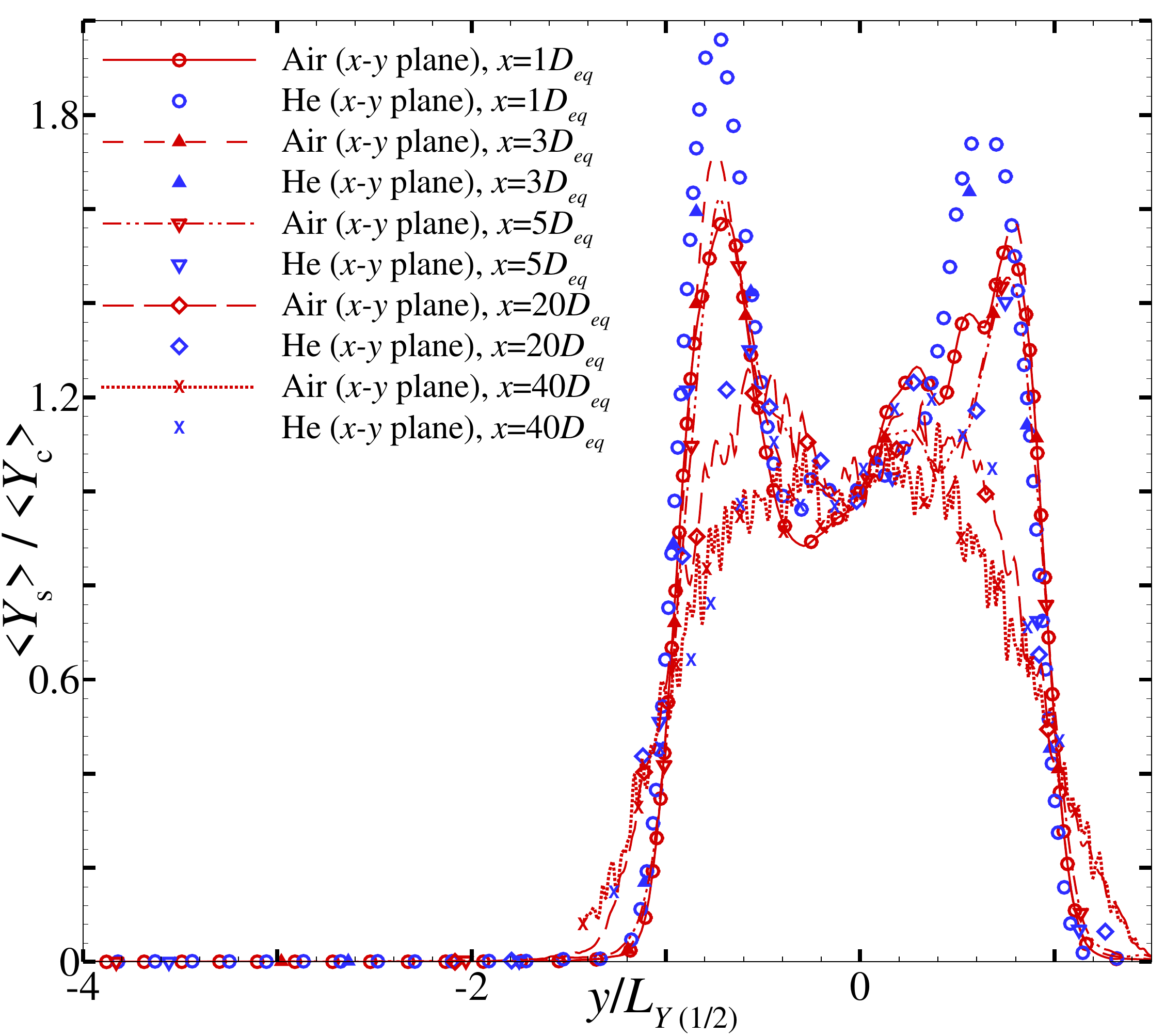}\\
	\caption{a) Normalized time-averaged velocity, and b) mass fraction profiles along jet centrelines ($y=0$) in $x$-$y$ plane, taken at various heights for both air and helium. Time-averaged velocity and mass fraction profiles are also compared to experimental vertical and horizontal 3D slot 1 jets \cite{Soleimaninia2018IJoHE,Soleimaninia2018Ae}.}
	\label{fig.Vs_Cs_Slot2}
\end{figure}

\begin{figure}[h!]
	\centering
	\raggedright \underline{\textbf{Slot 3, vertical:}}\\
	a)\includegraphics[scale=0.245]{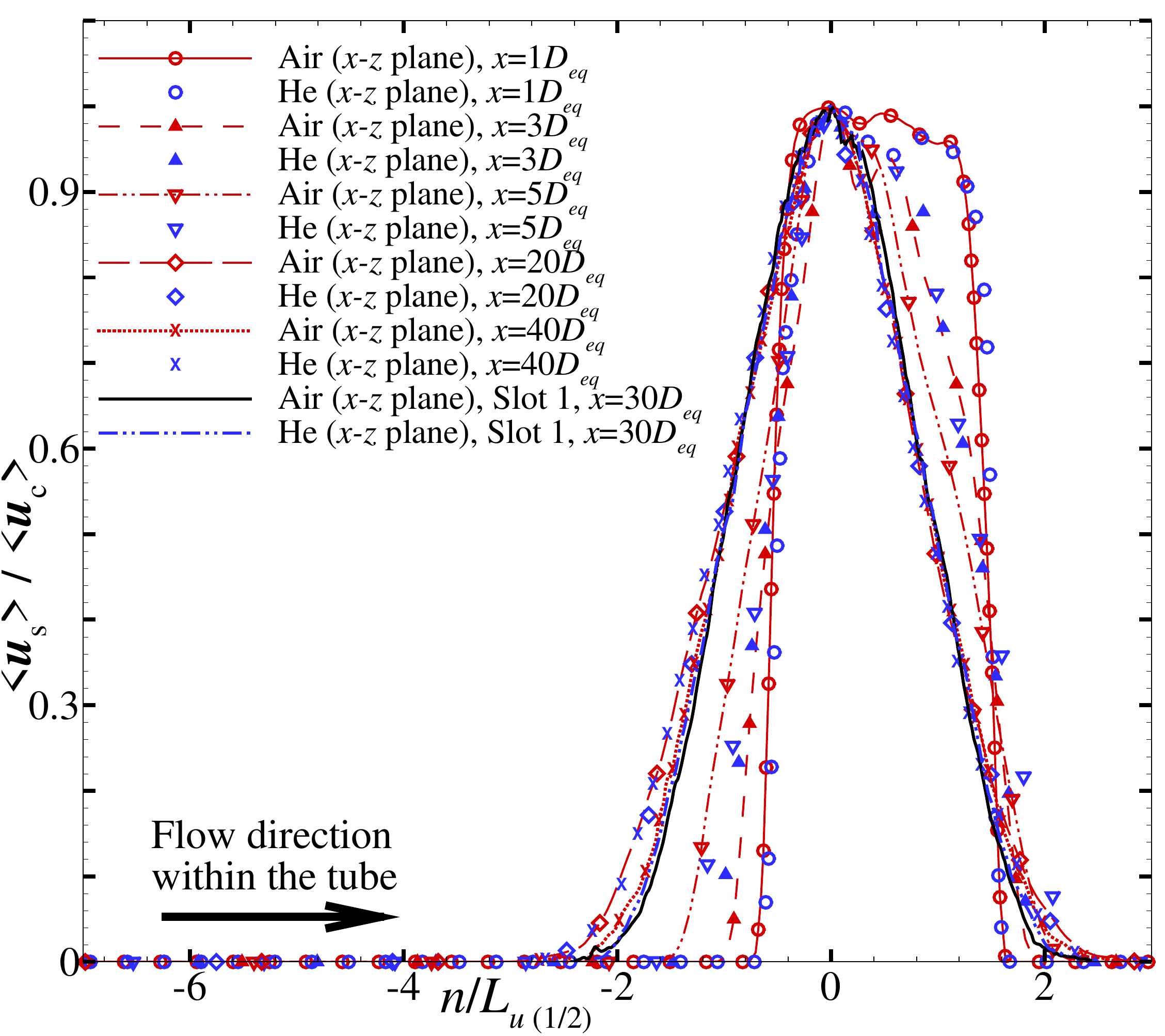} 
	b)\includegraphics[scale=0.245]{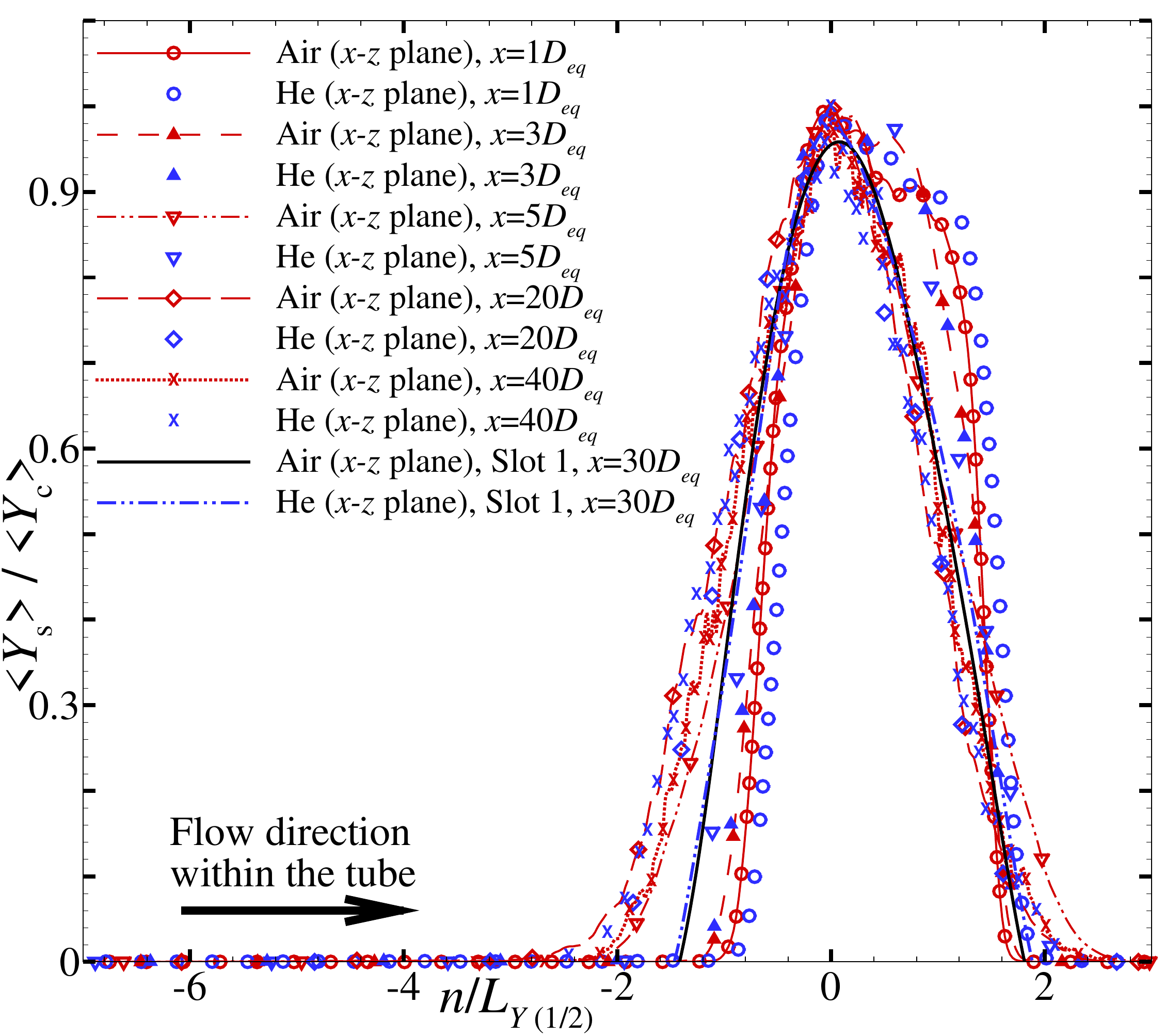}\\
	\raggedright \underline{\textbf{Slot 3, horizontal:}}\\
	a)\includegraphics[scale=0.245]{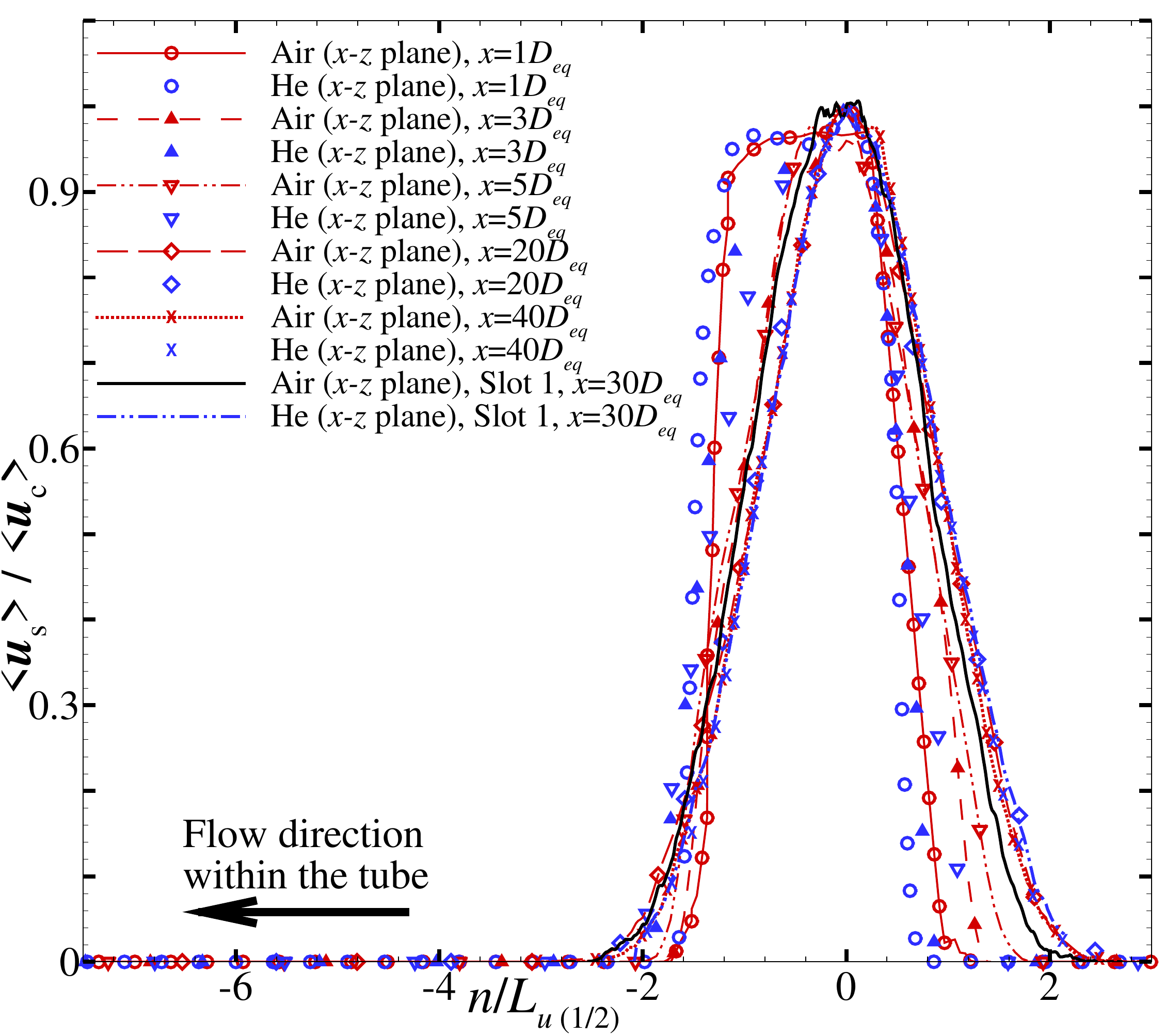} 
	b)\includegraphics[scale=0.245]{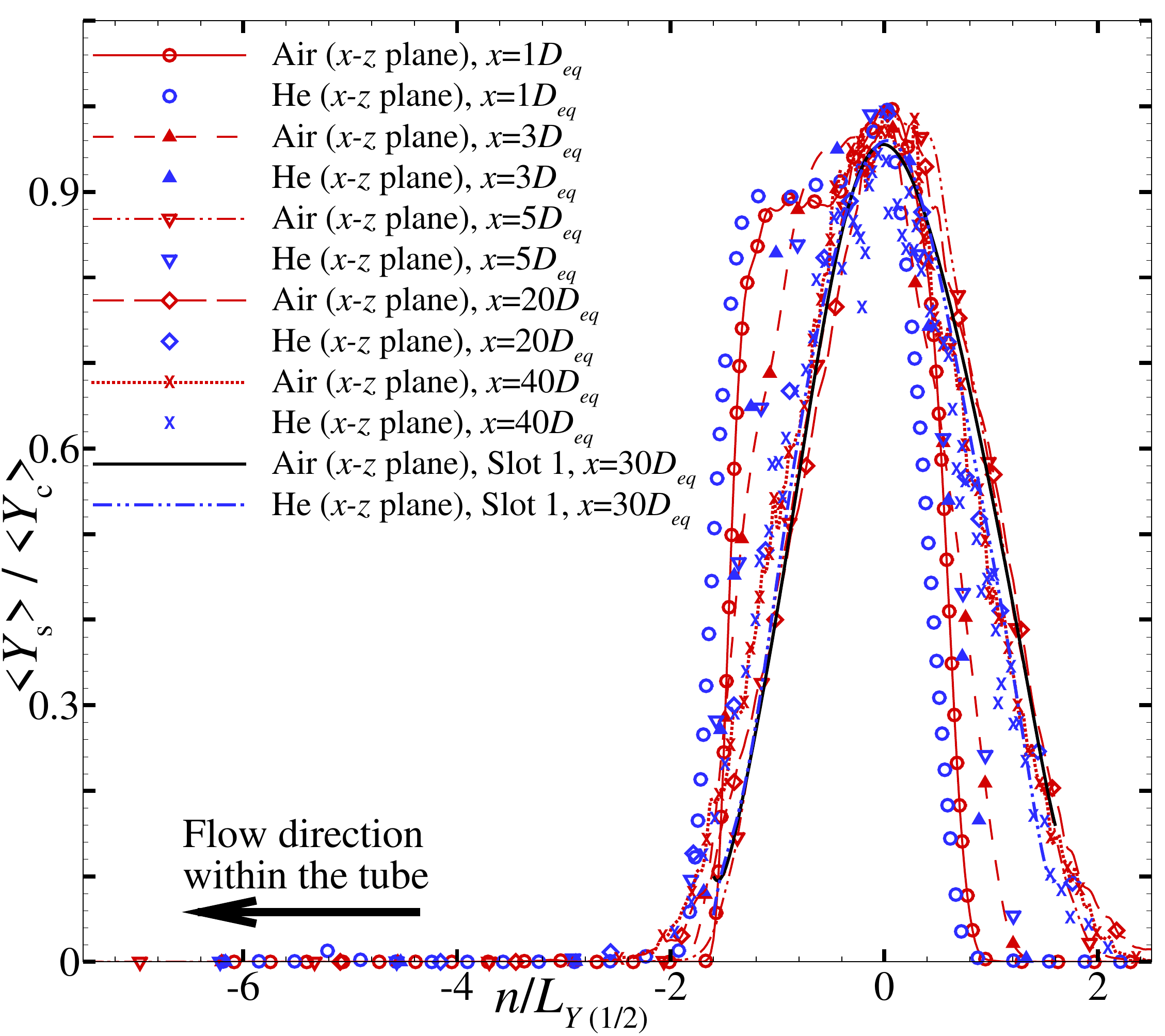}\\
	\caption{a) Normalized time-averaged velocity, and b) mass fraction profiles along jet centrelines in $x$-$z$ plane, taken at various heights for both air and helium. Time-averaged velocity and mass fraction profiles are also compared to experimental vertical and horizontal 3D slot 1 jets \cite{Soleimaninia2018IJoHE,Soleimaninia2018Ae}.}
	\label{fig.Vs_Cs_Slot3}
\end{figure}

The normalized time-averaged velocity profiles for all slot 2 \& 3 experiments are shown in Fig.\ \ref{fig.Vs_Cs_Slot2}a and Fig.\ \ref{fig.Vs_Cs_Slot3}a, respectively,  along the $y$ \& $n$ directions for several downstream locations along the jet centreline ($s$-curve Fig.\ \ref{fig.Experimental_Layout3}). Also shown for comparison are the velocity statistics obtained for the vertical and horizontal slot 1 experiments \cite{Soleimaninia2018IJoHE,Soleimaninia2018Ae}. It should be noted that the $s$-component velocities presented here were normalized by the local centreline velocity magnitudes,  $\langle{ \boldsymbol{u}}\rangle_{c}(s)$. Also, the $n$-coordinate, which was normal to the centreline $s$-curve, was normalized by the jet velocity half widths ($L_{u(1/2)}$) determined from Fig.\ \ref{fig.Velocity_JetDecay_SpreadingRate3}b.

For slot 2 in the $x$-$y$ plane, in both orientations for air and helium experiments (Fig.\ \ref{fig.Vs_Cs_Slot2}a), the jets emerged from the slot with a saddle-back profile; higher humps exhibited in the helium cases compared to the air, located at $\sim(y/L_{1/2}) \pm 0.75$.  This saddle-back profile was observed up to $x<40D_{eq}$, and beyond this location, all slot 2 cases developed into a self-similar Gaussian-like distribution of velocity. Whereas in the slot 1 experiments, the velocity profiles were previously found to develop to the self-similar Gaussian distribution much closer to the orifice, beyond $x>5D_{eq}$ \cite{Soleimaninia2018IJoHE,Soleimaninia2018Ae}. Finally, the curves obtained for both gases were found to be in agreement with each other in the far field region.

In the $x$-$z$ plane of slot 3, in all cases (Fig.\ \ref{fig.Vs_Cs_Slot3}a), the jets emerged from the slot with an initial semi top-hat profile, not shown here.  This behaviour was also previously observed in the slot 1 measurements \cite{Soleimaninia2018IJoHE,Soleimaninia2018Ae}. It should be noted that this semi top-hat profile was observed to deviate from the jet streamwise axis ($x$-axis) in the direction of flow inside the tube and was maintained up to $x\sim5D_{eq}$ distance from the slot. In both orientations, air and helium experiments were found to exhibit slightly more velocity spreading to the lower side of the jet centre (in the direction of flow inside the tube) compared to the other side, with more velocity spreading was observed for helium compared to air. Beyond $x>5D_{eq}$, in the far field, the experimental slot 3 jets developed into, and matched, the self-similar Gaussian distribution obtained from the slot 1 jets \cite{Soleimaninia2018IJoHE,Soleimaninia2018Ae}. Like the slot 2 experiments, the curves obtained for all gases were found to be in well agreement with each other in the far field.

The time-averaged mass fraction profiles for all slot 2 \& 3 jet experiments are shown in Fig.\ \ref{fig.Vs_Cs_Slot2}b and Fig.\ \ref{fig.Vs_Cs_Slot3}b, respectively. Here, time-averaged mass fraction ($\langle{Y}\rangle$), have been normalized by the local centreline mass fraction, $\langle{Y_c}\rangle(s)$. The $n$ and $y$ coordinates were normalized by the jet mass fraction half widths ($L_{Y(1/2)}$) determined from the locations where $\langle{Y/Y_c}\rangle=0.5$, Fig.\ \ref{fig.Concentration_JetDecay_SpreadingRate3}b.  In general, they were found to be qualitatively similar to the velocity profiles in all cases. For slot 2 in the $x$-$y$ plane (Fig.\ \ref{fig.Vs_Cs_Slot2}b), the saddle-back profile with the sharp humps was observed at $\sim(y/L_{1/2}) \pm 0.75$; higher humps exhibited in the helium cases compared to the air within the range of $x<5D_{eq}$ distance to the orifice. In all slot 2 cases, the saddle-back profile was developed into the self-similar Gaussian distribution beyond $x>40D_{eq}$, same as velocity profiles. In the $x$-$z$ plane (Fig.\ \ref{fig.Vs_Cs_Slot3}b), all slot 3 experiments exhibited initial top-hat mass fraction profile, with deviation from the jet streamwise axis ($x$-axis) in the direction of flow inside the tube. Where slightly more mass fraction spreading were found to exhibit to the lower side of the jet centre (in the direction of flow inside the tube), with more mass fraction spreading found for helium compared to air.  In the far field, beyond $x>5D_{eq}$,  the mass fraction profile was found to develop quickly into the self-similar solution as observed in the slot 1 experiments \cite{Soleimaninia2018IJoHE,Soleimaninia2018Ae}. 
\begin{figure}[h]
	\centering
	a)\includegraphics[scale=0.245]{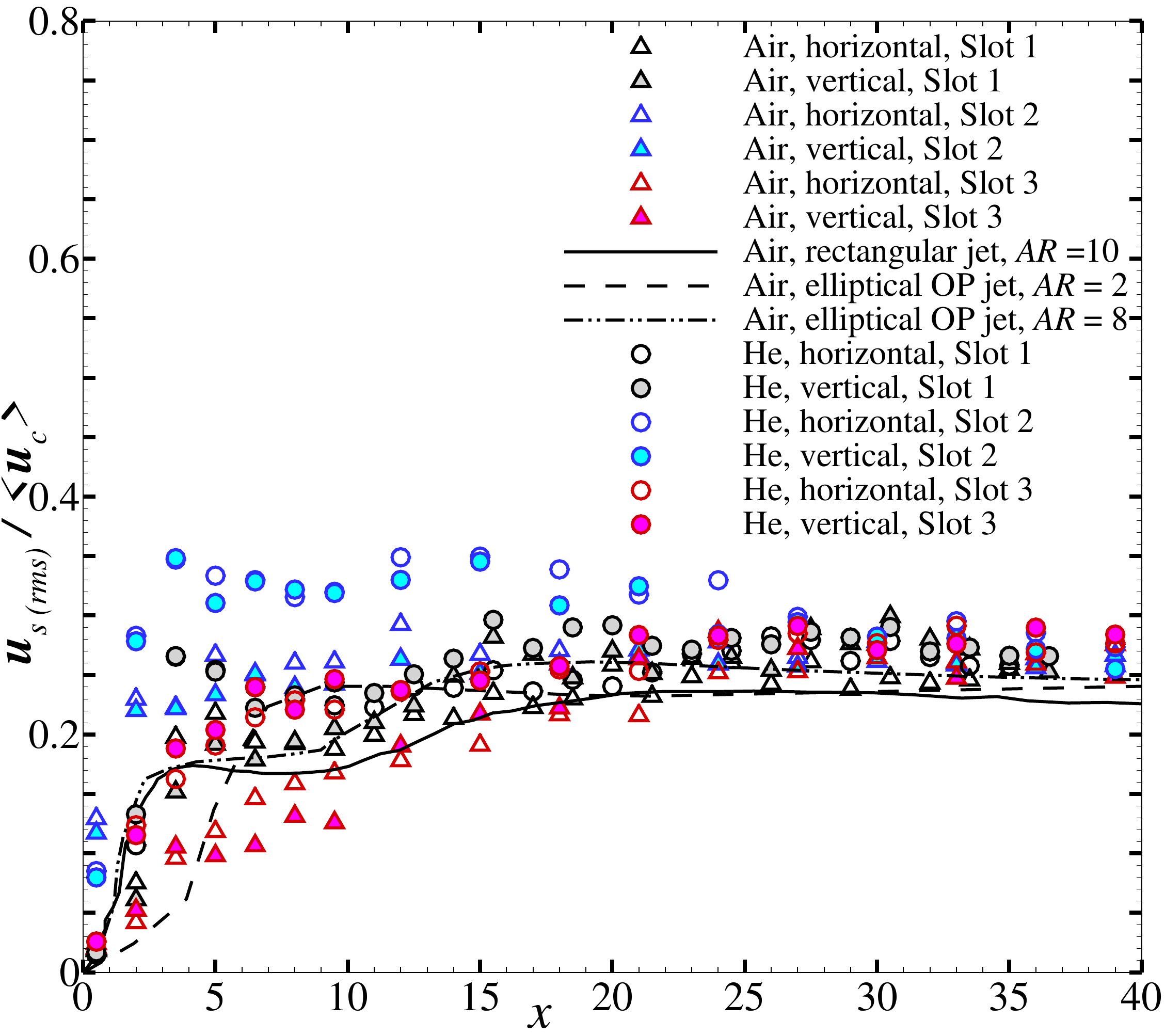} 
	b)\includegraphics[scale=0.245]{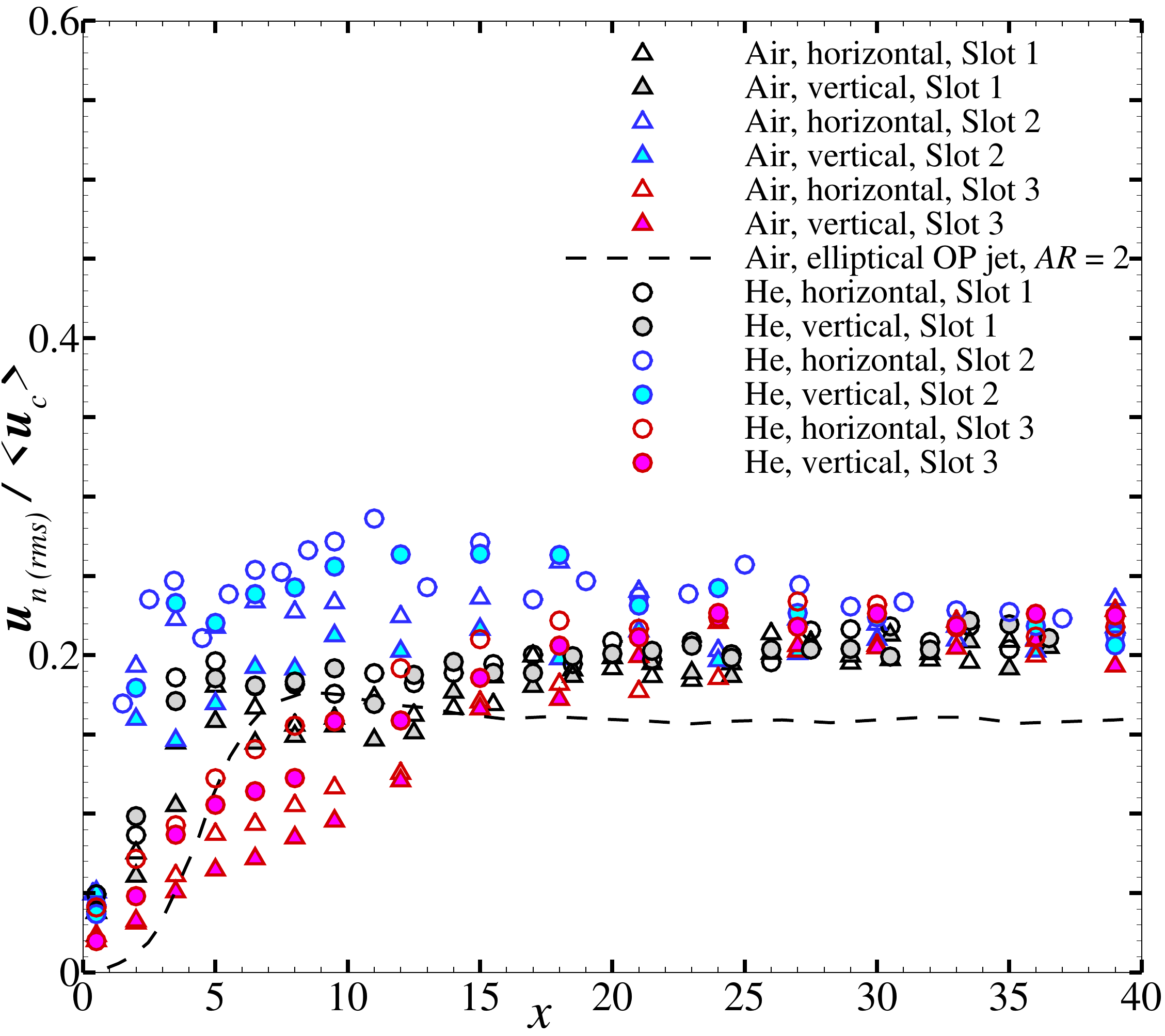}\\
	\caption{Axial development of turbulence intensities along jet centrelines, a) tangential turbulence intensity component ($\boldsymbol{u}_{s(rms)}/ \langle{\boldsymbol{u}_{c}}\rangle$) and b) orthogonal turbulence intensity component ($\boldsymbol{u}_{n(rms)}/ \langle{\boldsymbol{u}_{c}}\rangle$) for experiments. Also shown, for comparison are vertical and horizontal 3D Slot 1 experiments \cite{Soleimaninia2018IJoHE,Soleimaninia2018Ae}, vertical air sharp-edged rectangular and OP elliptical jet measurements \cite{Quinn1992ETaFS203,Quinn2007EJoM-B583,Hussain1989JoFM257}.}
	\label{fig.Velocity_fluctuation_rms_Slot1-3}
\end{figure} 

Fig.\ \ref{fig.Velocity_fluctuation_rms_Slot1-3} shows the normalized axial evolution of the r.m.s. velocity fluctuation components in the $s$ and $n$ directions, tangential and orthogonal to jet centreline, where  $\boldsymbol{u}_{(rms)}=\langle{\boldsymbol{u}^{\prime 2}}\rangle^{1/2}$. It should be noted that the prime ($^\prime$) represents the instantaneous fluctuating quantity ($\boldsymbol{u}^{\prime}=\boldsymbol{u}-\langle{\boldsymbol{u}}\rangle$). For helium vertical and horizontal slot 3 jets, the tangential turbulence intensity ($\boldsymbol{u}_{s(rms)}$) reached an asymptotic value of $\sim26\%$ at $x=15D_{eq}$, whereas such a value was not observed until $x=25D_{eq}$ and $x=20D_{eq}$ for air horizontal and vertical slot 3 jets, respectively. This trend was also observed in slot 1 measurements \cite{Soleimaninia2018IJoHE,Soleimaninia2018Ae} where helium reached the asymptotic value closer to the orifice, at $x=3D$, compared to horizontal and vertical air at $x=20D_{eq}$ and $x=15D_{eq}$, respectively. However, it appears that this asymptotic value $\sim26\%$ would be reached in the far field ($x>33D$) for all slot jets. On the other hand, the vertical air sharp-edged rectangular \cite{Quinn1992ETaFS203} and OP elliptical jet \cite{Hussain1989JoFM257} measurements reached the asymptotic values $\sim23\%$ and $\sim24\%$ in the far field, respectively. It should be noted that slightly lower turbulent intensities of the sharp-edged rectangular and OP elliptical jet, observed in both tangential and orthogonal components, are likely due to slightly higher initial turbulent intensities reported for those measurements (not shown)\cite{Mi2007EiF625}. The same remark is valid for the orthogonal turbulence intensity, as helium slot 3 jets reached the asymptotic value $\sim19-22\%$ more closer to orifice at $x=17D_{eq}$, compared to air at $x=27D_{eq}$. Also, the OP elliptical air jet reached the peak turbulence intensity ($\sim18\%$) at $x=8D_{eq}$ and then recovered the asymptotic value  $\sim16\%$ until $x=15D_{eq}$ for the rest of the measurement domain \cite{Quinn2007EJoM-B583}. In general, the intensity of tangential velocity fluctuations was higher than the orthogonal components, as observed in previous studies on 3D round \& elliptical jets \cite{Soleimaninia2018IJoHE,Soleimaninia2018Ae,Quinn2007EJoM-B583}.
\begin{figure}[h]
	\centering
	\includegraphics[scale=0.33]{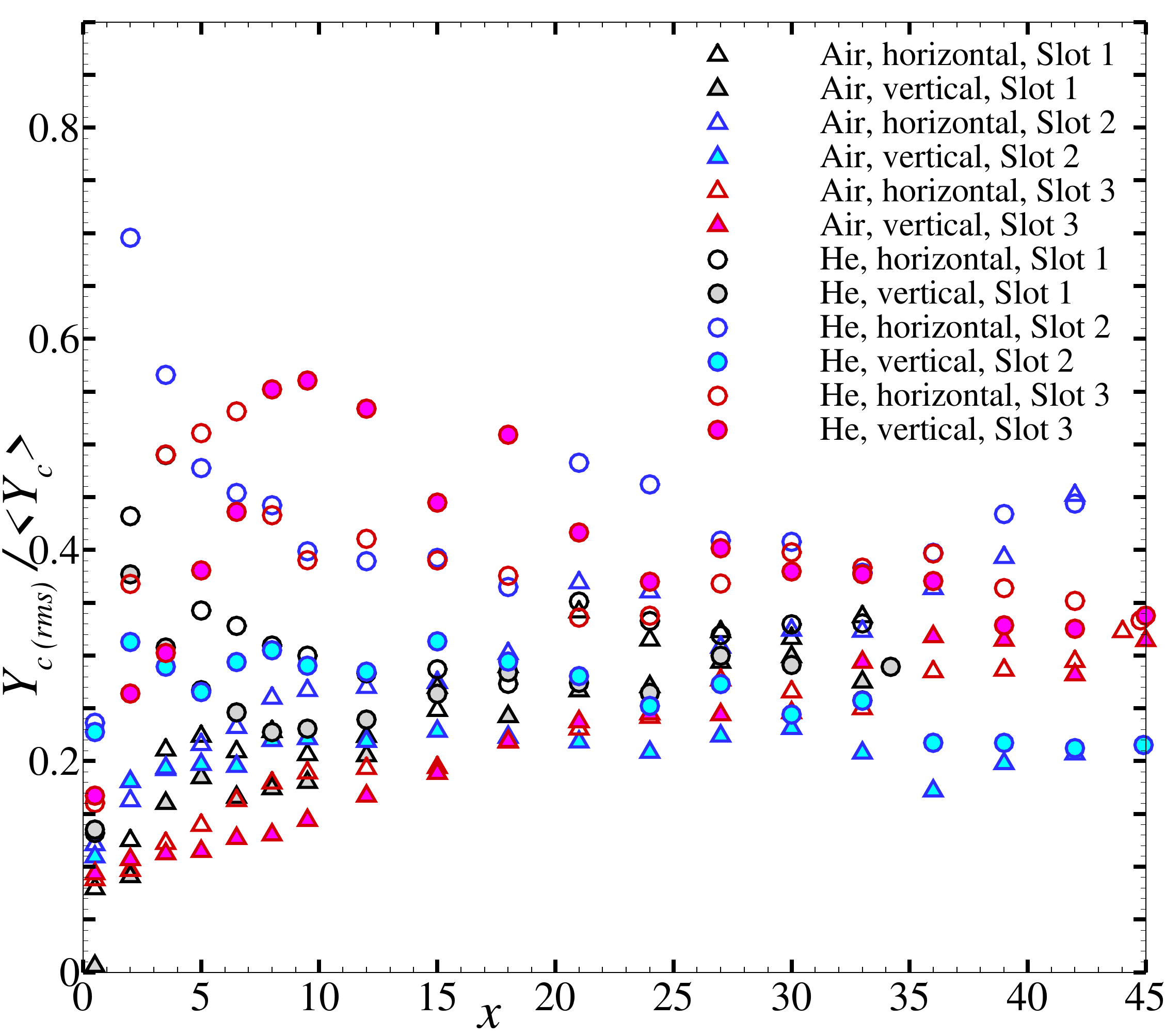} 
	\caption{Normalized axial evolution of mass fraction fluctuation intensities along jet centrelines, ${Y}_{c(rms)}/ \langle{{Y}_{c}}\rangle$, for experiments. Also shown, for comparison {are} vertical \& horizontal 3D slot 1 jets \cite{Soleimaninia2018IJoHE,Soleimaninia2018Ae}.}
	\label{fig.Concenration_fluctuation_rms_Slot1-3}
\end{figure} 

Fig.\ \ref{fig.Concenration_fluctuation_rms_Slot1-3} shows the normalized axial evolution of the r.m.s.\ jet gas mass fraction fluctuations (unmixedness), ${Y}_{c(rms)}/ \langle{{Y}_{c}}\rangle$, along the jet centreline, for all experiments. Also shown for comparison are vertical \& horizontal slot 1 jets \cite{Soleimaninia2018IJoHE,Soleimaninia2018Ae}. In the vertical jets, helium slot 2 reached the asymptotic value $\sim26\%$ at $x=5D_{eq}$, which is in agreement with the value reported for helium slot 1, which then decreased slightly to the asymptotic value $\sim21\%$ in the far field ($x>35D_{eq}$). In contrast, for air slot 2 such a value was recovered only when $x=8D_{eq}$ and it remained almost the same for the rest of data acquisition domain. It should be noted that this asymptotic value is in agreement with the values reported in literature for the far field of round axisymmetry jets ($\sim21-24\%$) \cite{Panchapakesan1993JoFM225, Chen1980, Richards1993JoFM417}. In contrast, for the vertical slot 3 cases, helium reached an unmixedness value $\sim55\%$ where $x=8D_{eq}$ and then recovered the asymptotic value $\sim33\%$ in the far field ($x>38D_{eq}$). The vertical air slot 3 also recovered this asymptotic value beyond the $x>34D_{eq}$ range. Helium horizontal slot 2 jet also reached a pick unmixedness value of $\sim70\%$ at $2D_{eq}$ but then decreased to the asymptotic value $\sim44\%$ beyond the $x>38D_{eq}$ for the rest of the measurement domain; whereas such a value was not recovered until $x=41D_{eq}$ for air.  For the horizontal slot 3 cases, helium reached an umixedness value $\sim55\%$ at $x=7D_{eq}$ and then recovered the asymptotic value $\sim33\%$ at the end of measurement domain ($x>43D_{eq}$). Similarly, air horizontal slot 3 jet recovered this value almost the same distance from the slot. This asymptotic value is in agreement with the previous measurement of horizontal slot 1 jets in the far field ($\sim33\%$)  \cite{Soleimaninia2018Ae}. 

\begin{figure}[h!]
	\centering
	\raggedright \underline{\textbf{Slot 2, vertical:}}\\
	a)\includegraphics[scale=0.245]{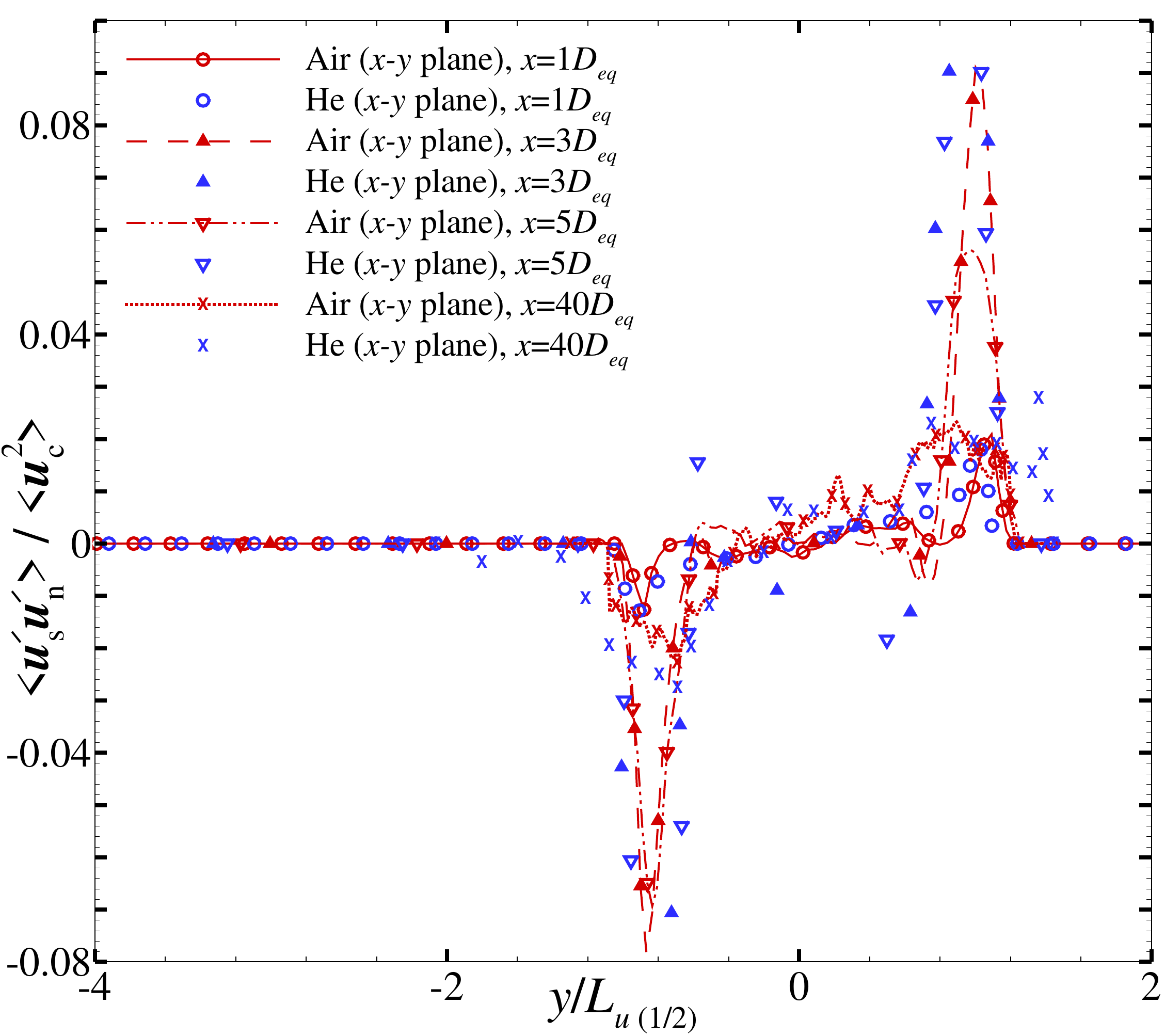} 
	b)\includegraphics[scale=0.245]{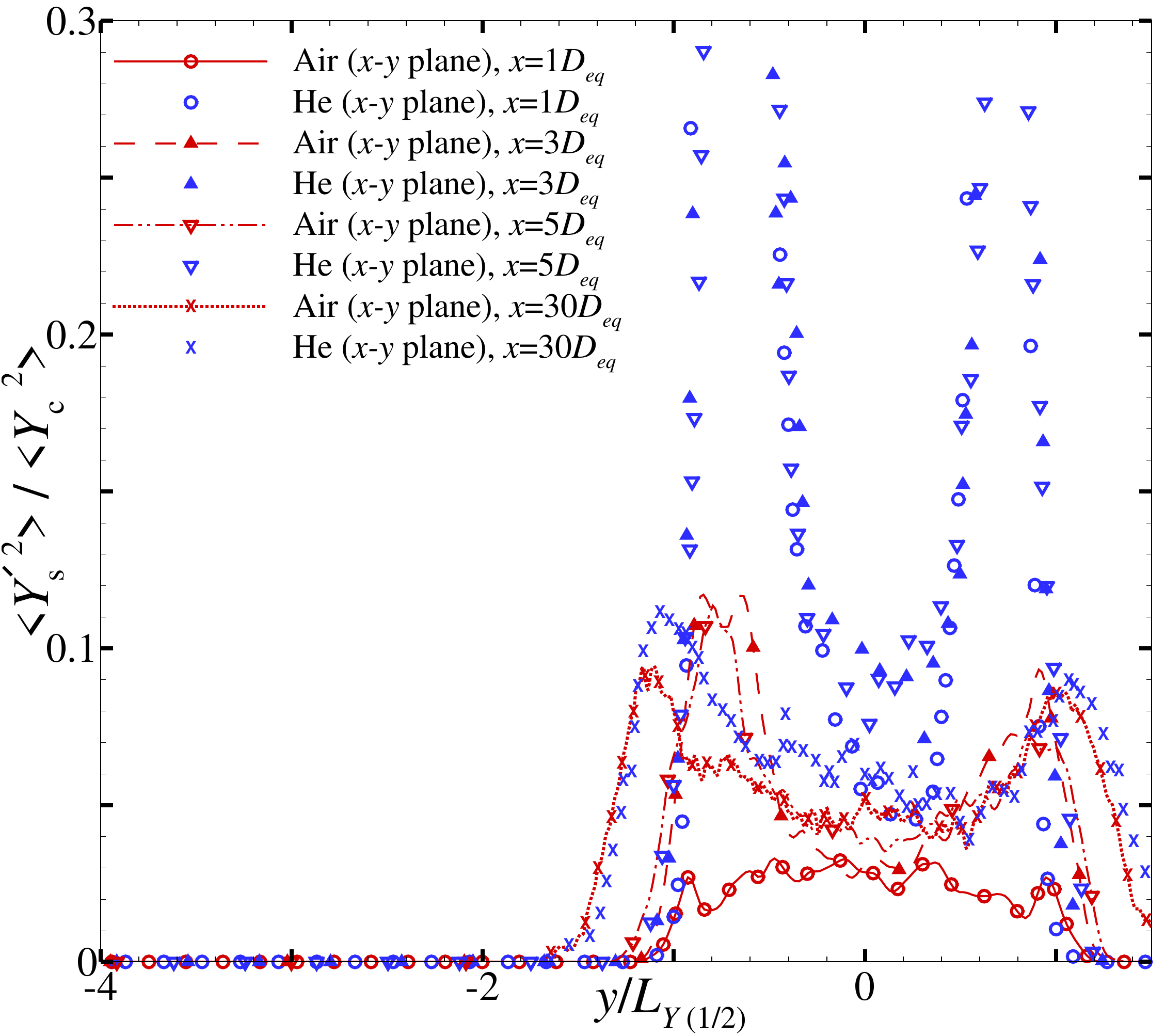}\\
	\raggedright \underline{\textbf{Slot 2, horizontal:}}\\
	a)\includegraphics[scale=0.245]{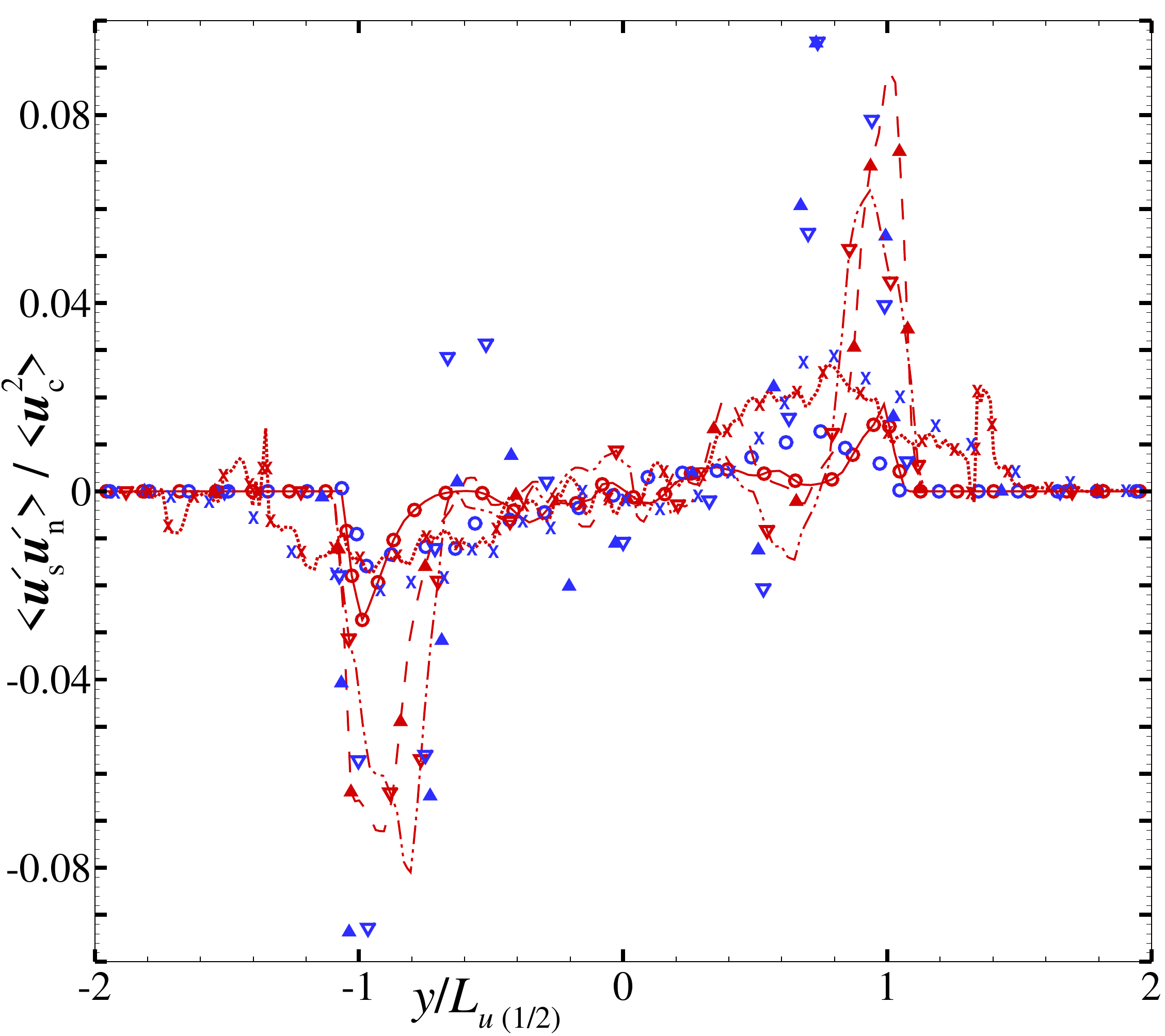} 
	b)\includegraphics[scale=0.245]{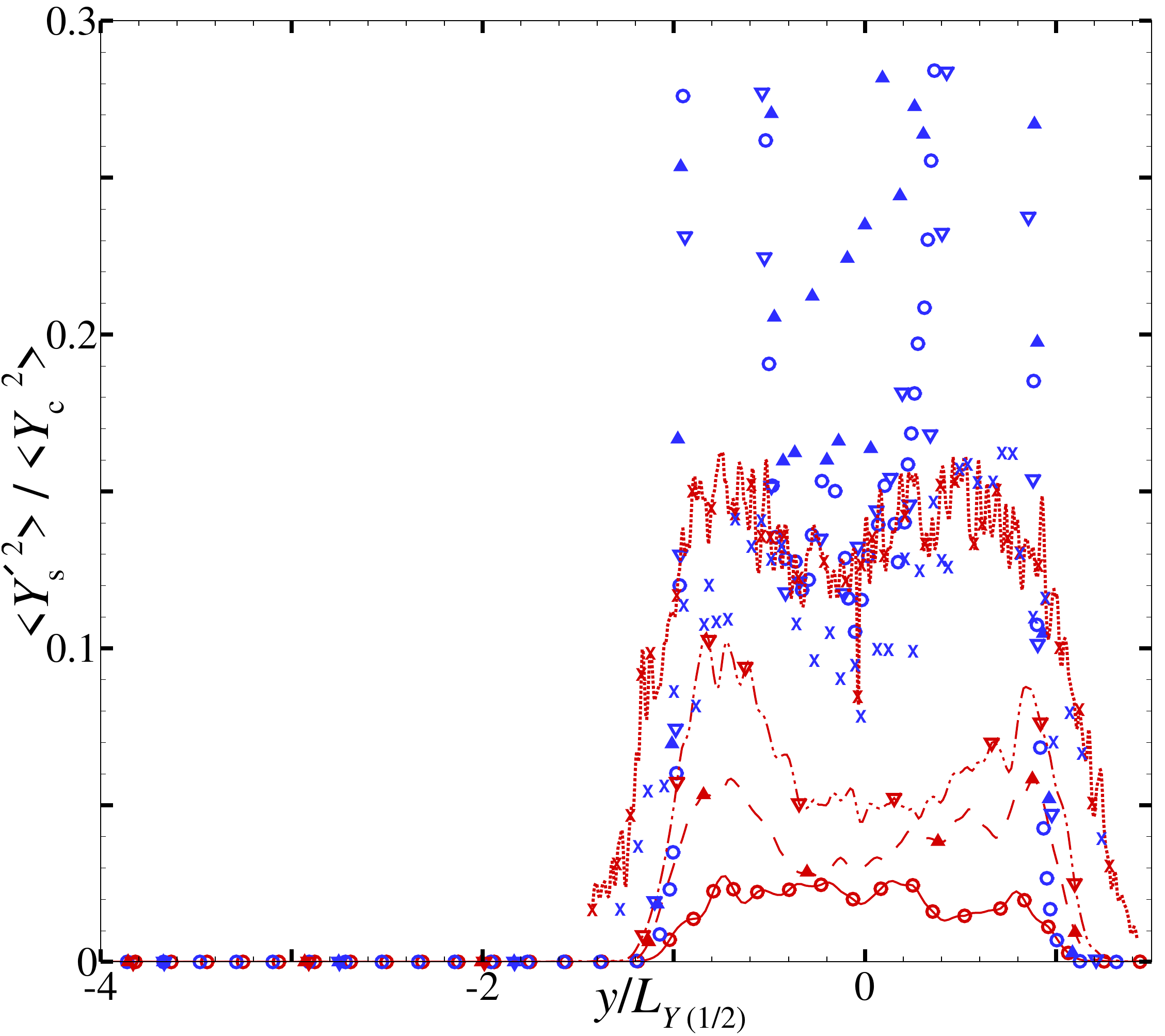}\\
	\caption{a) Normalized time-averaged Reynolds shear stress ($\langle{\boldsymbol{u}^{\prime}_{s} \boldsymbol{u}^{\prime}_{n}}\rangle/ \langle{\boldsymbol{u}^{2}_{c}}\rangle$) and b) concentration variance ($\langle{{Y}^{\prime2}_{s}}\rangle /\langle{{Y}^{2}_{c}}\rangle$) profiles along jet centrelines for air and helium experiments. Here, the profiles are taken at various heights for air and helium measurements in $x$-$y$ planes. Note, the legends in horizontal cases are same as the vertical experiments.}
	\label{fig.Reynolds_Stresses_C_Variances_Slot2}
\end{figure}
\begin{figure}[h!]
	\centering
	\raggedright \underline{\textbf{Slot 3, vertical:}}\\
	a)\includegraphics[scale=0.245]{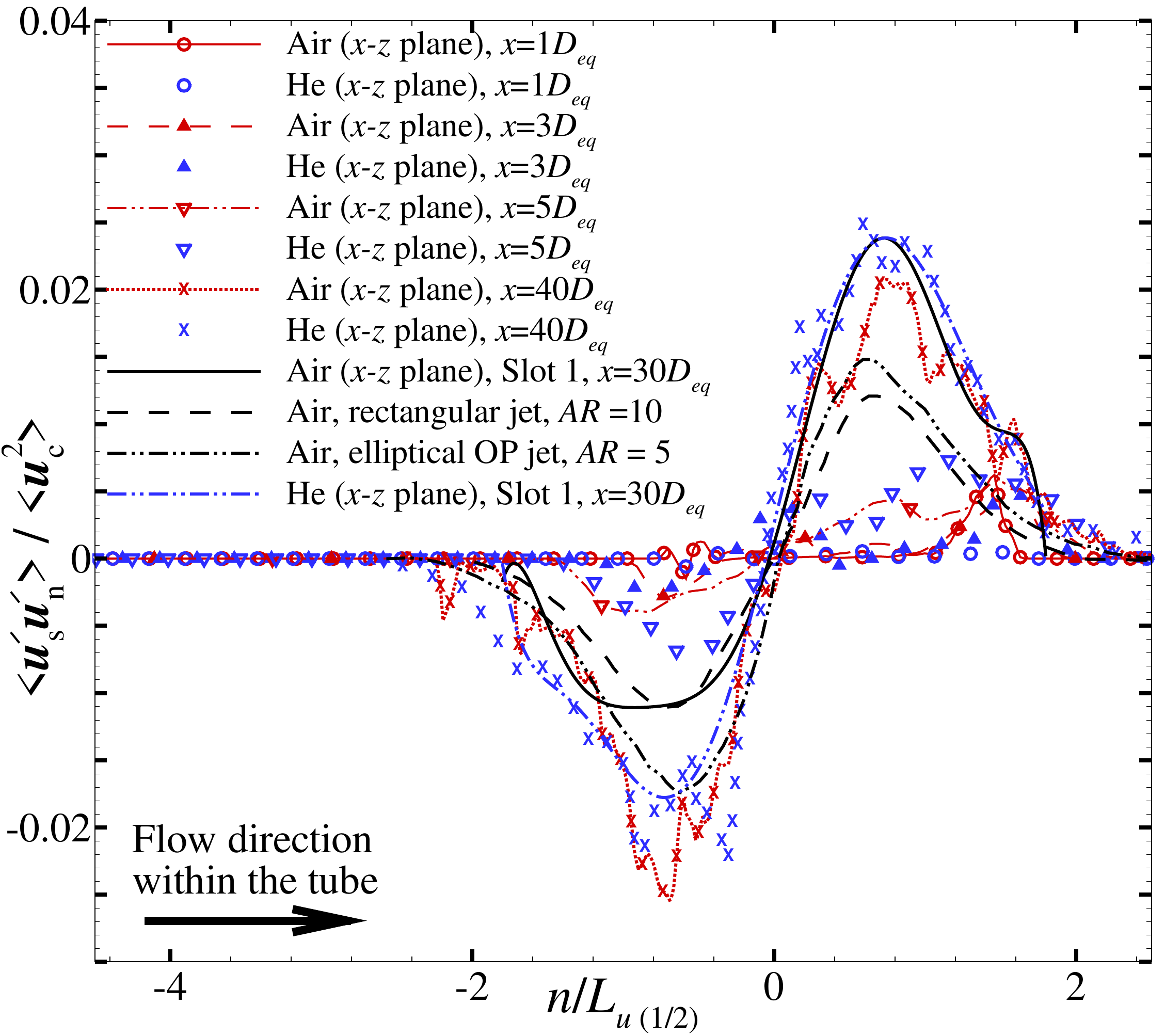} 
	b)\includegraphics[scale=0.245]{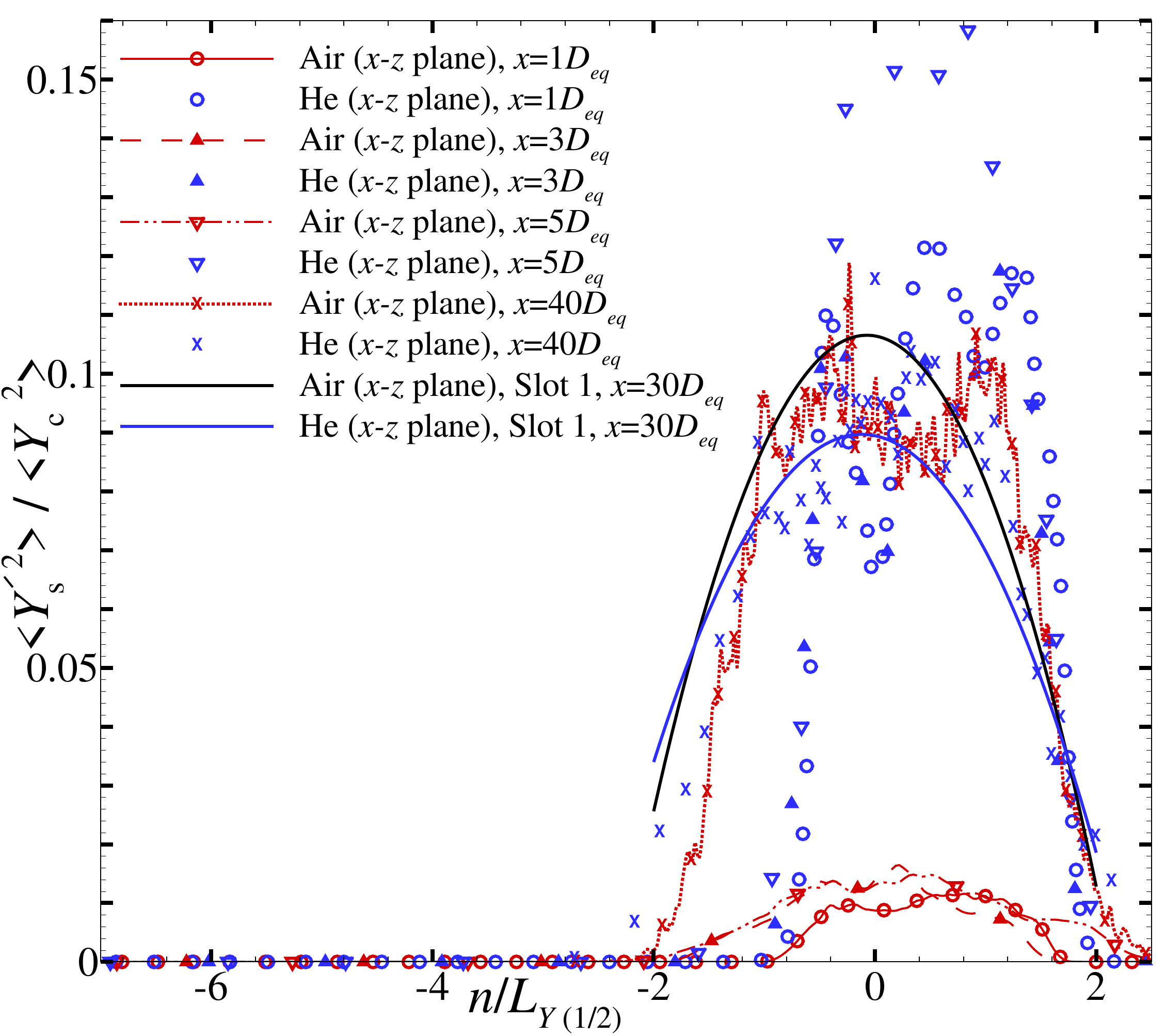}\\
	\raggedright \underline{\textbf{Slot 3, horizontal:}}\\
	a)\includegraphics[scale=0.245]{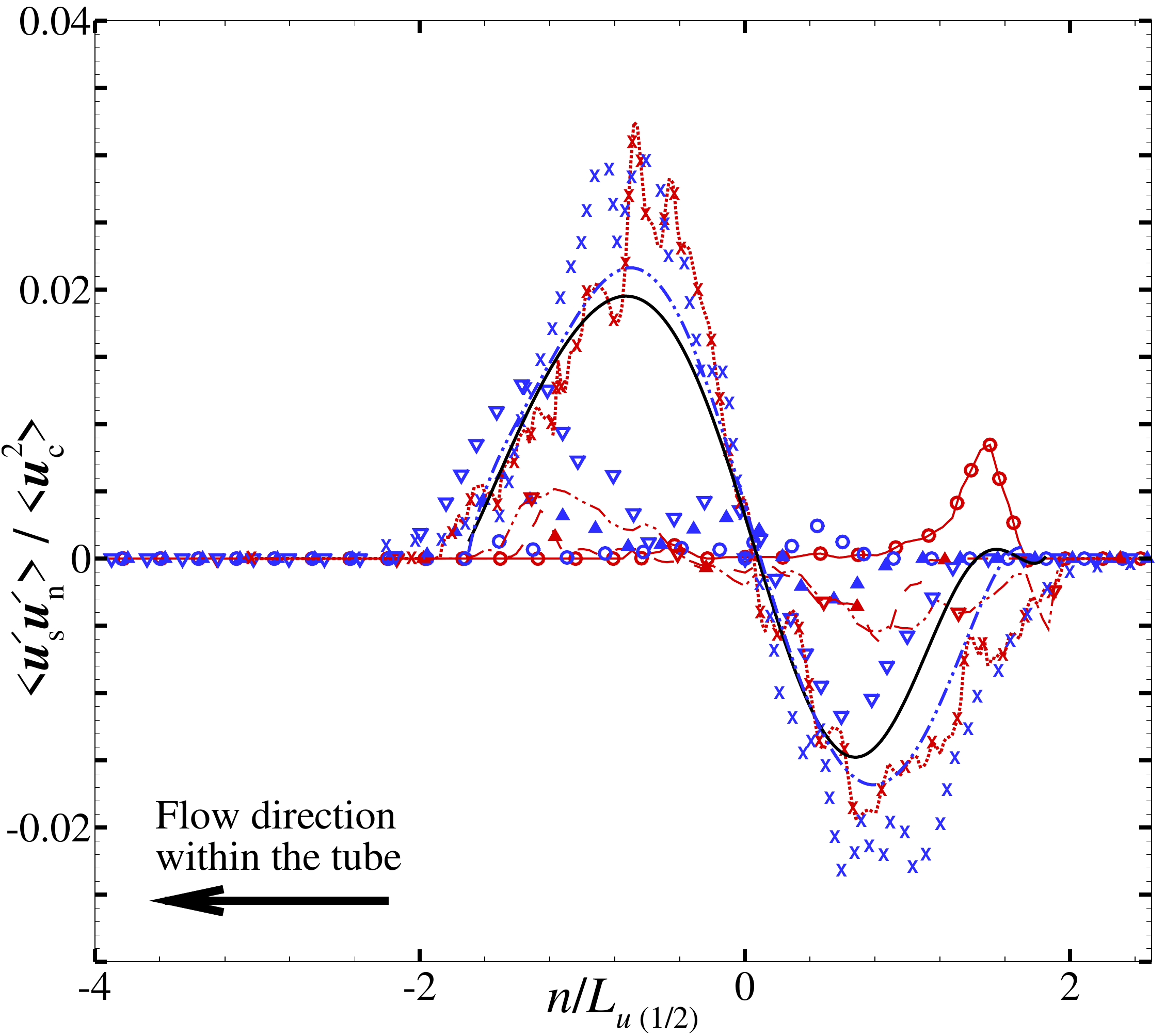} 
	b)\includegraphics[scale=0.245]{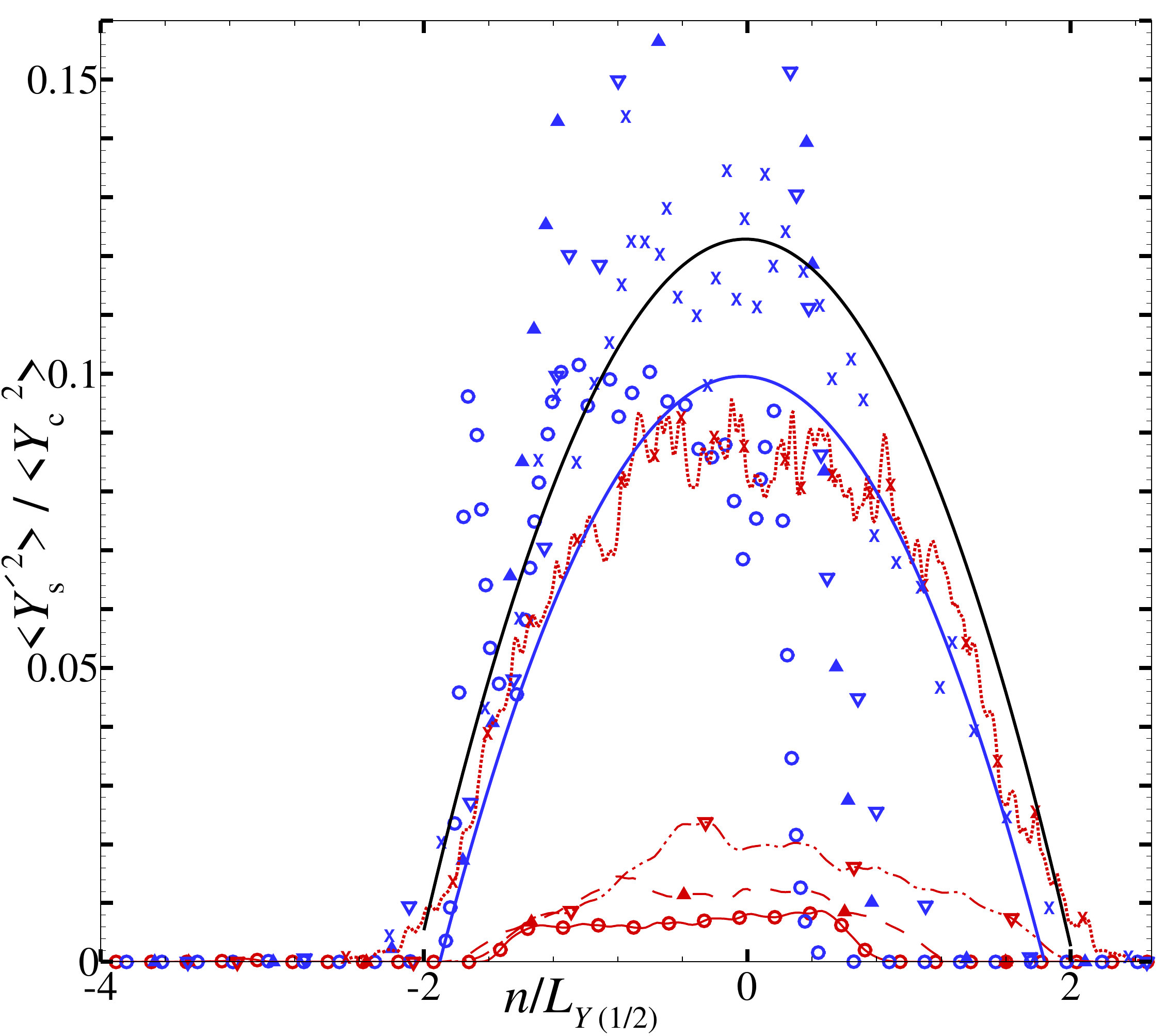}\\
	\caption{a) Normalized time-averaged Reynolds shear stress ($\langle{\boldsymbol{u}^{\prime}_{s} \boldsymbol{u}^{\prime}_{n}}\rangle/ \langle{\boldsymbol{u}^{2}_{c}}\rangle$) and b) concentration variance ($\langle{{Y}^{\prime2}_{s}}\rangle /\langle{{Y}^{2}_{c}}\rangle$) profiles along jet centrelines for air and helium experiments. Here, the profiles are taken at various heights for air and helium measurements in $x$-$z$ planes. Note, the flow direction inside the tube illustrated for both vertical and horizontal cases. Also, the legends in horizontal cases are same as the vertical experiments.}
	\label{fig.Reynolds_Stresses_C_Variances_Slot3}
\end{figure}

Higher order statistics were also acquired for the high-aspect-ratio slot experiments conducted here. The time-averaged Reynolds stress profiles ($\langle{\boldsymbol{u}^{\prime}_{s} \boldsymbol{u}^{\prime}_{n}}\rangle$) obtained from slots 2 \& 3 measurements, for both vertical and horizontal cases, are presented in Figs.\ \ref{fig.Reynolds_Stresses_C_Variances_Slot2} - \ref{fig.Reynolds_Stresses_C_Variances_Slot3} a) respectively. Also shown for comparison, in the $x$-$z$ plane, are vertical \& horizontal slots 1, and vertical sharp-edged rectangular ($AR$=10) and OP elliptical ($AR$=5) jet measurements \cite{Soleimaninia2018IJoHE,Soleimaninia2018Ae,Quinn1992ETaFS203,Quinn1989PoFA1716}. Here, the Reynolds stress quantities have been normalized by the local centreline velocity, $\langle{\boldsymbol{u}^{2}_{c}}\rangle(s)$. For both slot 2 orientations (Fig.\ \ref{fig.Reynolds_Stresses_C_Variances_Slot2} a), air and helium jets exhibit higher Reynolds stress compared to the axisymmetric jets in the range of $x\le25D_{eq}$; a peak magnitude observed at $x=3D_{eq}$ for both air and helium cases. Within this range, higher Reynolds stress was observed inside $\lvert n/L_{1/2}\rvert < 1$, with helium having a slightly higher Reynolds stress compared to air.  
However beyond this range, both air and helium measurements captured the far field self-similar profile well as seen before in slot 1 and the axisymmetric jets \cite{Soleimaninia2018IJoHE,Soleimaninia2018Ae,Panchapakesan1993JoFM225}. 

For both slot 3 orientations (Fig.\ \ref{fig.Reynolds_Stresses_C_Variances_Slot3} a) in the range of $x\le15D_{eq}$, air and helium experiments were found to have lower Reynolds stress compared to slot 1 \& 2 jets. Also, within $x\le5D_{eq}$, higher Reynolds stress was observed beyond $\lvert n/L_{1/2}\rvert < 1$. 
However, beyond $x\ge15D_{eq}$, air and helium experiments captured the far field self-similar profile well, with both helium and air having more Reynolds stress compared to the sharp-edged rectangular and OP elliptical jets \cite{Quinn1992ETaFS203,Quinn1989PoFA1716}. Upon comparison of slot 1 \& 3 jets, the vertical slot 3 cases recovered and matched the far field self-similar profile of slot 1; whereas, the horizontal slot 3 jets have higher a magnitude of the Reynolds stress compared to slot 1. In general, for all slot 2 \& 3 orientations, the magnitudes of Reynolds stress for helium were higher compared to air, more specifically in the near field.

Finally, the normalized mass fraction variance profiles ($\langle{{Y}^{\prime2}_{s}}\rangle /\langle{{Y}^{2}_{c}}\rangle$), obtained from slots 2 \& 3 measurements, for both vertical and horizontal cases, are presented in Figs.\ \ref{fig.Reynolds_Stresses_C_Variances_Slot2} - \ref{fig.Reynolds_Stresses_C_Variances_Slot3} b) respectively. In the $x$-$y$ plane for both slot 2 orientations, helium jet exhibited a semi symmetry saddle-back profile in the whole domain of measurements, whereas air jet showed the same trend only after $x=1D_{eq}$. In both vertical and horizontal orientations, the maximum magnitude of mass fraction variances were observed at $x\sim3D_{eq}$ for both gases. Also within  $x\le5D_{eq}$, the variance peaks were observed inside $\lvert n/L_{1/2}\rvert < 1$. However, both gases return to self-similarity in the far field, with helium having slightly higher magnitude of mass fraction variance compared to air. Also, the magnitude of mass fraction variances were higher in the far field of horizontal cases compared to the vertical orientation. In general, the magnitudes of mass fraction variances for helium were higher compared to air, specially in the near field.

For the vertical slot 3 jets in the $x$-$z$ plane (Fig.\ \ref{fig.Reynolds_Stresses_C_Variances_Slot3} b), the initial development of slot jets had a higher variance of mass fraction to the right of the jet centre (in the $+n$ direction aligned with the flow direction within the tube) within the ranges of $x\le10D_{eq}$ and $x\le6D_{eq}$ for the air and helium jets, respectively. Beyond this range, air jet exhibited semi symmetry saddle-back profile up to $x\le25D$ and after this point the variances recovered the semi Gaussian profile; whereas, helium (beyond $x\ge6D_{eq}$) revealed semi Gaussian profile with a maximum magnitude at $x\sim10D_{eq}$. In contrast, the initial development of horizontal slot 3 jets had a higher variance of mass fraction to the left of the jet centre (in the $-n$ direction) within the same range found for the vertical cases ($x\le10D_{eq}$ and $x\le6D_{eq}$ for the air and helium jets, respectively). It should be noted that the initial shifts in variance profiles to the right or left of the jet centre, are consistent between both vertical and horizontal orientation, and aligned with the direction of flow within the tube (see Fig.\ \ref{fig.Reynolds_Stresses_C_Variances_Slot3}). Beyond the range of $x\le10D_{eq}$, the horizontal air slot 3 jet revealed semi symmetry saddle-backed profile up to $x\le20D$, after this point variances recovered the semi Gaussian profile; whereas, helium (beyond $x\le6D_{eq}$) exhibited a semi Gaussian profile with a maximum magnitude at $x\sim7D_{eq}$. For all slot 3 experiments in the far field, beyond $x\ge30D_{eq}$ jet heights, mass fraction variances revealed self-similar profile. In general, helium jets exhibited higher magnitudes of mass fraction variances compared to air, more specifically in the near field.

\section{Discussion}
\subsection{Aspect Ratio effects}
Velocity and scalar statistical properties obtained from vertical and horizontal high-aspect-ratio slots 2 \& 3 ($AR=10$) were reported in the previous section. The results were  compared to round 3D jet counterparts (slot 1 with $AR=1$) \cite{Soleimaninia2018IJoHE,Soleimaninia2018Ae} and with results available in the literature for high-aspect-ratio rectangular ($AR=10$) and elliptical OP ($AR=2-8$) jets issuing through flat surfaces. It was found that the perpendicular nature of the slot, relative to the direction of flow within the tube, caused a deflection of the jet away from the jet streamwise axis ($x$-axis). This deflection is demonstrated in Fig.\ \ref{fig.Jet_Centerline_Trajectory3}, where the jet centreline trajectories, are presented for all slots 1 \& 3 jets. Significant deflection, in the direction of flow within the tube were observed in both orientations of slot 1 compared to vertical and horizontal slot 3 experiments. Also, the jet trajectory for the horizontal helium slot 1 jet was found to be described by a power law exponent $\sim1.3$, whereas the horizontal helium slot 3 jet's trajectory was found to be described by a nearly linear relation (i.e. power law exponent $\sim1$). Upon these observations, it became clear that the aspect ratio is the main player in governing the deflection angle in 3D jets; an increase in the aspect ratio results in a decrease of deflection angle of jets. However, it is not clear yet how the deflection angle scales with the aspect ratio; more measurements with a broad range of aspect ratios are needed to quantify such scales. 

As previously observed from the time-averaged velocity and molar concentration contours (Figs.\ \ref{fig.Velocity_Concentartion_Contours_Slot2-3_V} - \ref{fig.Velocity_Concentartion_Contours_Slot2-3_H}), slightly shorter potential core length were found for helium ($x\sim3D_{eq}$) and air ($x\sim4D_{eq}$) in slot 3 jets compared to $x\sim4D_{eq}$ and $x\sim5D_{eq}$ for helium \& air slot 1 jets \cite{Soleimaninia2018IJoHE,Soleimaninia2018Ae}, respectively. This is consistent with the very thin vortical structure of high-aspect-ratio slot 3 near the jet exit compared to slot 1, which allows the dynamics of the rolled-up vortical structure to govern the initial development of jet's centre region. In addition, due to the higher entrainment rate that occur in the higher aspect ratio jets, a shorter potential core length and faster decay rate of the jet centreline velocity is expected.

Further interesting observations are made from Fig.\ \ref{fig.Velocity_JetDecay_SpreadingRate3} a), which shows that all slot 1 experiments have a slightly higher velocity decay compared to slot 3 after their potential core region (in the far field beyond $x\le5D_{eq}$), as seen previously in OP elliptical jet experiments \cite{Hussain1989JoFM257,Quinn1989PoFA1716,Quinn2007EJoM-B583}. But within the range of $x\le4D_{eq}$ for helium and $x\le5D_{eq}$ for air, all slot 3 experiments demonstrates slightly higher velocity decay rate compared to slot 1 jets. The higher velocity decay rate in the near field region, is more clear in the results reported previously for the sharp-edged rectangular \cite{Quinn1992ETaFS203} and OP elliptical jets \cite{Hussain1989JoFM257} compared to the current slot jets. The non-linear decay region within the near field, called the ``characteristic decay'' region \cite{Sforza1979JoHT353}, is significantly influenced by the geometry and aspect ratio of a nozzle. That might be the main reason for the higher velocity decay rates are reported in OP elliptical or rectangular jets compared to slot jets. It should be noted that the reported measurements in rectangular and elliptical jets are limited to the flow emerging through flat surface, where the large curvature on the major axis of the vortical structure plays the main role on the initial development of jet. Whereas in 3D jet, due to the curvature of tube surface (where orifice is machined), dynamics of initial vortical structure may depend on both curvatures of major and minor axes. Three dimensional measurement is required to reveal such complicated vortical dynamics in 3D jets.  

The same remarks were also observed from mass fraction decay profiles in Fig.\ \ref{fig.Concentration_JetDecay_SpreadingRate3} a), where slower decay rate was observed in centreline mass fraction for slot 3 in the far field compared to slot 1. The similar observation can also be made from the jet velocity and the scalar width growth presented in Figs.\ \ref{fig.Velocity_JetDecay_SpreadingRate3} - \ref{fig.Concentration_JetDecay_SpreadingRate3} b), where all slot 1 experiments experience much higher growth rates in both velocity and scalar fields compared to the slot 3 in the far field. But within the range of $x\sim\le8D_{eq}$ faster growth were observed for all slot 3 experiments in the major plane ($x$-$z$ plane). In general, upon comparison of only two sets of slot aspect ratio ($AR$= 1 vs $AR$= 10), higher aspect ratio resulted in a decrease on the jet centreline decay rates and a growth rates on both velocity and scalar fields in the 3D jets, more specifically in the far field.

Upon comparison of the far field's axial evolution of turbulent intensities and unmixedness of slot 3 and 1 along the jet centreline (Figs.\ \ref{fig.Velocity_fluctuation_rms_Slot1-3}-\ref{fig.Concenration_fluctuation_rms_Slot1-3}), all slot 3 jets recovered the asymptotic value of those previously reported for slot 1 jets \cite{Soleimaninia2018IJoHE,Soleimaninia2018Ae}. These remarks indicate that the far field's development of turbulent intensities and unmixedness in 3D slot jets, same as other jet flows \cite{Quinn1992ETaFS203,Hussain1989JoFM257}, are independent of the slot aspect ratio.

Asymmetric flow pattern were previously reported in the lower side of slot 1 jet's centreline (in the opposite direction of flow within the tube), where more velocity and mass fraction spreading were exhibited near the tail ends of radial profiles ($1\pm<(n/L_{1/2})$) within the range of $x\le5D_{eq}$ \cite{Soleimaninia2018IJoHE,Soleimaninia2018Ae}. It was previously found that the flow separation at the exit of the orifice and the resulting deficits along with the curvature of the tube, relative to the size of the orifice, have contributed to the asymmetric flow patterns observed in the near field of slot 1 jet. Although the flow separation is still persists in all slot 3 jets due to the un-parallelism nature of 3D jet flows, the higher aspect ratio ($AR=10$) reduced the effects of flow separation and resulting deficits; as a result, the asymmetric patterns toward the edge of jet boundaries were not observed in slot 3 jets. However, the asymmetric patterns were shifted towards the jet centreline ($(n/L_{1/2})<\pm1$) within the range of $x\le5D_{eq}$ in slot 3 jets, as observed in both radial velocity and scalar profiles (Fig.\ \ref{fig.Vs_Cs_Slot3}).

\subsection{Buoyancy effects}
As previously observed from the jet centreline trajectories (Fig.\ \ref{fig.Jet_Centerline_Trajectory3}), all slot 3 jets experienced a deflection in the jet away from its streamwise axis ($x-$axis). This deflection was also previously reported in slot 1 jets \cite{Soleimaninia2018IJoHE,Soleimaninia2018Ae}, and the major player was found to be the perpendicular streamwise axis of the slot, relative to the flow direction within the tube. As a result, both vertical and horizontal air (heavier gas) slot 3 jets were found to deflect more than their helium (lighter gas) jet counterparts after $x\sim3D_{eq}$ and $x\sim4D_{eq}$, respectively. Even though other factors might play a role, the higher deflection of the vertical air jets compared to the vertical helium jets is consistent with the relative strength of the corresponding vertical buoyancy force. However, buoyancy force was clearly the main contributor in the significant deflection of horizontal helium case from the horizontal $x-$axis compared to vertical helium jet in the far field ($x\ge17D_{eq}$); where such effect were not observed in neither of orientations for air slot 3 jet. 

Further interesting observations are made upon comparison of the axial evolution of the turbulent intensities and the unmixedness of slot 3 along the jet centreline (Figs.\ \ref{fig.Velocity_fluctuation_rms_Slot1-3}-\ref{fig.Concenration_fluctuation_rms_Slot1-3}). It was found that helium jets reached the asymptotic values more closer to the orifice at $x=15D_{eq}$ and $x=17D_{eq}$ compared to air jets at $x\ge20D_{eq}$ and $x=27D_{eq}$ in tangential and orthogonal turbulent intensities, respectively. In axial evolution of unmixedness, while all slot 3 jets  would recovered the asymptotic value of $\sim33\%$ at some distance from the orifice; but the initial increase of the mass fraction fluctuation intensities in the near field occurred more rapidly with lower density gases. These remarks would might be solely associated with the buoyancy effect, which is become the major player in helium jets compared to air jets. It should be noted that such buoyancy effect were also previously observed in slot 1 measurements \cite{Soleimaninia2018IJoHE,Soleimaninia2018Ae}. Other parameters such as co-flow, initial conditions, Reynolds number, and measurement uncertainty could also influence the evolution of the turbulent intensities and the unmixedness \cite{Pitts1991EiF125}. However, their effects are  negligible since the similar experimental system and parameters have been used in current measurements.

\section{Conclusions}
In this study, simultaneous velocity and concentration measurements were conducted in order to investigate vertical and horizontal turbulent jets, of varying gas densities and Reynolds numbers, issuing from a high-aspect-ratio slots machined into the side of a round tube. Two slots with the same aspect ratio ($AR=10$), as possible crack geometries were considered in this study; although their configurations was aligned parallel to and perpendicular with the direction of flow inside the tube. The fluids considered were air and helium. The results were compared to the previous studies of the vertical and horizontal 3D round jets (slot 1), issuing from the same experimental system and pipeline geometry but with aspect ratio of $AR=1$ \cite{Soleimaninia2018IJoHE,Soleimaninia2018Ae}.  Comparison was also made to the non-circular jets, OP elliptical and rectangular jets issuing through flat surfaces \cite{Quinn1992ETaFS203,Quinn2007EJoM-B583,Hussain1989JoFM257}.

By considering flow emerging through a slot, machined on the side of a tube, it was found that the flow arrangement caused a significant deflection from the axis normal to the orifice (streamwise axis, $x$) as was previously observed in vertical and horizontal slot 1 jets with similar arrangement \cite{Soleimaninia2018IJoHE,Soleimaninia2018Ae}.  For the horizontal helium slot 3 jet, this deflection was found to be initially governed by the density of the gas in the near field, and it has experienced further deflection due to the buoyancy in the far field regardless of the high Froude number ($Fr= 2.8\times 10^6$). Such deviation due to the buoyancy in the far field was also previously observed for the horizontal helium slot 1. However, such deviation for the horizontal helium slot 3 jet was found to be well described by a nearly linear relation (i.e. power law exponent $\sim1$), whereas the horizontal helium slot 1 jet was previously found to have a power law exponent $\sim1.3$. This difference may arise due to the higher aspect ratio in slot 3 ($AR=10$) compared to $AR=1$ in slot 1. Although other factors might contribute, the lower deflection of all slot 3 jets compared to slot 1 jets is consistent with relative importance of aspect ratio effect in development of 3D jet flows. In general, higher aspect ratio resulted in less deviation for all jets from their streamwise axis ($x$-axis) and also reduced the order of power law exponent from $\sim1.3$ to $\sim1$ in the centreline trajectory correlations of horizontal helium slots 1 \& 3, respectively. Upon comparison of the only two sets of aspect ratios ($AR$=1 \& $AR$=10), it also can be generalized that an increase in aspect ratio causes a reduction in the jet centreline decay rates and growth rates on both velocity and scalar fields in the 3D jets, more specifically in the far field. In contrast, the axial evolutions of turbulent intensities and unmixedness of slot 3 jets were found to be independent of the aspect ratio effects, and were reached the similar asymptotic values which are previously reported for slot 1 jets. In addition, a higher aspect ratio reduced the effects of the flow separation at the orifice exit and the resulting deficits in the near field development of slot 3 jets; consequently the asymmetric pattern, which was previously observed close to the edge of slot 1 jet's boundary ($1\pm<(n/L_{1/2})$), was shifted toward the slot 3 jet's centreline ($(n/L_{1/2})<\pm1$) within the range of $x\le5D_{eq}$ in both radial velocity and scalar profiles. Finally, significant discrepancies were found between the evolution of current realistic jets issuing from curved surfaces and those conventional high-aspect-ratio jets originating from flat surfaces. Should the development of realistic jet for hydrogen behave the same, caution is required when using conventional jet assumptions to predict the correct dispersion, velocity and concentration decay rates and, consequently, the extent of flammability envelope of a realistic gas leaks.

\section{Acknowledgments}
The authors would like to acknowledge the support of the Natural Sciences and Engineering Research Council of Canada (NSERC).

\section*{References}

\bibliography{Literature_Review}

\end{document}